\documentclass[preprintnumbers,prd,singlecolumn,floatfix,preprintnumbers,amsmath,amssymb,nofootinbib,superscriptaddress]{revtex4}
\usepackage{hyperref}

\usepackage{times,graphicx,amssymb, amsmath, natbib}
\usepackage[T1]{fontenc}

\usepackage{epsfig}

\newcommand{\beq}{\begin{equation}}
\newcommand{\eeq}{\end{equation}}

\newcommand{\ber}{\begin{eqnarray}}
\newcommand{\eer}{\end{eqnarray}}

\def\beq{\begin{equation}}
\def\eeq{\end{equation}}

\def\ber{\begin{eqnarray}}
\def\eer{\end{eqnarray}}

\pdfoutput=1 

\begin{document}

\title{\boldmath Probing dark energy using convergence power spectrum and bi-spectrum}

\author{Bikash R. Dinda}
\email{bikash@ctp-jamia.res.in} 
\affiliation{Centre for Theoretical Physics, Jamia Millia Islamia, New Delhi-110025, India}

\begin{abstract}
Weak lensing convergence statistics is a powerful tool to probe dark energy. Dark energy plays an important role in the structure formation and the effects can be detected through the convergence power spectrum, bi-spectrum etc. One of the most promising and simplest dark energy models is the $ \Lambda $CDM. However, it is worth investigating different dark energy models with evolving equation of state of the dark energy. In this work, detectability of different dark energy models from $ \Lambda $CDM model has been explored through convergence power spectrum and bi-spectrum.
\end{abstract}

\maketitle
\date{\today}

\section{Introduction}

A combination of different cosmological observations now points towards a concordance model for our Universe where $1/3$rd of the energy budget of the Universe is in non-relativistic matter comprising baryons and dark matter and the rest $2/3$rd is cosmological constant $\Lambda$ with a constant equation of state $w = -1$ (\cite{Planck}). This model, as popularly called $\Lambda$CDM model, although is consistent with majority of cosmological measurements, some recent observations suggest potentially important discrepancies in $\Lambda$CDM model (\cite{inctny1} - \cite{inctny4}). This is in addition to the theoretical problems in $\Lambda$CDM, e.g, the fine tuning and cosmic coincidence problem (\cite{fntn}).

This is why the major goal for upcoming high precision cosmological experiments is to determine the evolution of the equation of state for the dark energy at the level of percentage level accuracy. Among these, experiments related to weak lensing measurements are particularly promising in determining the nature of dark energy because of the high sensitivity of weak lensing effect on both the background evolution of the Universe as well as on the growth of structures (\cite{WLp1} - \cite{WLp14}).

Weak lensing is the distortions of galaxy images due to gravitational bending of light by the intervening large scale structures along the photon propagation. The measurement of weak lensing around massive halos was first measured in nineties \cite{tyson, brainerd}; but the first detection of weak lensing by large scale structures was done independently by four groups in 2000 (\cite{WLSF1} - \cite{WLSF4}). After that weak lensing has become one of the most accurate probes for our observable Universe.

The main advantage of using weak lensing as cosmological probe compared to other probes related to large scale structures is due to the fact that it solely depends on the underlying total matter distribution and hence one can avoid the complicated bias modeling. The other advantage of using weak lensing as a cosmological probe is due to the relatively straightforward measurement of the galaxy shear which can be observed in millions of galaxies in latest surveys. Correlation of galaxy shear across the sky together with the redshift information of individual galaxies provides a 3-dimensional information of our Universe which is a powerful probe for dark energy.

Given the current and future surveys like Dark Energy Survey (DES \cite{des}), Large Synoptic Survey Telescope (LSST \cite{lsst}), Euclid (\cite{euclid}) and the Wide-Field Infrared Survey Telescope (WFIRST \cite{wfirst}), the prospects of accurately measure the evolution of dark energy density and its equation of state using weak lensing is extremely bright.

In this paper, we study the prospects of distinguishing any individual dark energy model from $\Lambda$CDM model using weak lensing power spectrum and bi-spectrum. We consider the parametrization of dark energy equation of state e.g the CPL \cite{CPL1, CPL2} and GCG \cite{GCG} parametrization as well as thawing class of quintessence model for dark energy with power-law potentials (\cite{ThQ1} - \cite{ThQ5}) and compare their weak lensing signal with that from $\Lambda$CDM model. This gives a broad picture of how far we can expect to distinguish different dark energy models from $\Lambda$CDM using weak lensing.

The paper is organised as: in section II background evolution has been discussed; in section III perturbation in the matter has been studied with the above mentioned dark energy models using Newtonian perturbations; in section IV linear solutions to the perturbation and linear matter power spectrum have been studied; In section V second order solutions to the perturbation and tree-level matter bi-spectrum have been studied; in section VI corresponding convergence power spectrum and bi-spectrum have been discussed; and finally in section VII conclusion has been given.

\section{Background evolution with dark energy models}

Considering flat Friedman-Robertson-Walker (FRW) background of the Universe with two components, non-relativistic matter (baryons + dark matter) and dark energy, the Hubble parameter, $ H $ can be expressed as

\begin{equation}
H^{2} = H_{0}^{2} \Big{[} \Omega_{m}^{(0)} a^{-3} + (1-\Omega_{m}^{(0)}) \exp[-3(\int_{1}^{a} da^{'} \frac{1+w(a^{'})}{a^{'}})]\Big{]},
\label{eq:Hubble}
\end{equation}

\noindent
where $ a $ is the scale factor, $ w(a) $ is the general time dependent equation of state of dark energy, $ \Omega_{m}^{(0)} $ and $ H_{0} $ are the present day matter density parameter and Hubble parameter respectively. Different dark energy models have different equations of state (e.o.s), which is defined as $ w(a) \equiv \frac{\bar{p}_{q} (a)}{\bar{\rho}_{q} (a)} $, where $ \bar{\rho}_{q} $ and $ \bar{p}_{q} $ are background energy density and pressure for dark energy respectively. We consider three types of dark energy models: (I) models where the e.o.s for the dark energy is given by Chevallier-Polarski-Linder (CPL) parametrization, (II) thawing class of minimally coupled canonical scalar field models (thawing quintessence) and (III) models where the e.o.s for the dark energy is given by the GCG (generalized Chaplygin gas) parametrization. For CPL and GCG parametrizations, e.o.s $ w(a) $ have analytical expressions whereas, for quintessence models, background quantities have to be evaluated numerically. Let us briefly discuss these models below.

\section*{(I) CPL parametrization:}

In CPL parametrization, the e.o.s of dark energy is given by

\begin{equation}
w(a) = w_{0} + w_{a} (1-a),
\label{eq:CPLw}
\end{equation}

\noindent
where $ w_{0} $ and $ w_{a} $ are two model parameters. $ w_{0} $ represents the present day e.o.s of dark energy whereas $ w_{a} $ gives its evolution with time \cite{CPL1, CPL2}. For any time, $ w < -1 $ and $ w > -1 $ correspond to phantom and non-phantom behaviours of dark energy respectively. For $ w_{0} < -1 $ and $ w_{a} < 0 $ ($ w_{0} > -1 $ and $ w_{a} > 0 $), dark energy shows phantom (non-phantom) behaviour for all time. The special case $ w_{0} = -1 $ and $ w_{a} = 0 $ corresponds to the exact $ \Lambda$CDM model.

\section*{(II) Quintessence:}

The Lagrangian density for a minimally coupled scalar field can be written as (\cite{ThQ1} - \cite{ThQ5})

\noindent
\begin{equation}
\mathcal{L} = \dfrac{1}{2} (\partial^{\mu}\phi) (\partial_{\mu}\phi) - V(\phi),
\end{equation}

\noindent
where $ \phi $ is the field and $ V $ is the potential. The background energy density and pressure of the quintessence become

\begin{eqnarray}
\bar{\rho}_{q} = \frac{1}{2} \dot{\phi}^{2} + V(\phi), \nonumber\\
\bar{p}_{q} = \frac{1}{2} \dot{\phi}^{2} - V(\phi),
\end{eqnarray} 

\noindent
respectively, where overdot represents derivative with respect to the cosmic time $ t $. The equation of motion for the scalar field is given by

\begin{equation}
\ddot{\phi} + 3 H \dot{\phi} + V_{\phi} = 0,
\end{equation}

\noindent
where subscript $ \phi $ is the derivative w.r.t the field $ \phi $. To study the background evolution equations, it is a standard procedure to define few dimensionless quantities as given below:

\begin{eqnarray}
x &=& \frac{\Big{(} \dfrac{d \phi}{d N} \Big{)}}{\sqrt{6} M_{Pl}}, \hspace{1 cm}  y = \frac{\sqrt{V}}{\sqrt{3} H M_{Pl}}, \nonumber\\
\lambda &=& - M_{Pl} \Big{(} \frac{V_{\phi}}{V} \Big{)}, \hspace{1 cm}  \Gamma = V \Big{(} \frac{V_{\phi \phi}}{V_{\phi}^{2}} \Big{)}, \nonumber\\
\Omega_{q} &=& x^{2} + y^{2}, \hspace{1 cm} w = \gamma - 1 = \frac{x^{2} - y^{2}}{x^{2} + y^{2}},
\end{eqnarray} 

\noindent
where $ N =lna $ is the e-folding, $ M_{Pl} $ is the reduced Planck mass and $ \Omega_{q} = 1 - \Omega_{m} $ is the energy density parameter of the scalar field. Now, the background evolution can be studied through an autonomous system of equations given by

\begin{eqnarray}
\dfrac{d \gamma}{d N} &=& 3 \gamma (\gamma - 2) + \sqrt{3 \gamma \Omega_{q}} (2 - \gamma) \lambda, \nonumber\\
\dfrac{d \Omega_{q}}{d N} &=& 3 (1 - \gamma) \Omega_{q} (1 - \Omega_{q}), \nonumber\\
\dfrac{d \lambda}{d N} &=& \sqrt{3 \gamma \Omega_{q}} \lambda^{2} (1 - \Gamma).
\end{eqnarray}

\noindent
To solve the above-coupled differential equations, we consider the thawing class of initial conditions where the dark energy is initially frozen at the flat part of the potential (similar to what happens in the early inflationary epoch). This gives $ \gamma_{in} \thickapprox 0 $ at the initial time (initial redshift, $ z_{in} = 1100 $ is considered in our calculations). In the subsequent calculations, we assume $ \gamma_{in} = 10^{-10} $ and the results are not sensitive to the values of $ \gamma_{in} $ as long as $ \gamma_{in} \ll 1 $. The initial value for $\Omega_{q}$ is chosen such that one gets its required value at present. The initial value of $\lambda_{in}$ controls the equation of state of the scalar field at present. For $\lambda_{in} << 1$, the equation of state of the scalar field is always close to $w = -1$ (the cosmological constant); for higher values of $\lambda_{i}$, the equation of state starts deviating from the cosmological constant behaviour. The quintessence models with this type of initial condition is called thawing class of quintessence models where the e.o.s of the dark energy is $ -1 $ initially in early matter dominated era and with the expansion of the Universe the slope of e.o.s increases slowly and at late times the e.o.s becomes greater than $ -1 $. We fix $ \lambda_{in} = 0.7 $ for all the subsequent calculations.

\section*{(III) GCG parametrization:}

In GCG parametrization, the e.o.s of dark energy is given by \cite{GCG}

\begin{equation}
w(a) = - \frac{A_{s}}{A_{s}+(1-A_{s})a^{-3(1+\alpha)}},
\label{eq:GCGw}
\end{equation}

\noindent
where $ \alpha $ and $ A_{s} $ are two model parameters. The special case $ A_{s} =1 $ corresponds to $ \Lambda$CDM model. It is interesting to notice that the GCG parametrization incorporates both the thawing and tracker behaviors of dark energy \cite{GCG}. Thawing model corresponds to $1+\alpha <0$ whereas the tracker model (where e.o.s of the dark energy initially in matter dominated era mimics the background with a value nearly $ 0 $ and with the expansion of the Universe it decreases towards $ -1 $ value at very late times and finally freezes to $ w \approx -1 $ in future) corresponds to $1+\alpha >0$ (\cite{ThQ1} - \cite{ThQ3}).

\noindent

\begin{figure}[tbp]
\centering
\includegraphics[width=.45\textwidth]{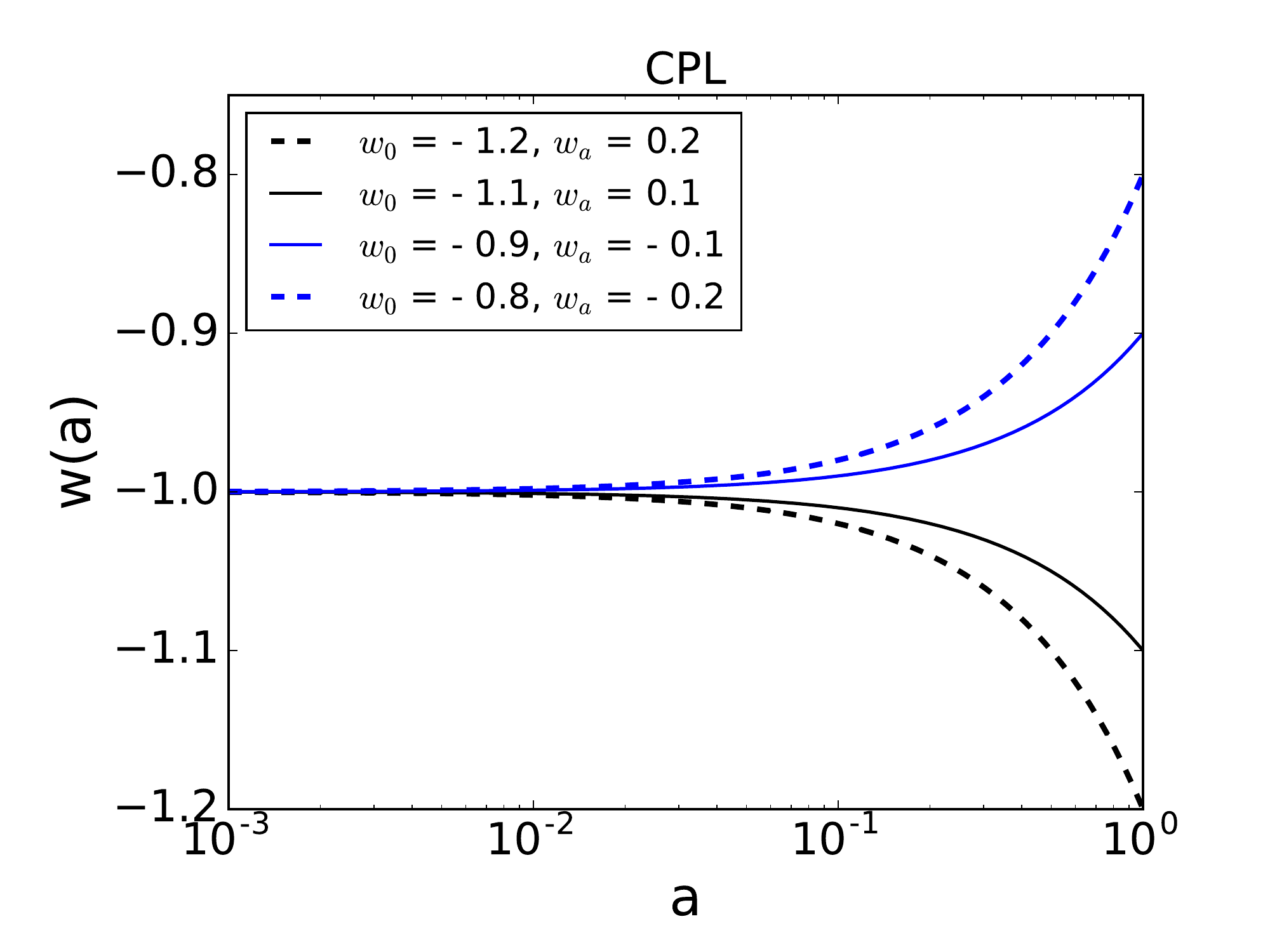}
\includegraphics[width=.45\textwidth]{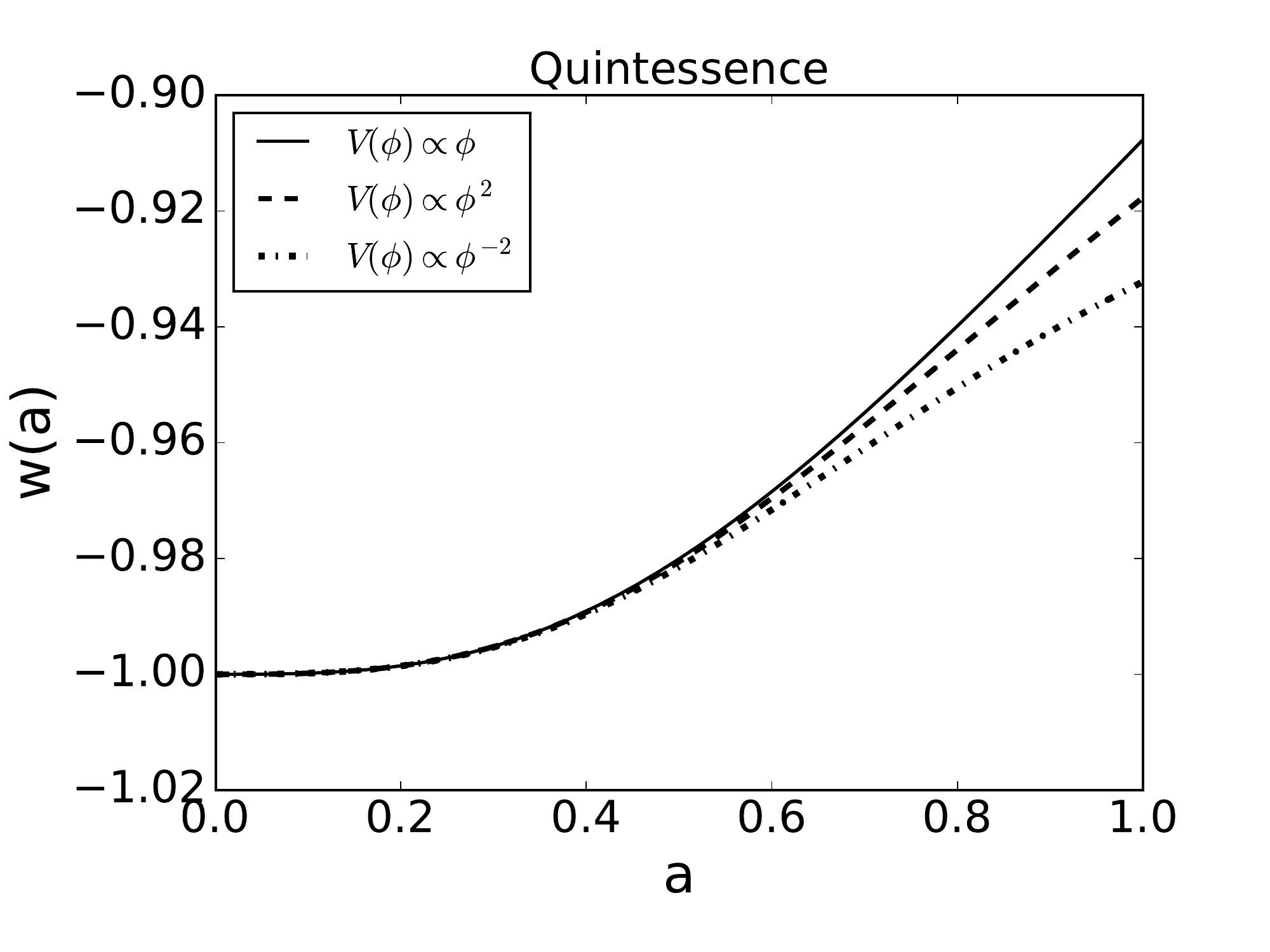}\\
\includegraphics[width=.45\textwidth]{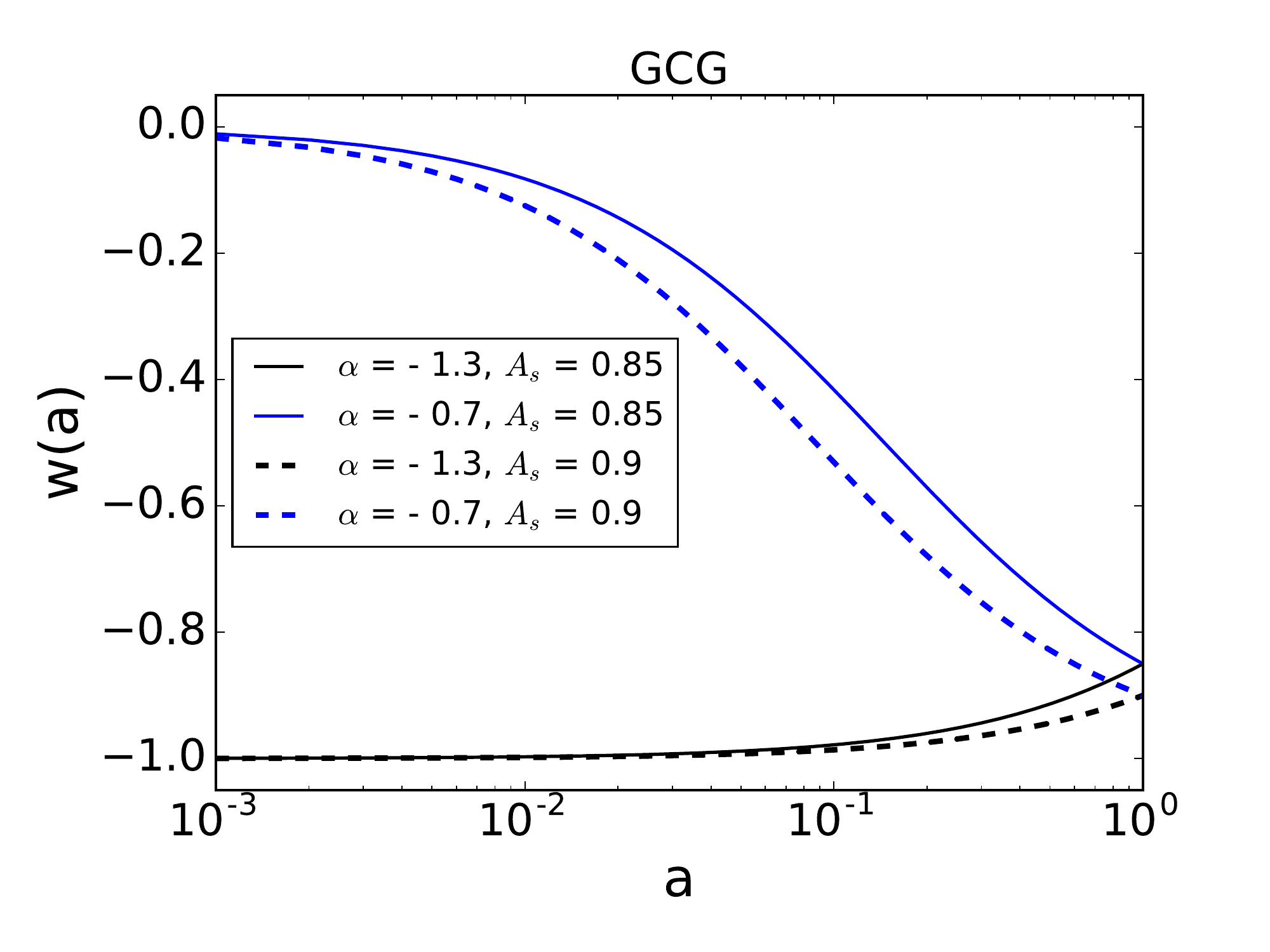}
\caption{\label{fig:eos} $ w(a) $ vs. $ a $ plots for different dark energy models.}
\end{figure}

\section*{Background evolution:}

\noindent
The initial conditions of the background quantities are chosen such that for all the dark energy models at present $ \Omega_{m}^{(0)} = 0.308 $. The initial condition for $ \Omega_{q} $ is fixed by this normalisation for thawing quintessence models and we consider linear, squared and inverse-squared potentials. For CPL parametrization four models are chosen given by $ w_{0} = -1.2 $, $ w_{a} = 0.2 $; $ w_{0} = -1.1 $, $ w_{a} = 0.1 $; $ w_{0} = -0.9 $, $ w_{a} = -0.1 $ and $ w_{0} = -0.8 $, $ w_{a} = -0.2 $. The chosen models for GCG parametrization are $ \alpha = -1.3 $, $ A_{s} = 0.85 $; $ \alpha = -0.7 $, $ A_{s} = 0.85 $; $ \alpha = -1.3 $, $ A_{s} = 0.9 $ and $ \alpha = -0.7 $, $ A_{s} = 0.9 $.

The equation of states of the above mentioned models for all three types of dark energy models have been plotted in Fig.~\ref{fig:eos} to show how the equation of states evolve with expansion of the Universe. The four models of the CPL parametrization are chosen such a way that the e.o.s of the dark energy is $ -1 $ initially in matter dominated era ($ a<<1 $) and for the models $ w_{0} = -1.2 $, $ w_{a} = 0.2 $ and $ w_{0} = -1.1 $, $ w_{a} = 0.1 $ (which are phantom models) e.o.s slowly decreases with expansion of the Universe whereas for the models $ w_{0} = -0.9 $, $ w_{a} = -0.1 $ and $ w_{0} = -0.8 $, $ w_{a} = -0.2 $ (which are non-phantom models) e.o.s slowly increases with expansion of the Universe. The model $ w_{0} = -1.2 $, $ w_{a} = 0.2 $ has larger negative slope compared to the model $ w_{0} = -1.1 $, $ w_{a} = 0.1 $ and the model $ w_{0} = -0.8 $, $ w_{a} = -0.2 $ has larger positive slope compared to the model $ w_{0} = -0.9 $, $ w_{a} = -0.1 $. As thawing class of quintessence models have been considered the equation of states of all three quintessence models start from $ -1 $ initially and slowly slopes increase towards higher values at present time. Note that thawing class of quintessence models are always non-phantom. Although the differences between three quintessence models are very small but it is clear to see that the e.o.s has the highest slope for linear potential whereas the lowest slope for inverse-squared potential. The present day e.o.s of the dark energy becomes nearly $ -0.9 $, $ -0.92 $ and $ -0.93 $ for the linear, squared and inverse-squared potentials respectively. In GCG parametrization two thawing ($ \alpha = -1.3 $, $ A_{s} = 0.85 $ and $ \alpha = -1.3 $, $ A_{s} = 0.9 $) and two tracker models ($ \alpha = -0.7 $, $ A_{s} = 0.85 $ and $ \alpha = -0.7 $, $ A_{s} = 0.9 $) have been considered. So, the two thawing models corresponding to $ \alpha = -1.3 $ have the similar behaviour as the thawing quintessence models but the tracker models have different behaviour where e.o.s has negative slopes. The e.o.s of the dark energy for two tracker models starts from nearly  $ 0 $ initially at early matter dominated era (by which the dark energy tracks the background initially) and at late times e.o.s decreases towards lower values. At present time for all the GCG models e.o.s of the dark energy is fixed to a value according to the chosen $ A_{s} $ value. Note that the GCG models are also always non-phantom like quintessence models. The above mentioned models have been considered throughout all the subsequent sections.

\begin{figure}[tbp]
\centering
\includegraphics[width=.45\textwidth]{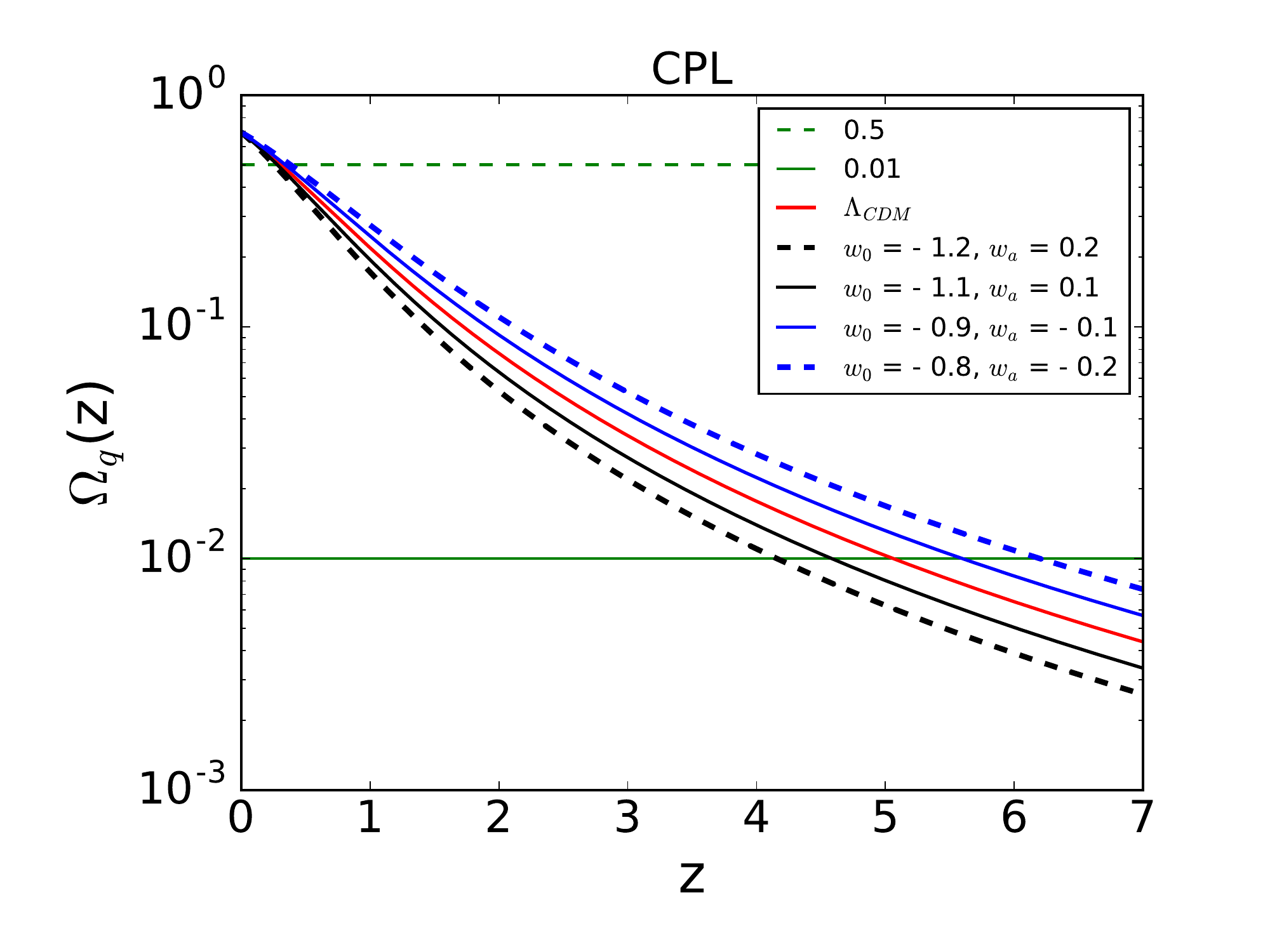}
\includegraphics[width=.45\textwidth]{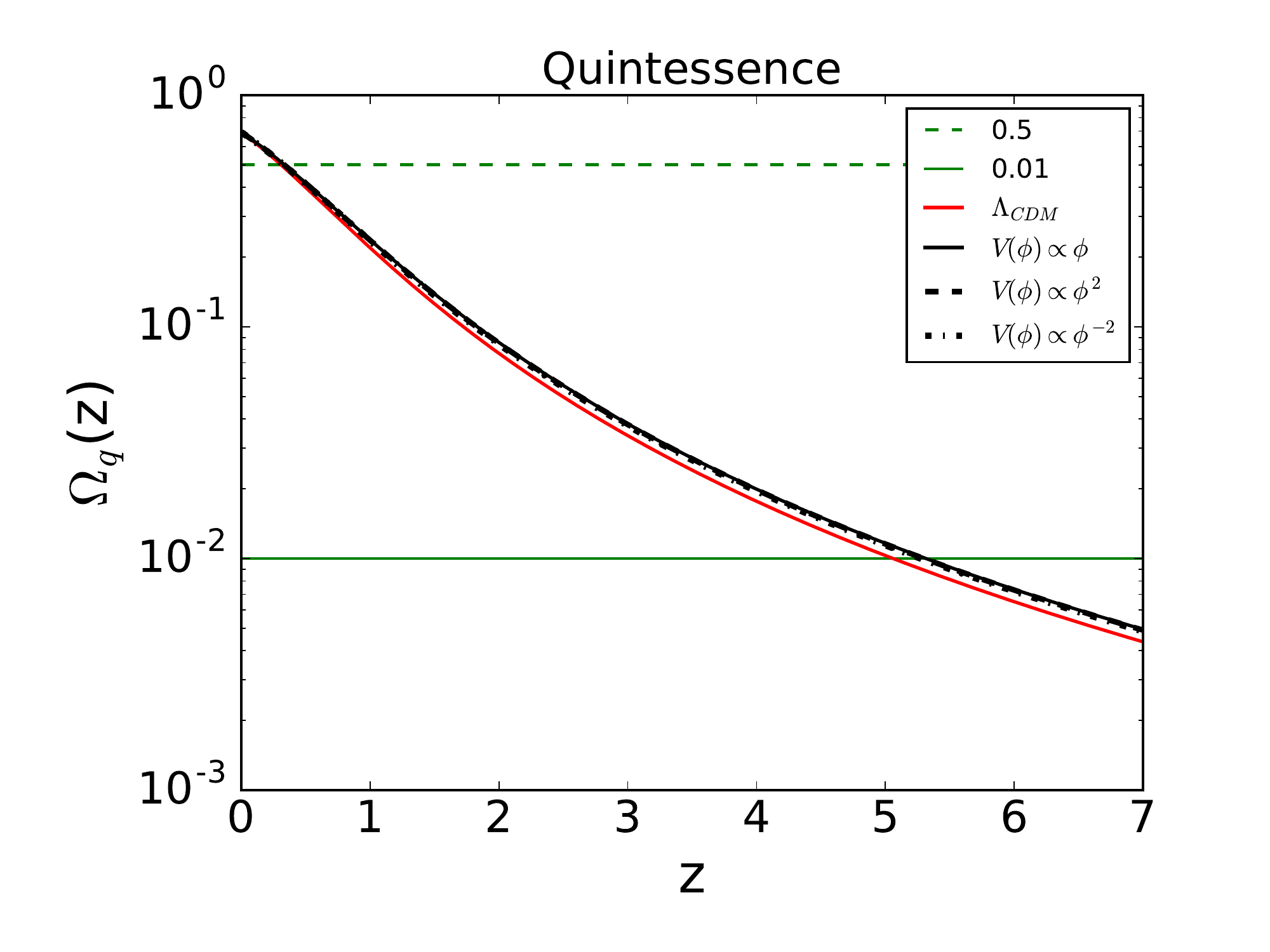}\\
\includegraphics[width=.45\textwidth]{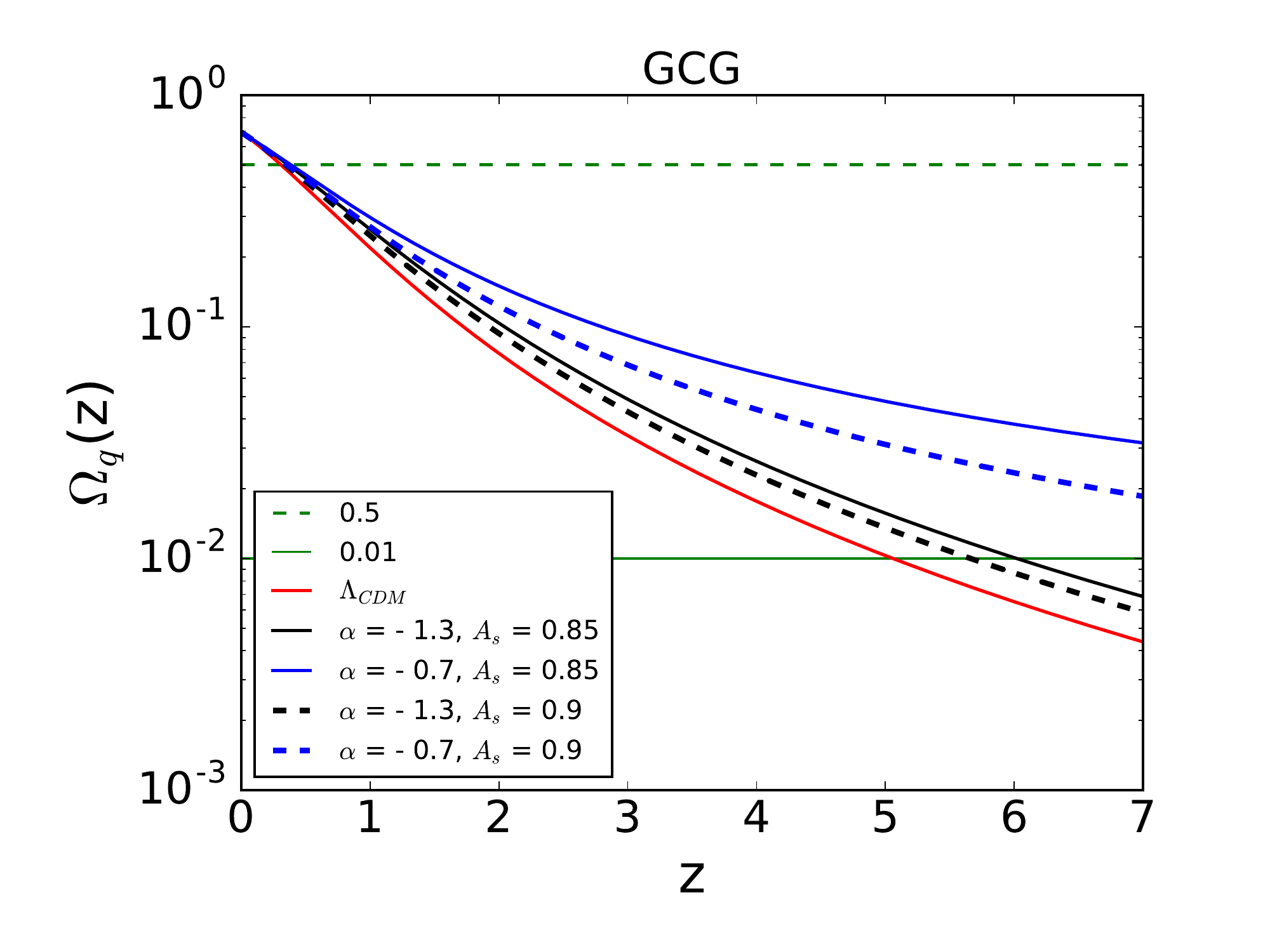}
\includegraphics[width=.45\textwidth]{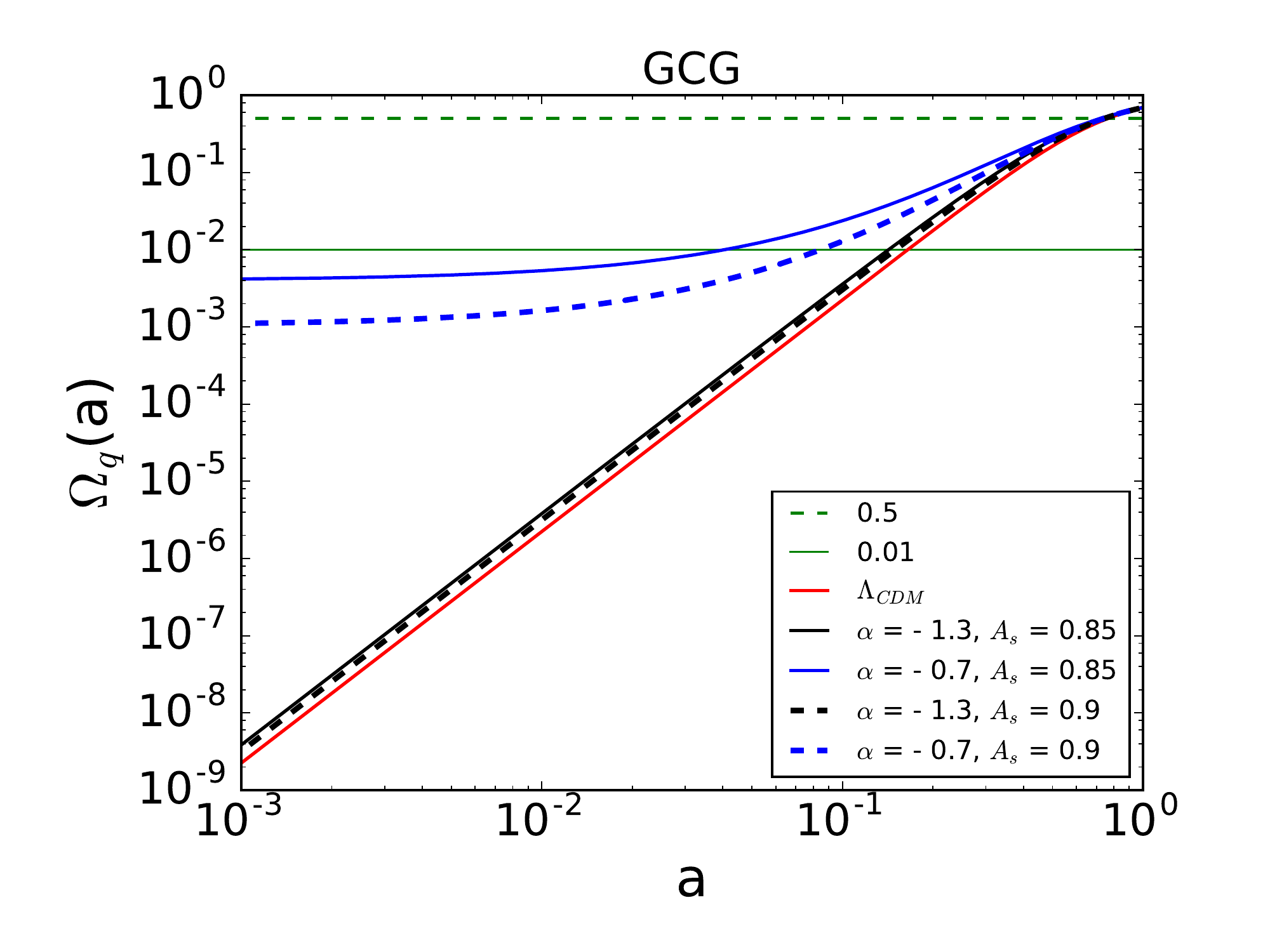}
\caption{\label{fig:Omegaq} $ \Omega_{q} $(z) vs. z plots for the same models as in Fig.~\ref{fig:eos} including $ \Lambda $CDM (including $ \Omega_{q} $(a) vs. a for GCG to show tracker behaviour especially).}
\end{figure}

In Fig.~\ref{fig:Omegaq} we have plotted $ \Omega_{q} $ for the above mentioned dark energy models. In all the panels horizontal continuous and dashed green lines represent the value of $ \Omega_{q} $ to be $ 0.01 $ and $ 0.5 $ respectively which correspond to the contribution of the energy density of the dark energy are $ 1\% $ and $ 50\% $ to the total energy density in the Universe respectively. Since we have fixed $ \Omega_{m}^{(0)} = 0.308 $, at present for all the dark energy models $ \Omega_{q}^{(0)} $ to be $ 0.692 $. As redshift increases the value of $ \Omega_{q} $ decreases monotonically except for the GCG tracker models. We have included an extra panel (bottom-right panel) for the GCG models to show exclusively the behavior of the tracker models. One can see that at early matter dominated era the values of $ \Omega_{q} $ for GCG tracker models are nearly constant which means $ \Omega_{q} $ and $ \Omega_{m} $ scales to each other i.e. for tracker models dark energy initially tracks the background. This behavior is the main difference for tracker models compared to the thawing models. After certain redshift (nearly $ 100 $ corresponding to the scale factor $ 10^{-2} $) the energy density parameter of the dark energy starts to increase. Due to the tracker behavior, the dark energy domination starts much earlier compared to the other models. In the CPL parametrization, for the non-phantom models, the dark energy starts to dominate earlier compared to the phantom models. More the non-phantom behavior earlier the domination of the dark energy budget. This same behavior is applicable for thawing quintessence and thawing GCG models too since they are also non-phantom models.

\begin{figure}[tbp]
\centering
\includegraphics[width=.45\textwidth]{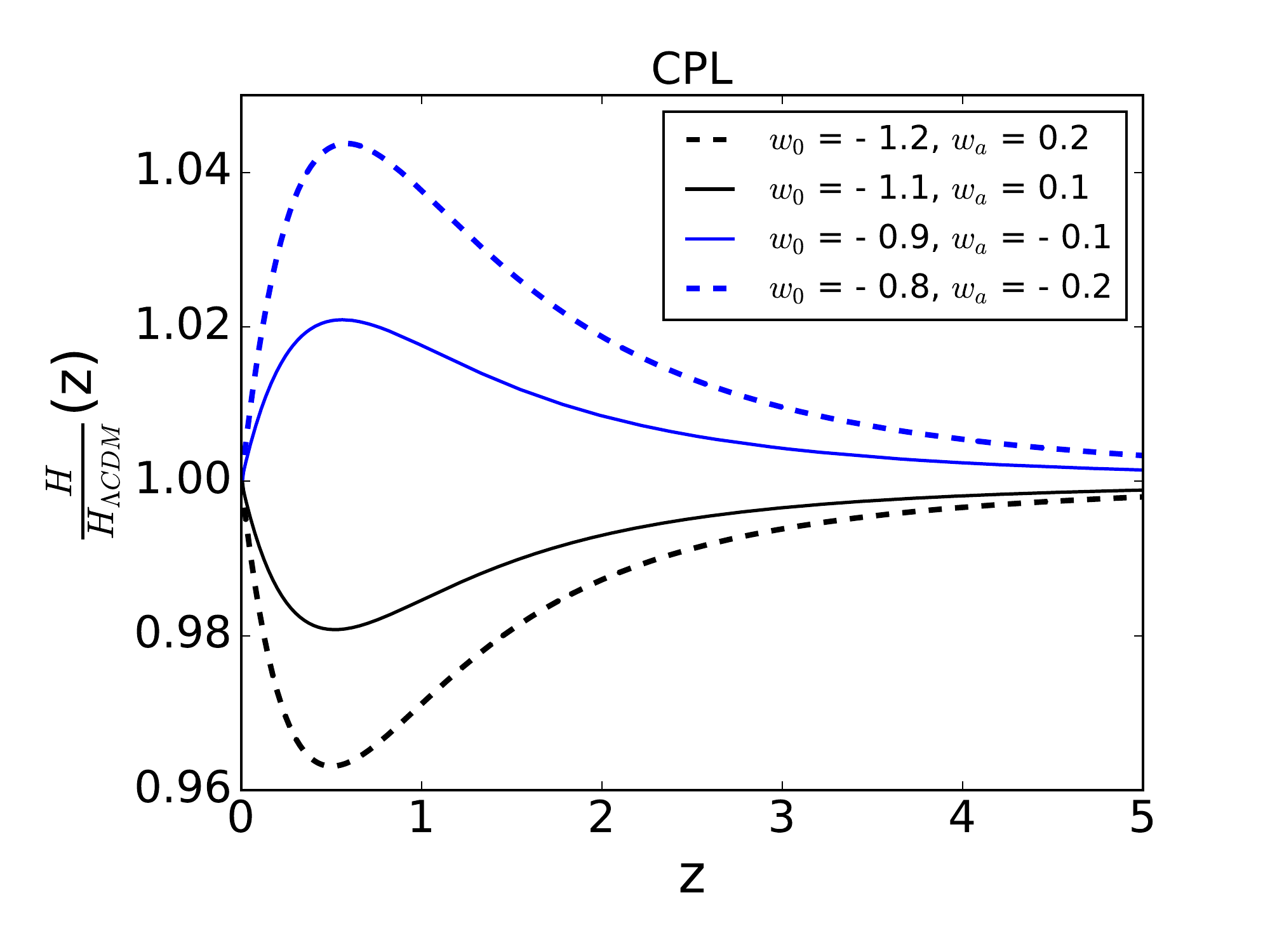}
\includegraphics[width=.45\textwidth]{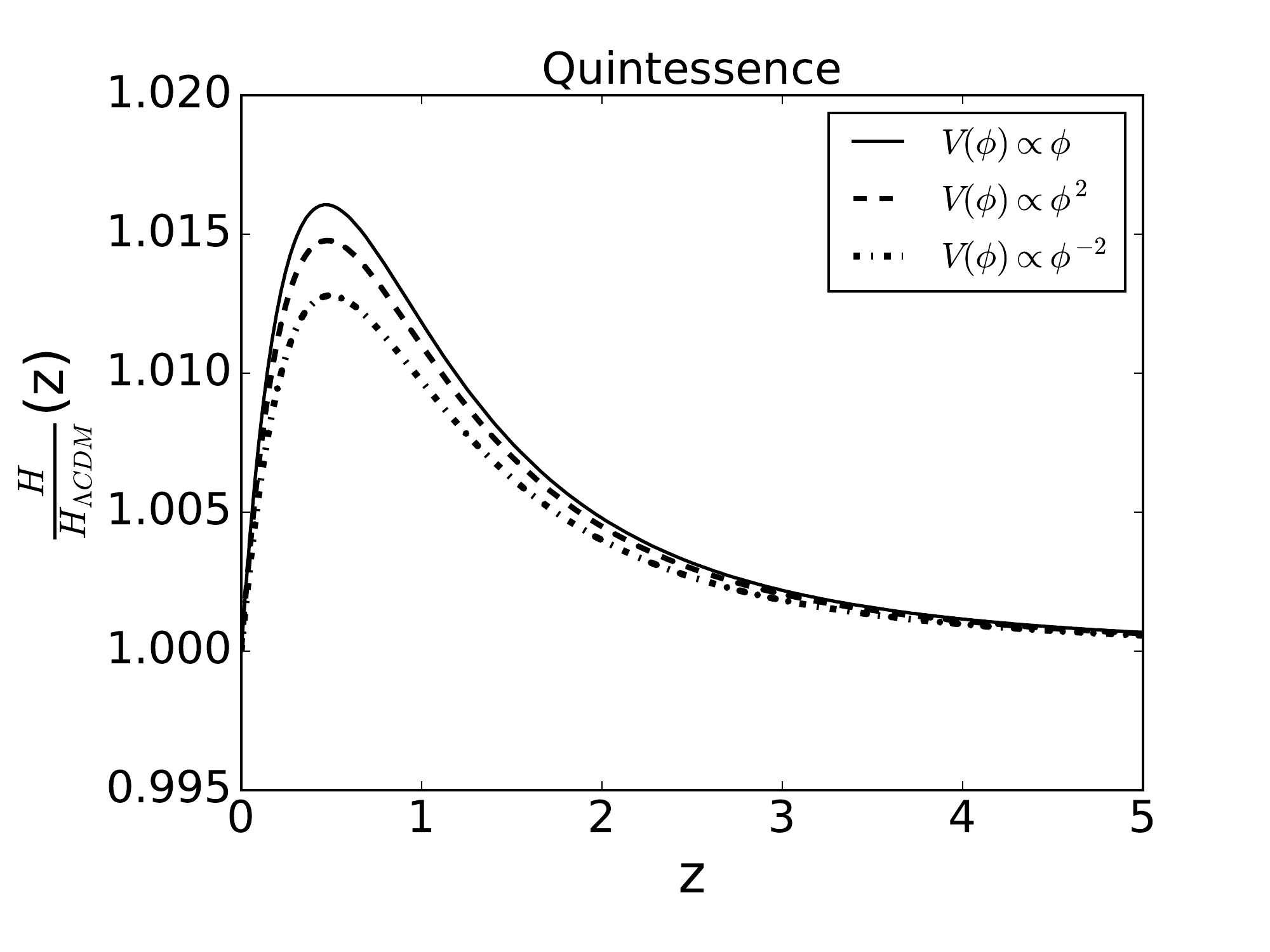}\\
\includegraphics[width=.45\textwidth]{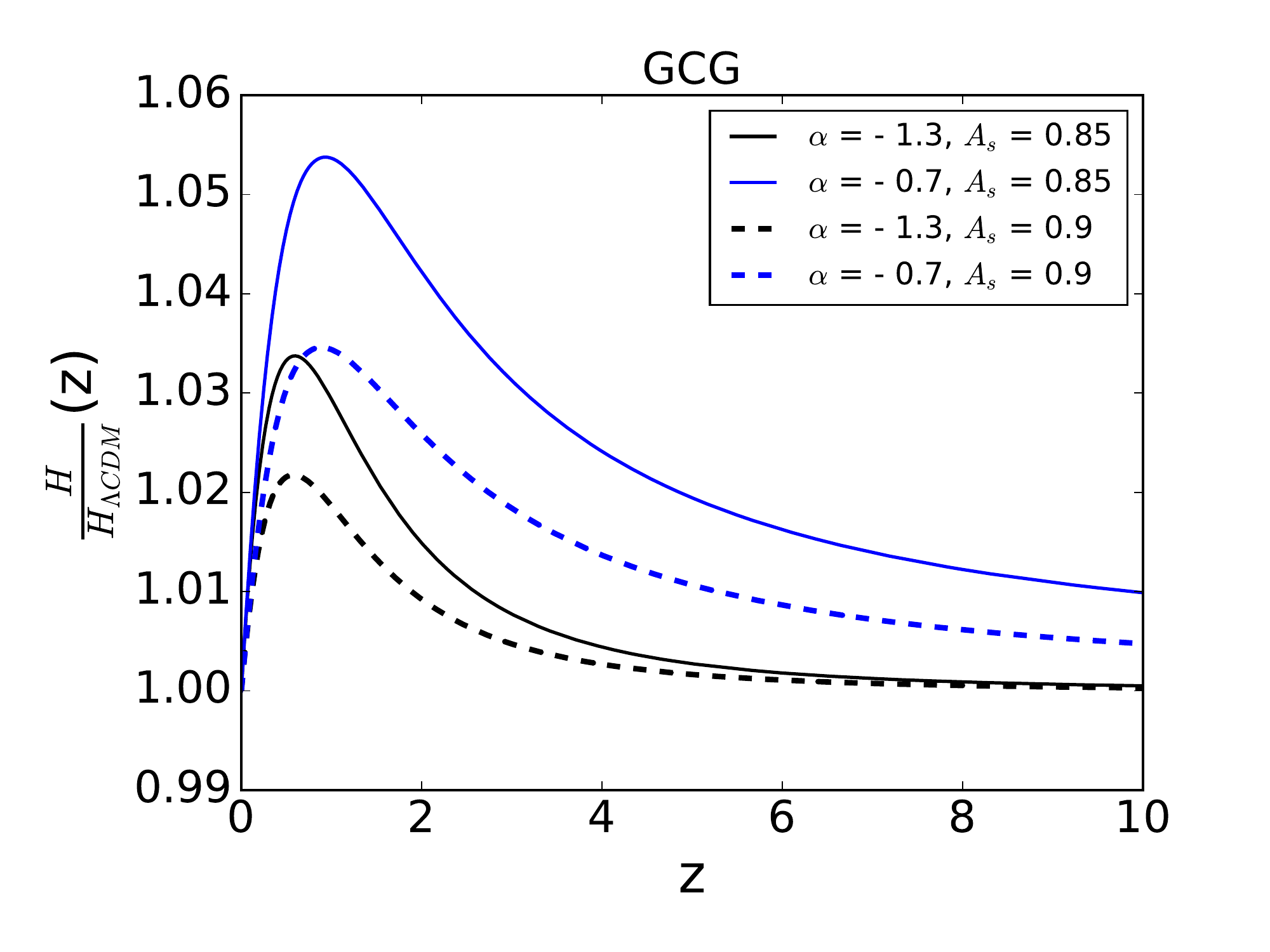}
\caption{\label{fig:hubble} Deviations in the Hubble parameter for the same models as in Fig.~\ref{fig:eos} from the $ \Lambda $CDM model.}
\end{figure}

In Fig.~\ref{fig:hubble} we have plotted deviations in the Hubble parameter for the same models as in Fig.~\ref{fig:eos} from the $ \Lambda $CDM model to compare the effects on the expansion history of the Universe relative to $ \Lambda $CDM. Since for the non-phantom models, dark energy dominates over matter earlier than the $ \Lambda $CDM model, the expansion of the scale factor enhances for the non-phantom models compared to $ \Lambda $CDM model. Therefore the deviation in the Hubble parameter compared to $ \Lambda $CDM model is positive for all the non-phantom models and the deviation is negative for the phantom models. More the non-phantom (phantom) behavior more the positive (negative) deviation. These behaviors are same for all the dark energy models. Only further comment for GCG tracker models is that the deviations in the expansion for the GCG tracker models compared to $ \Lambda $CDM model become significant at higher redshifts ($ 1\% $ nearly at redshifts $ 5 $ and $ 10 $ for the models $ \alpha = -0.7 $, $ A_{s} = 0.9 $ and $ \alpha = -0.7 $, $ A_{s} = 0.85 $ respectively) compared to other models. For all the other models $ 1\% $ deviation is at redshift $ z < 3 $. This behavior of GCG tracker models is because of the fact that the dark energy dominates much earlier compared to other models (see the bottom-right panel of Fig.~\ref{fig:Omegaq}). 

For all the models there is a change in the slope at particular redshift ranges from $ 0.3 $ to $ 1 $ depending on the model parameters. This is because of the fact that the transition of the deceleration to the acceleration in the expansion of the Universe happens when the dark energy starts to dominate. Since the dark energy with phantom e.o.s has more repulsive gravitational force compared to the $ \Lambda $CDM, the dark energy enhance the acceleration in the expansion for the phantom models compared to the $ \Lambda $CDM model. More the phantom behavior more the acceleration in the expansion. As opposed to it, more the non-phantom behavior lesser the acceleration in the expansion. Because of this behavior after the dark energy domination (more precisely after the slope turn over) at the late time the slopes are positive for phantom models whereas slopes are negative for non-phantom models. More the phantom (non-phantom) behavior larger the positive (negative) slope.

\section{Evolution of perturbations}

On sub-horizon scales, the evolution of fluctuations in the matter can be studied under Newtonian perturbation theory. On these scales, one can also ignore the fluctuations in the dark energy component as dark energy only clusters on horizon/super-horizon scales. Hence on sub-horizon scales, the dark energy affects the clustering of matter through the background evolution only. Under these assumptions the evolution of fluctuations are governed by the continuity, Euler and Poisson equations which are given below: 

\begin{equation}
\delta_{m}' + \vec{\nabla}.[(1+\delta_{m})\vec{v}_{m}]=0,
\label{eq:continuity}
\end{equation}

\begin{equation}
\vec{v}'_{m} + \mathcal{H} \vec{v}_{m} + (\vec{v}_{m}.\vec{\nabla})\vec{v}_{m} = - \vec{\nabla} \Phi,
\label{eq:Euler}
\end{equation}

\noindent
and

\begin{equation}
\nabla^{2} \Phi = 4\pi G a^{2} (\delta \rho_{m}) = \dfrac{3}{2} \mathcal{H}^{2} \Omega_{m} \delta_{m},
\label{eq:Poisson}
\end{equation}

\noindent
where prime ("$ ' $") denotes derivative w.r.t conformal time, $ \tau $, $ \delta_{m} $ is the matter energy density contrast defined as $ \delta_{m} \equiv \frac{\rho_{m} - \bar{\rho}_{m}}{\bar{\rho}_{m}} $ with $ \bar{\rho}_{m} $ and $ \rho_{m} $ be the background and perturbed matter energy densities respectively, $ \vec{v}_{m} $ is the velocity field of matter, $ \mathcal{H} $ is the conformal Hubble parameter defined as $ \mathcal{H} \equiv \frac{a'}{a} $, $ \Phi $ is the Newtonian gravitational potential, $ G $ is the Newtonian gravitational constant and $ \Omega_{m} $ is the matter energy density parameter (\cite{LSS1, LSS2}).
\\
Assuming matter velocity field is irrotational, it can be completely described by its divergence $ \theta_{m} = \vec{\nabla} . \vec{v}_{m} $ \cite{LSS1}. Using this, continuity equation \eqref{eq:continuity} can be rewritten as 

\begin{equation}
\delta_{m}' + \theta_{m} = - \vec{\nabla}.(\delta_{m} \vec{v}_{m}).
\label{eq:delmprime}
\end{equation}

\noindent
Now, taking divergence of the Euler equation \eqref{eq:Euler} and using Poisson equation \eqref{eq:Poisson} into it, the evolution equation of $ \theta_{m} $ becomes 

\begin{equation}
\theta'_{m} + \mathcal{H} \theta_{m} + \dfrac{3}{2} \mathcal{H}^{2} \Omega_{m} \delta_{m} = - \vec{\nabla}.[(\vec{v}_{m}.\vec{\nabla})\vec{v}_{m}].
\label{eq:thetamprime}
\end{equation}

\noindent
In Fourier space, eqs. \eqref{eq:delmprime} and \eqref{eq:thetamprime} can be written as (\cite{LSS1} - \cite{LSS9})

\begin{equation}
\delta'_{\vec{k}} + \theta_{\vec{k}} = - \int d^{3}\vec{q}_{1} \int d^{3}\vec{q}_{2} \;\; \delta_{D}^{(3)}(\vec{k}-\vec{q_{1}}-\vec{q_{2}}) \alpha(\vec{q_{1}}, \vec{q_{2}}) \theta_{\vec{q_{1}}} \delta_{\vec{q_{2}}},
\label{eq:delkprime}
\end{equation}

\begin{equation}
\theta'_{\vec{k}} + \mathcal{H} \theta_{\vec{k}} +\dfrac{3}{2} \mathcal{H}^{2} \Omega_{m} \delta_{\vec{k}} = - \int d^{3}\vec{q_{1}} \int d^{3}\vec{q_{2}} \;\; \delta_{D}^{(3)}(\vec{k}-\vec{q_{1}}-\vec{q_{2}}) \beta(\vec{q_{1}},\vec{q_{2}}) \theta_{\vec{q_{1}}} \theta_{\vec{q_{2}}},
\label{eq:thetakprime}
\end{equation}

\noindent
where

\begin{equation}
\alpha(\vec{q_{1}},\vec{q_{2}})=1+\dfrac{\vec{q_{1}}.\vec{q_{2}}}{q_{1}^2},
\label{eq:alpha}
\end{equation}

\begin{equation}
\beta(\vec{q_{1}},\vec{q_{2}})=\dfrac{(\vec{q_{1}}+\vec{q_{2}})^{2} \;\; (\vec{q_{1}}.\vec{q_{2}})}{2q_{1}^{2}q_{2}^{2}},
\label{eq:beta}
\end{equation}

\noindent
and $ \vec{k} $, $ \vec{q_{1}} $ $ \& $ $ \vec{q_{2}} $ correspond to different wave modes in the Fourier space. Here, subscript "m" has been omitted for the sake of simplified notation and in the subsequent sections this notation will be used.

\section{Linear solutions and linear matter power spectrum}

In linear theory, the 2nd and higher order terms in perturbation equations can be neglected. So, in the linear regime, eqs. \eqref{eq:delkprime} and \eqref{eq:thetakprime} becomes

\begin{equation}
\dfrac{\partial \delta^{lin}_{\vec{k}}}{\partial \tau} + \theta^{lin}_{\vec{k}} = 0,
\label{eq:dellineq}
\end{equation}

\begin{equation}
\dfrac{\partial \theta^{lin}_{\vec{k}}}{\partial \tau} + \mathcal{H} \theta^{lin}_{\vec{k}} + \dfrac{3}{2} \mathcal{H}^{2} \Omega_{m} \delta^{lin}_{\vec{k}}= 0,
\label{eq:thetalineq}
\end{equation}

\noindent
where superscript 'lin' stands for linear theory. Taking derivative of equation \eqref{eq:dellineq} and using equation \eqref{eq:thetalineq} into it, evolution equation for $ \delta^{lin}_{\vec{k}} $ becomes

\begin{equation}
\dfrac{\partial^{2} \delta^{lin}_{\vec{k}}}{\partial \tau^{2}} + \mathcal{H} \dfrac{\partial \delta^{lin}_{\vec{k}}}{\partial \tau} - \dfrac{3}{2} \mathcal{H}^{2} \Omega_{m} \delta^{lin}_{\vec{k}} = 0.
\label{eq:deltalineqn}
\end{equation}

\noindent
This is the standard evolution equation for matter energy density contrast in linear regime in the presence of smooth dark energy.
\\
In Fourier space, linear matter energy density contrast can be described through linear growth function, $ D $ which is defined as

\begin{equation}
\delta^{lin}_{\vec{k}}(\tau) = D(\tau) \delta^{in}_{\vec{k}},
\label{eq:delklin}
\end{equation} 

\noindent
where $ \delta^{in}_{\vec{k}} $ is the initial density contrast at a sufficient initial time. Using above definition of the linear growth function into equation \eqref{eq:dellineq}, linear velocity field of the matter can be obtained as  

\begin{equation}
\theta^{lin}_{\vec{k}}= - \mathcal{H}(\tau) f(\tau) D(\tau) \delta^{in}_{\vec{k}}.
\label{eq:thetaklin}
\end{equation}

\noindent
where $ f $ is the linear growth rate which is defined as

\begin{equation}
f \equiv \dfrac{d \; lnD}{d \; lna}.
\end{equation}

\noindent 
Putting equation \eqref{eq:delklin} into equation \eqref{eq:deltalineqn} and using an identity $ \frac{\mathcal{H}'}{\mathcal{H}^{2}} = - \frac{1}{2} (1 + 3 w \Omega_{q}) $, evolution equation for the linear growth function becomes

\begin{equation}
\dfrac{d^{2} D}{d N^{2}} + \frac{1}{2} \Big{(} 1 - 3 w \Omega_{q} \Big{)} \dfrac{d D}{d N} - \frac{3}{2} \Omega_{m} D = 0,
\label{eq:growtheq}
\end{equation}

\noindent
where conformal time derivative has been transferred to the derivative w.r.t e-folding, $ N $. Being a 2nd order differential equation, it has two solutions; one solution is same as the Hubble parameter which is the decaying mode and another one is the growing mode solution which has to be computed numerically.
\\
Let us denote growing and decaying mode growth functions as $ D_{+} $ and $ D_{-} $ respectively and the corresponding growth rate becomes $ f_{-} = \dfrac{d lnD_{-}}{d lna} $ and $ f_{+} = \dfrac{d lnD_{+}}{d lna} $ respectively. Since the decaying mode growth function decays away with the expansion of the Universe, to study the structure formation history the growing modes are of main interest. So, in the subsequent sections, growing mode solution has been considered.
\\
In the matter dominated era one can show that the growing mode solution of the growth function varies with the scale factor i.e. $ D_{+} \propto a $. So, we take $ D_{+}(a) = \frac{a}{a_{in}} $ at early matter dominated era. Using this fact we get $ \frac{d D_{+}}{d N}(a) = \frac{a}{a_{in}} $. So, our initial conditions become $ D_{+}^{in} = 1 $ and $ \frac{d D_{+}}{d N} \Big{|}_{in} = 1 $.
\\
Note that in the literature it is more conventional to take $ D_{+}^{in} = a_{in} $ and $ \frac{d D_{+}}{d N} \Big{|}_{in} = a_{in} $ in the matter dominated era. Since for both the cases (either $ D_{+}^{in} = 1 $ or $ D_{+}^{in} = a_{in} $) the fact that $ D_{+} \propto a $ is intact, the evolution of the growth function is same. Only the value of $ D_{+} $ scales accordingly. One can check that the ratios of $ D_{+} $ at two different times are same for both the cases. And since in all the subsequent calculations the ratio of $ D_{+} $ is involved (for example in equation \eqref{eq:PSlinear} the ratio $ \frac{D_{+}^{2}(a)}{D_{+}^{2}(a_{in})} $ is involved) the results do not depend on the exact value of $ D_{+}^{in} $. So, the results will always be the same (both for $ D_{+}^{in} = 1 $ and $ D_{+}^{in} = a_{in} $).

\begin{figure}[tbp]
\centering
\includegraphics[width=.45\textwidth]{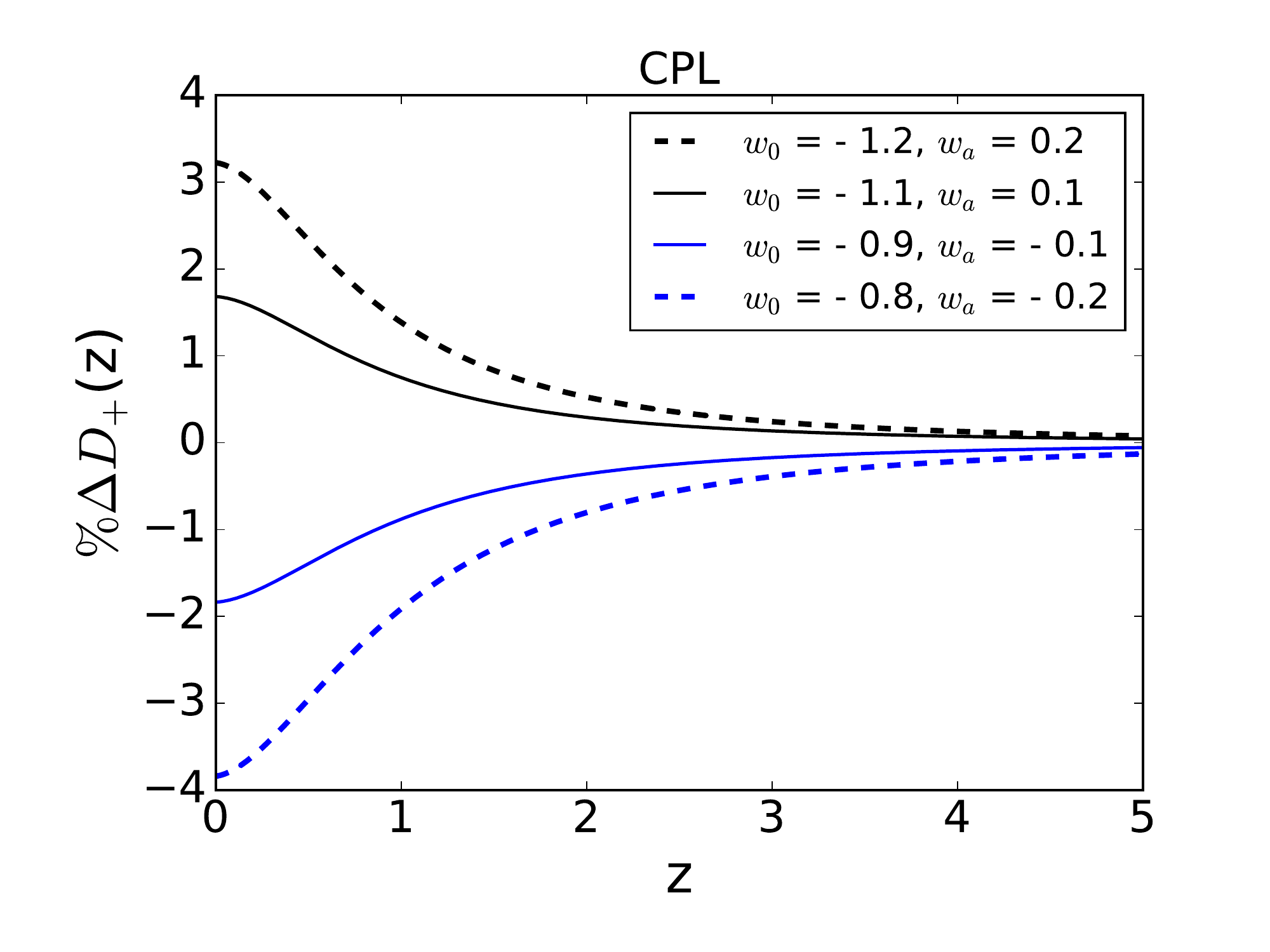}
\includegraphics[width=.45\textwidth]{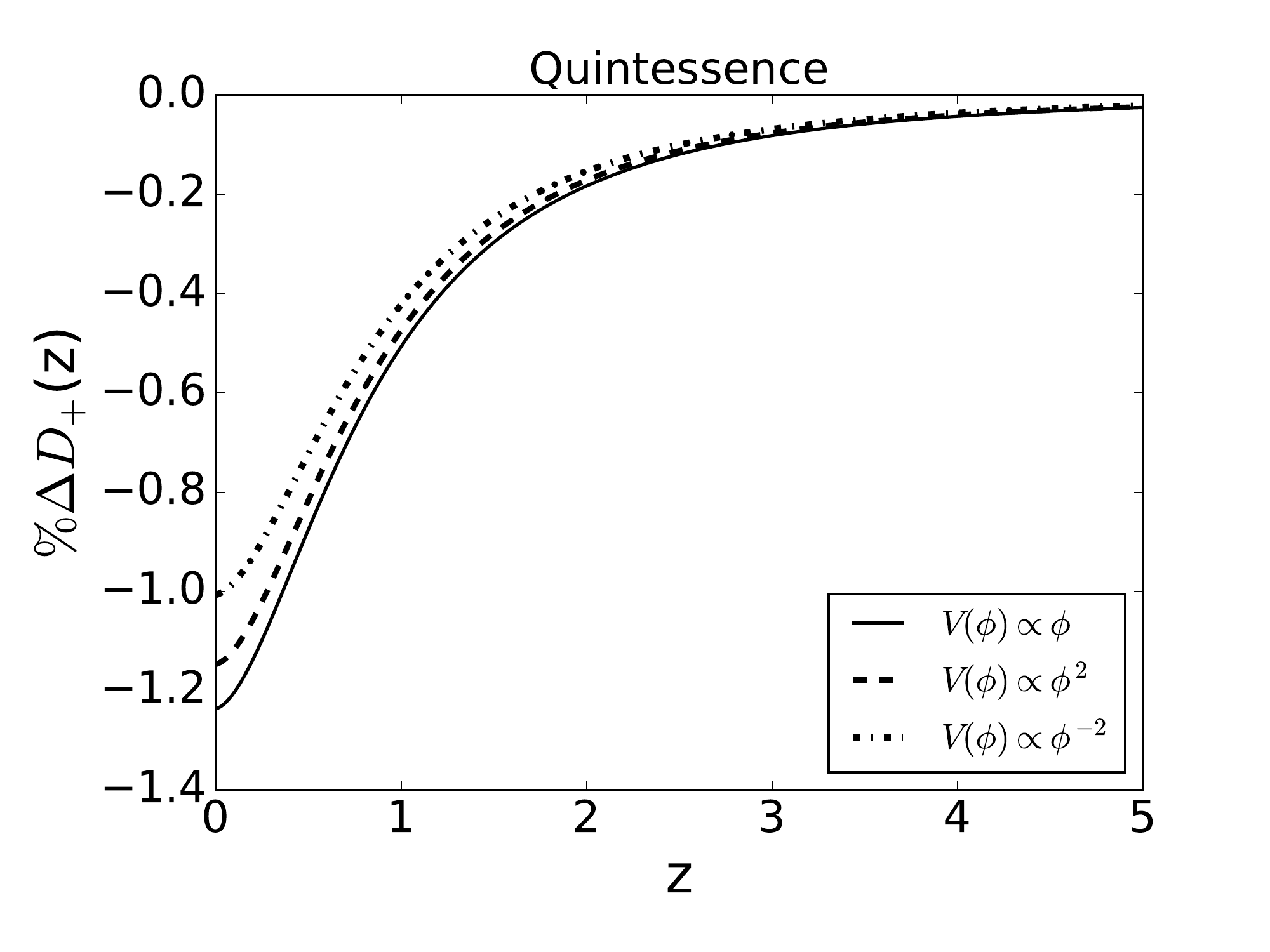}\\
\includegraphics[width=.45\textwidth]{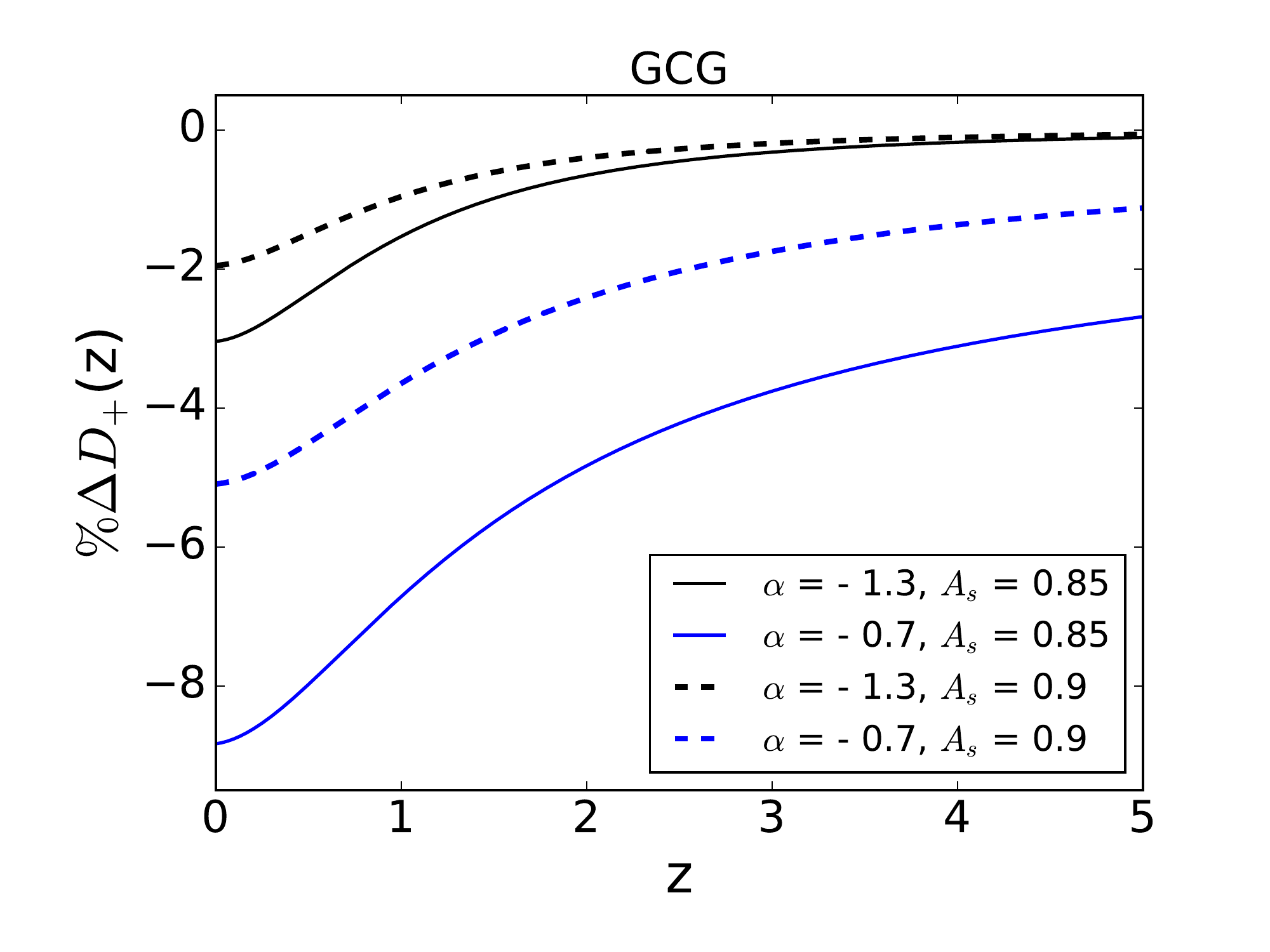}
\caption{\label{fig:growth} Percentage deviation in growing mode linear growth function from $ \Lambda $CDM model for the same models as in Fig.~\ref{fig:eos}.}
\end{figure}

In Fig.~\ref{fig:growth} we have plotted the percentage deviation in the growing mode linear growth function for the same models as in Fig.~\ref{fig:eos} compared to $ \Lambda $CDM model. In this figure and in all the subsequent figures the definition of $ \% \Delta X $ is given by $ \% \Delta X \equiv \frac{X_{M}-X_{\Lambda CDM}}{X_{\Lambda CDM}} \times 100 $, for any quantity $ X $ and for any model $ M $. Because of the attracting nature of the gravitational force of the matter contents (here cold dark matter and baryons together) whatever initial overdensity is present it will grow with the expansion. In contrast due to the repulsive nature of the gravitational force of the dark energy, the rate of the growth of the overdensity slows down. Since for the non-phantom models, dark energy dominates the Universe earlier i.e. the accelerated expansion starts earlier relative to $ \Lambda $CDM model (see Fig.~\ref{fig:Omegaq}) growth function is smaller for non-phantom models compared to $ \Lambda $CDM model. Due to the same reason, growth function is larger for phantom models compared to $ \Lambda $CDM model. More the phantom (non-phantom) behavior larger the positive (negative) deviation in growth function compared to $ \Lambda $CDM. Since quintessence and GCG models are always non-phantom the deviations in the growth function are always negative compared to $ \Lambda $CDM model. And since in the GCG tracker models the dark energy dominates the Universe significantly earlier the deviations are larger (negative) compared to other models.

\section*{Power spectrum:}

Using statistical homogeneity and isotropy the matter power spectrum can be defined as 

\begin{equation}
<\delta_{\vec{k}}(\tau) \delta_{\vec{k}'}(\tau)> = \delta_{D}^{(3)}(\vec{k} + \vec{k}') P(k, \tau).
\label{eq:PSdef}
\end{equation}

\noindent
Putting equation \eqref{eq:delklin} into equation \eqref{eq:PSdef}, one can see that $ P_{lin} \propto D_{+}^{2} $ and then the linear matter power spectrum related to the initial matter power spectrum is given by

\begin{equation}
P_{lin}(k,\tau) = \frac{D_{+}^{2}(\tau)}{D_{+}^{2}(\tau_{in})} P_{in}(k),
\label{eq:PSlinear}
\end{equation}

\noindent
where $ P_{in}(k) = P(k,\tau_{in}) $ is the initial matter power spectrum with $ D_{+}(\tau_{in}) = 1 $. For all the dark energy models we have taken same initial matter power spectrum where we have fixed baryonic energy density parameter at present $ \Omega_{b}^{(0)} = 0.05 $, total matter energy density parameter at present $ \Omega_{m}^{(0)} = 0.308 $, Hubble parameter at present $ H_{0} = 67.8 km/s/Mpc $, scalar spectral index $ n_{s} = 0.968 $ and scalar power spectrum amplitude $ A_{s} = 2.2 \times 10^{-9} $ at pivot scale $ k_{*} = 0.05 Mpc^{-1} $ related to the primordial curvature perturbation. These values are consistent with the Planck 2015 results (\cite{Planck},\cite{Planck2}). To get the initial matter power spectrum we use the publicly available CAMB code \cite{camb}. In CAMB we have computed the linear matter power spectrum at $ z = 0 $ for the $ \Lambda $CDM model with the above mentioned parameter values and by using equation \eqref{eq:PSlinear} we get the initial matter power spectrum by evolving back using the linear growth function of the $ \Lambda $CDM model.

\noindent

\begin{figure}[tbp]
\centering
\includegraphics[width=.45\textwidth]{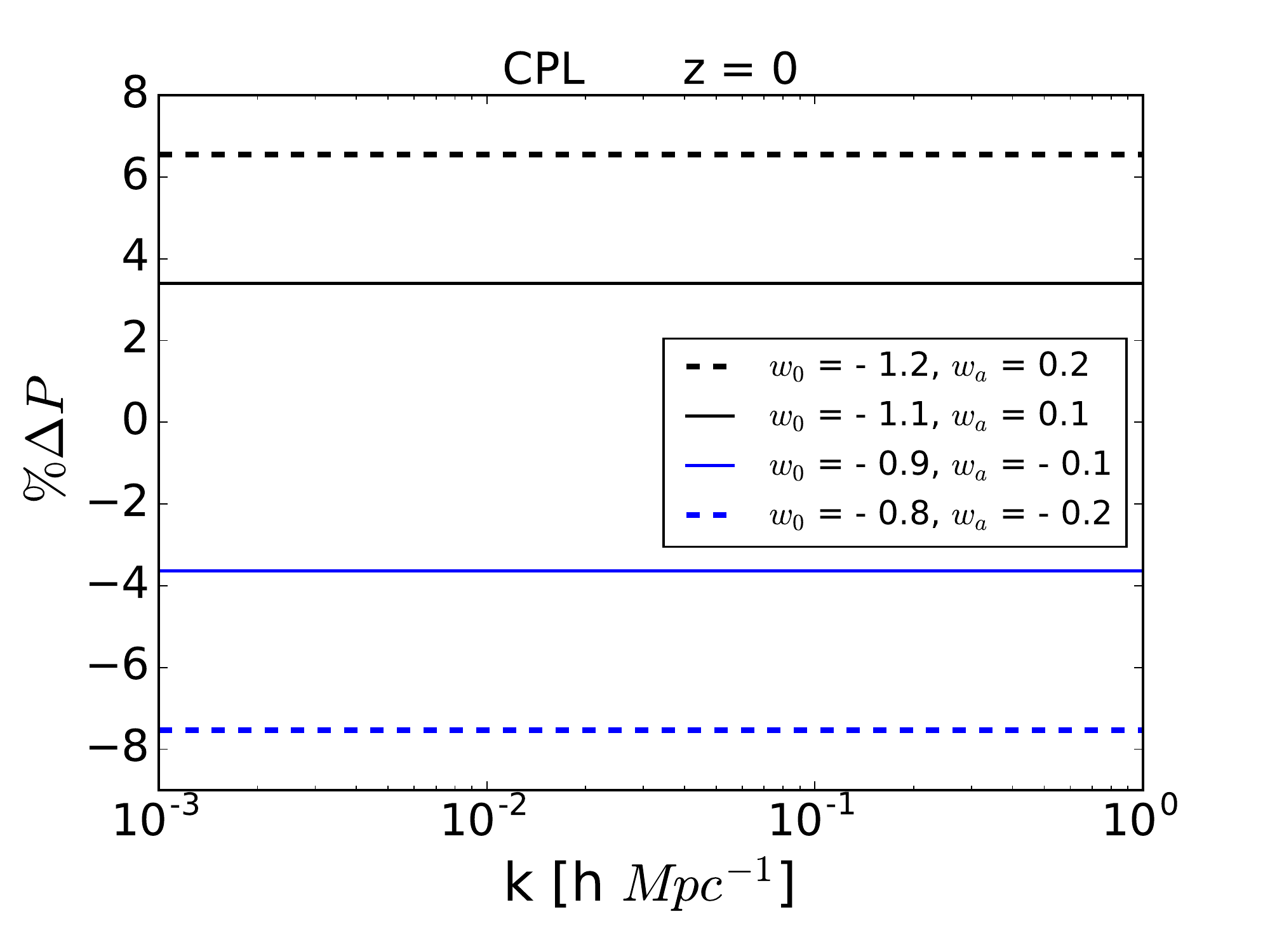}
\includegraphics[width=.45\textwidth]{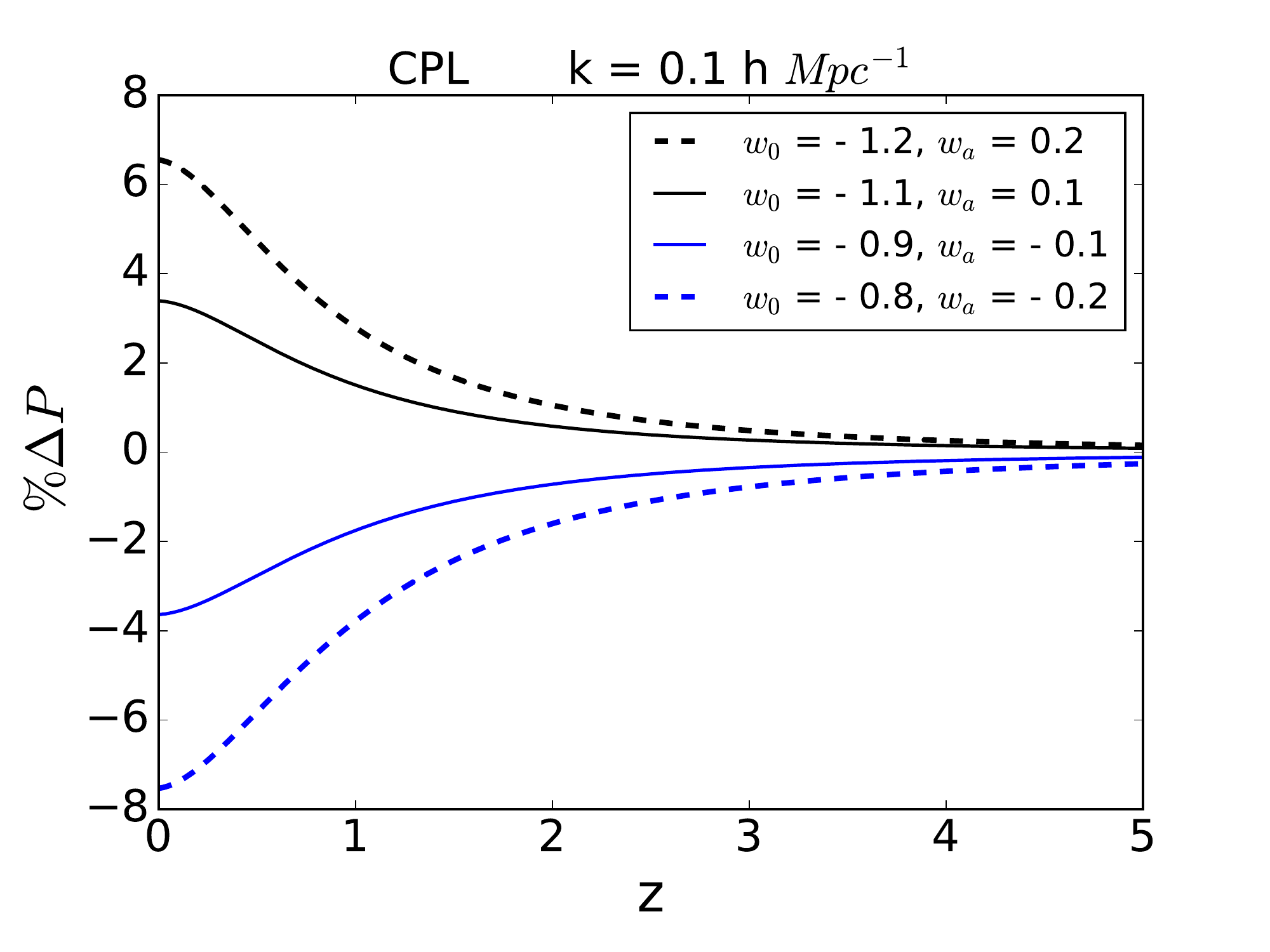}\\
\includegraphics[width=.45\textwidth]{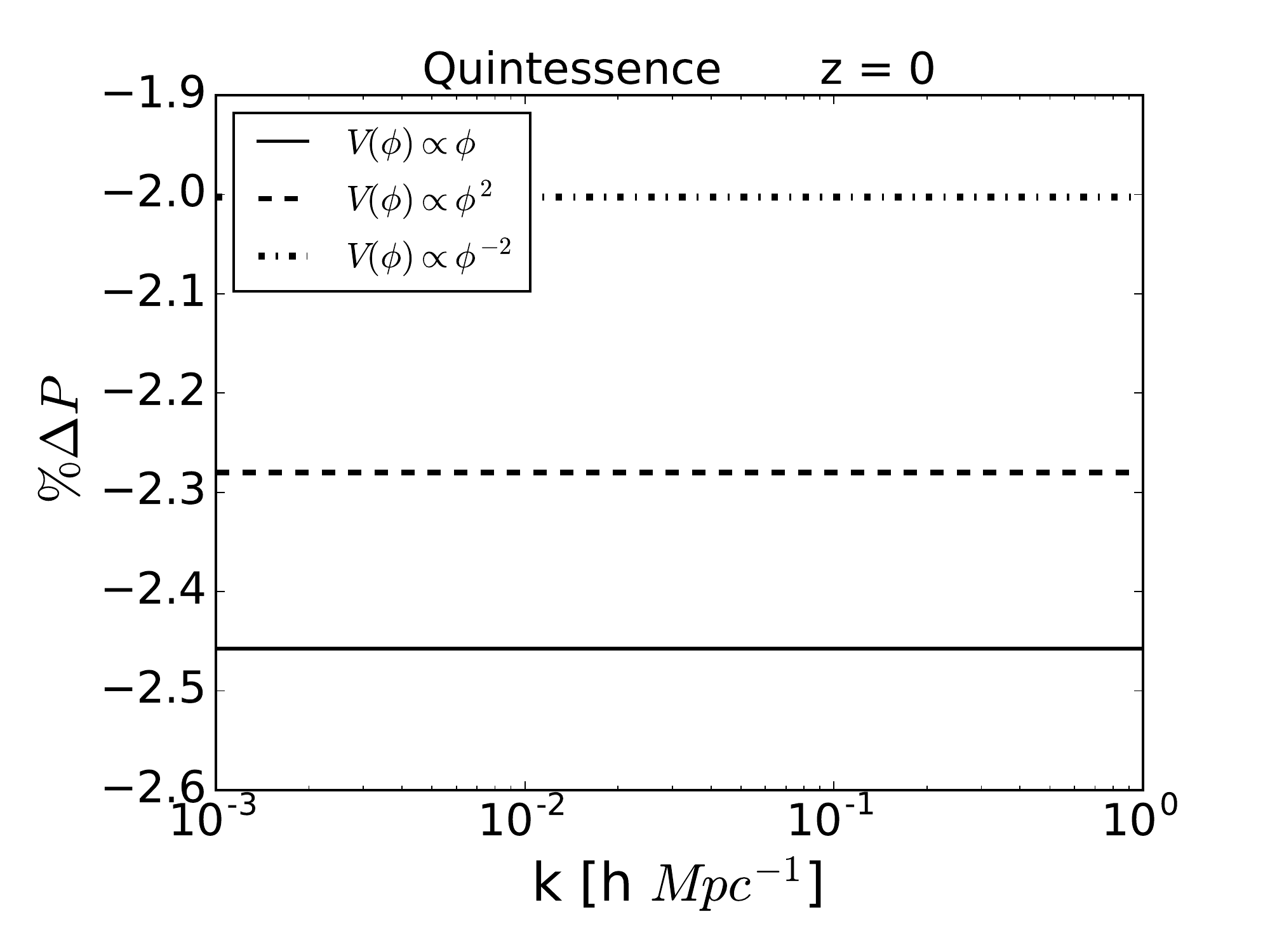}
\includegraphics[width=.45\textwidth]{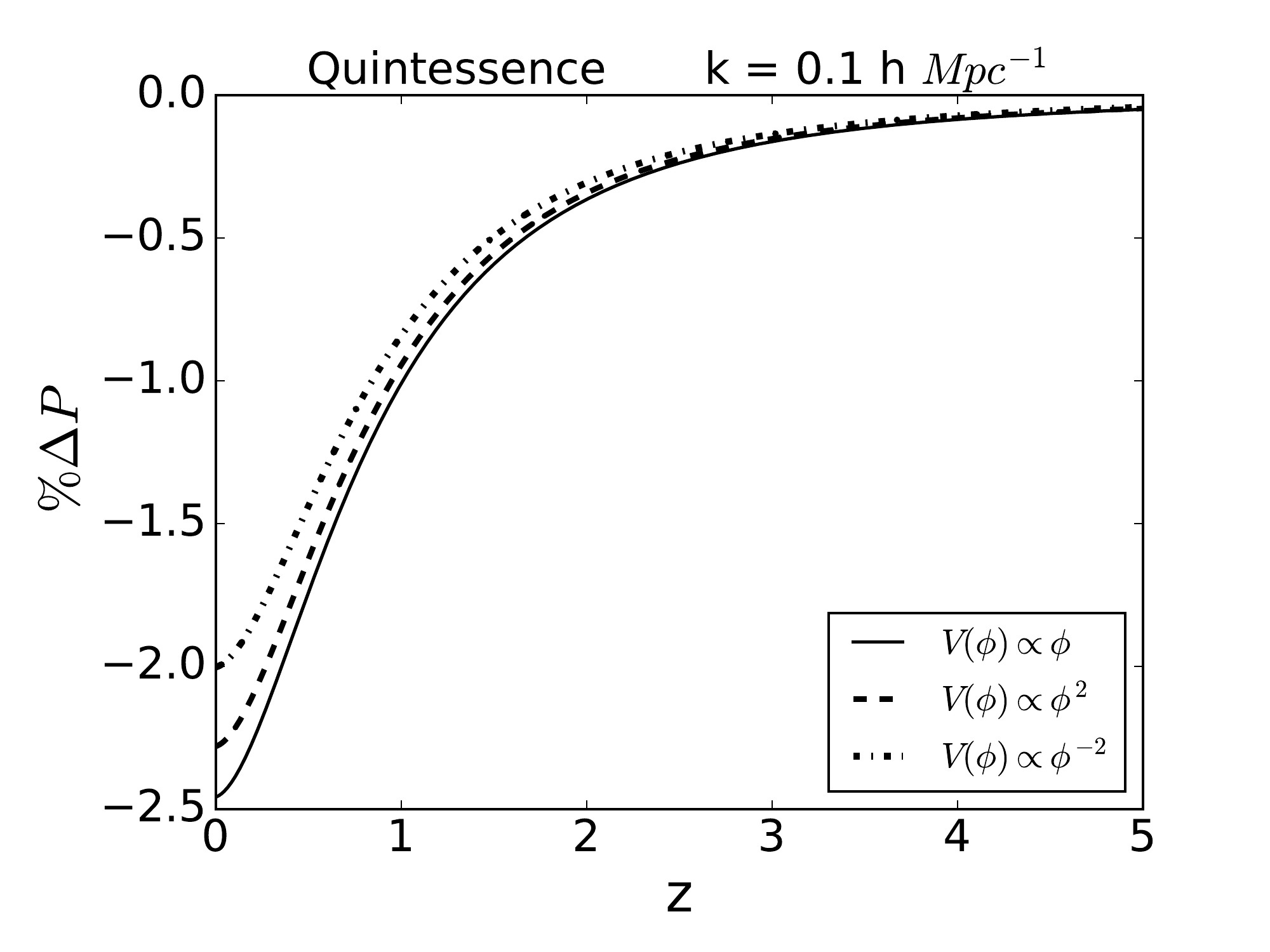}\\
\includegraphics[width=.45\textwidth]{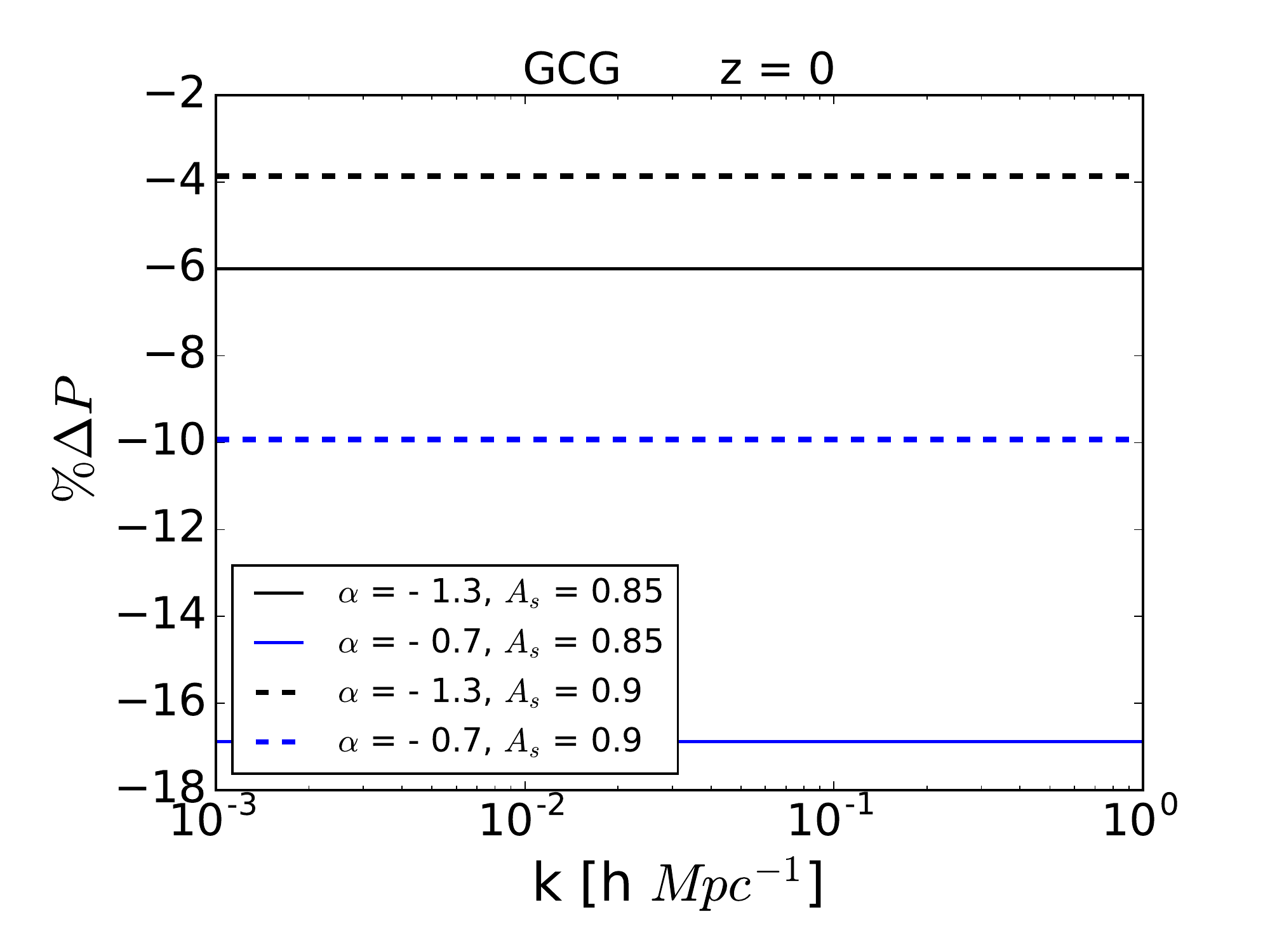}
\includegraphics[width=.45\textwidth]{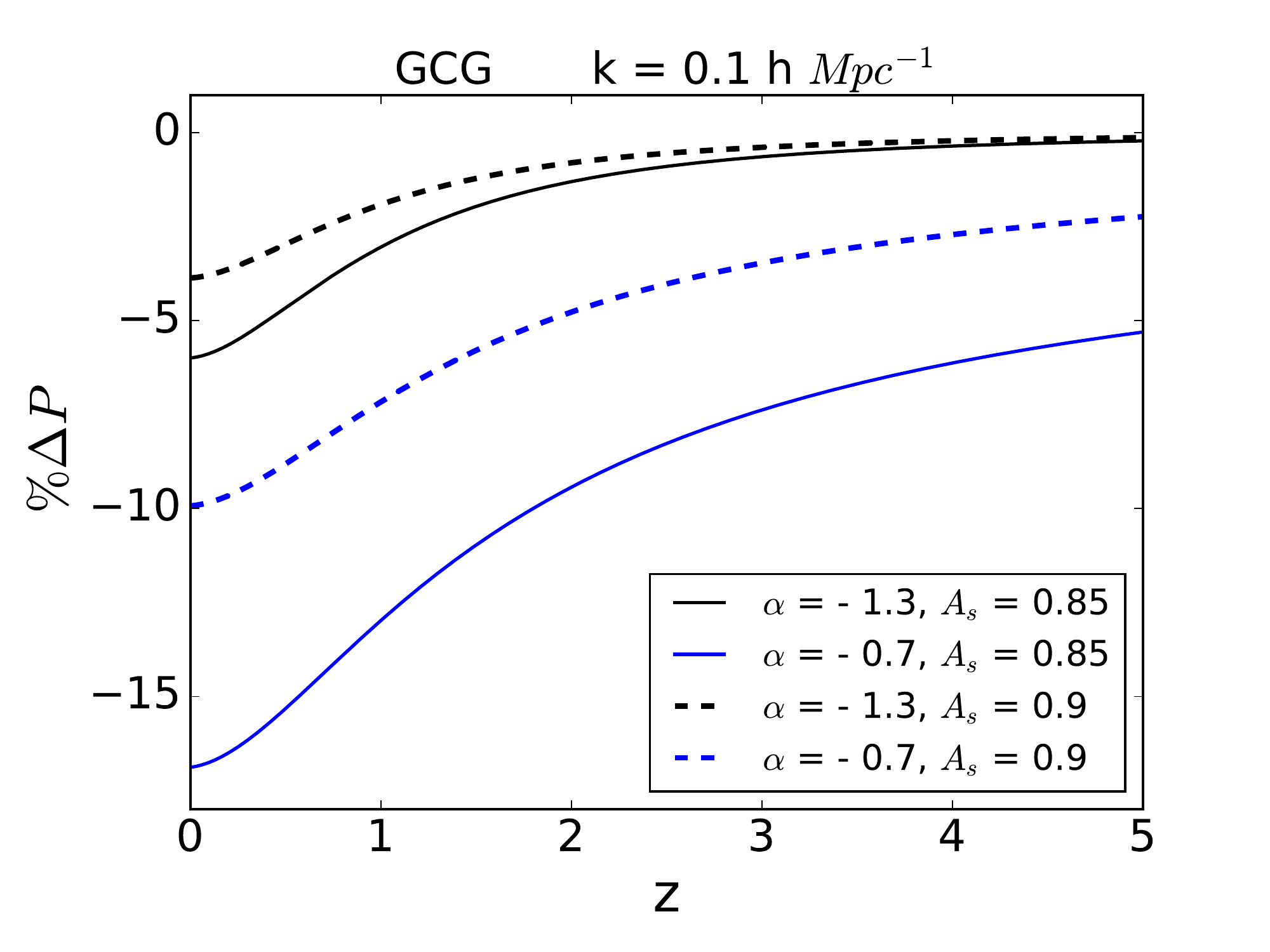}
\caption{\label{fig:ps} Percentage deviation in the linear matter power spectrum from $ \Lambda $CDM model for the same models as in Fig.~\ref{fig:eos}.}
\end{figure}

\noindent

In Fig.~\ref{fig:ps}, percentage deviation in linear matter power spectrum from $ \Lambda $CDM model has been plotted for the same models as in Fig.~\ref{fig:eos} using equation \eqref{eq:PSlinear}. As expected, for all the models these deviations are scale independent as there is no dark energy clustering at sub-horizon scales. Since the initial matter power spectrum is same for all the models, the deviations in the linear matter power spectrum represent the deviations in the linear growth function. As because of $ P_{lin} \propto D_{+}^{2} $ the deviations in the linear matter power spectrum are roughly twice the deviations in the linear growth function corresponding to the same model considered. In CPL parametrization the non-phantom dark energy models have slightly larger deviations ($ 0.5-1\% $) compared to the phantom models but this difference is not significant. In our analysis, we have considered thawing type of quintessence models and in these models matter power spectrum does not deviate much from the $\Lambda$CDM model for any of the potential considered (nearly $ 3 \% $) and the differences between three potentials are very small (sub-percentage level). In the GCG parametrization the tracker models have significantly larger deviations ($5-10\%$) from $\Lambda$CDM model compared to the thawing models.

\section{Non-linear solutions and tree-level bi-spectrum}

Having studied the dark energy effect on linear matter power spectrum, we move on to study its effect on the nonlinear matter spectrum, more specifically on the three point correlation function, the bi-spectrum.

Let us define a quantity $ \eta $ which is given by (\cite{LSS2},\cite{LSS3},\cite{LSS4})

\begin{equation}
\eta = ln \Big{[} \frac{D_{+}}{D_{+}^{in}} \Big{]} = ln D_{+}.
\label{eq:eta}
\end{equation}

\noindent
Note that the value of $ \eta $ remains same either for the normalisation $ D_{+}^{in} = 1 $ or $ D_{+}^{in} = a_{in} $ as long as the fact $ D_{+} \propto a $ remains intact initially at matter dominated era. This is because of the same reason mentioned before that the ratio of $ D_{+} $ is involved in the definition of $ \eta $. Using this definition eqs. \eqref{eq:delkprime} and \eqref{eq:thetakprime} can be rewritten as \cite{LSS2}

\begin{equation}
\dfrac{\partial \delta_{\vec{k}}}{\partial \eta} - \Theta_{\vec{k}} = \int d^{3}\vec{q}_{1} \int d^{3}\vec{q}_{2} \;\; \delta_{D}^{(3)}(\vec{k}-\vec{q_{1}}-\vec{q_{2}}) \alpha(\vec{q_{1}}, \vec{q_{2}}) \Theta_{\vec{q_{1}}} \delta_{\vec{q_{2}}},
\label{eq:deltak2ndeqn}
\end{equation}

\begin{equation}
\dfrac{\partial \Theta_{\vec{k}}}{\partial \eta} - \Theta_{\vec{k}} + \frac{3}{2} \dfrac{\Omega_{m}}{f_{+}^{2}} (\Theta_{\vec{k}} - \delta_{\vec{k}}) = \int d^{3}\vec{q_{1}} \int d^{3}\vec{q_{2}} \;\; \delta_{D}^{(3)}(\vec{k}-\vec{q_{1}}-\vec{q_{2}}) \beta(\vec{q_{1}},\vec{q_{2}}) \Theta_{\vec{q_{1}}} \Theta_{\vec{q_{2}}},
\label{eq:Capthetak2ndeqn}
\end{equation}

\noindent
where a new quantity $ \Theta $ is introduced which is related to $ \theta $ given by

\begin{equation}
\Theta_{\vec{k}} = -\frac{\theta_{\vec{k}}}{\mathcal{H} f_{+}}.
\label{eq:Captheta}
\end{equation}

\noindent
Defining this quantity has an advantage that in linear regime it is exactly same as the matter energy density contrast which can be seen through eqs. \eqref{eq:delklin} and \eqref{eq:thetaklin}, and hence

\begin{equation}
\Theta_{\vec{k}}^{lin} = \delta_{\vec{k}}^{lin}.
\label{eq:Capthetalin}
\end{equation}

\section*{Solutions:}

Equations \eqref{eq:deltak2ndeqn} and \eqref{eq:Capthetak2ndeqn} can be solved order by order using perturbative approach given by \cite{LSS2}

\begin{eqnarray}
\delta_{\vec{k}} = \sum_{n = 1}^{\infty} \delta_{\vec{k}}^{(n)} \hspace{0.5 cm} and \hspace{0.5 cm} \Theta_{\vec{k}} = \sum_{n = 1}^{\infty} \Theta_{\vec{k}}^{(n)},
\label{eq:nthordertrial}
\end{eqnarray}

\noindent
where the nth order terms $ \delta_{\vec{k}}^{(n)} $ and $ \Theta_{\vec{k}}^{(n)} $ are given by

\begin{equation}
\delta_{\vec{k}}^{(n)} (\eta) = \int d^{3}\vec{q_{1}} ... \int d^{3}\vec{q_{n}} \;\; \delta_{D}^{(3)}(\vec{k} - \vec{q_{1}} - ... - \vec{q_{n}}) F_{n} (\vec{q_{1}}, ..., \vec{q_{n}}, \eta) D_{+}^{n}(\eta) \delta_{\vec{q}_{1}}^{in} ... \delta_{\vec{q}_{n}}^{in},
\label{eq:nthorderdelta}
\end{equation}

\begin{equation}
\Theta_{\vec{k}}^{(n)} (\eta) = \int d^{3}\vec{q_{1}} ... \int d^{3}\vec{q_{n}} \;\; \delta_{D}^{(3)}(\vec{k} - \vec{q_{1}} - ... - \vec{q_{n}}) G_{n} (\vec{q_{1}}, ..., \vec{q_{n}}, \eta) D_{+}^{n}(\eta) \delta_{\vec{q}_{1}}^{in} ... \delta_{\vec{q}_{n}}^{in},
\label{eq:nthorderCaptheta}
\end{equation}

\noindent
respectively.

\section*{Second order solutions:}

Putting equation \eqref{eq:nthordertrial} into eqs. \eqref{eq:deltak2ndeqn} and \eqref{eq:Capthetak2ndeqn}, the 2nd order perturbation equations become

\begin{equation}
\dfrac{\partial \delta_{\vec{k}}^{(2)}}{\partial \eta} - \Theta_{\vec{k}}^{(2)} = D_{+}^{2}(\eta) \int d^{3}\vec{q}_{1} \int d^{3}\vec{q}_{2} \;\; \delta_{D}^{(3)}(\vec{k}-\vec{q_{1}}-\vec{q_{2}}) \alpha_{s}(\vec{q_{1}}, \vec{q_{2}}) \delta_{\vec{q_{1}}}^{in} \delta_{\vec{q_{2}}}^{in},
\label{eq:delta2ndorder}
\end{equation}

\begin{equation}
\dfrac{\partial \Theta_{\vec{k}}^{(2)}}{\partial \eta} - \Theta_{\vec{k}}^{(2)} + \frac{3}{2} \dfrac{\Omega_{m}}{f_{+}^{2}} (\Theta_{\vec{k}}^{(2)} - \delta_{\vec{k}}^{(2)}) = D_{+}^{2}(\eta) \int d^{3}\vec{q_{1}} \int d^{3}\vec{q_{2}} \;\; \delta_{D}^{(3)}(\vec{k}-\vec{q_{1}}-\vec{q_{2}}) \beta(\vec{q_{1}},\vec{q_{2}}) \delta_{\vec{q_{1}}}^{in} \delta_{\vec{q_{2}}}^{in}.
\label{eq:Captheta2ndorder}
\end{equation}

\noindent
The right hand side of equation \eqref{eq:delta2ndorder} is obtained after symmetrizing the right hand side of equation \eqref{eq:deltak2ndeqn}. This is because, at 2nd order, the source terms $ \Theta_{\vec{q_{1}}} $ and $ \delta_{\vec{q_{2}}} $ in equation \eqref{eq:deltak2ndeqn} have to be linear and they are also equal (see equation \eqref{eq:Capthetalin}). Interchanging these two linear source terms will not affect the evolution equation. Hence we need to symmetrize it by introducing a quantity, $ \alpha_{s}(\vec{q_{1}}, \vec{q_{2}}) = \frac{1}{2} [\alpha(\vec{q_{1}}, \vec{q_{2}}) + \alpha(\vec{q_{2}}, \vec{q_{1}})] $. From eqs. \eqref{eq:nthorderdelta} and \eqref{eq:nthorderCaptheta}, the 2nd order solutions are given by

\begin{equation}
\delta_{\vec{k}}^{(2)} (\eta) = D_{+}^{2}(\eta) \int d^{3}\vec{q_{1}} \int d^{3}\vec{q_{2}} \;\; \delta_{D}^{(3)}(\vec{k} - \vec{q_{1}} - \vec{q_{2}}) F_{2} (\vec{q_{1}}, \vec{q_{2}}, \eta) \delta_{\vec{q}_{1}}^{in} \delta_{\vec{q}_{2}}^{in},
\label{eq:deltak2ndordersoln}
\end{equation}

\begin{equation}
\Theta_{\vec{k}}^{(2)} (\eta) = D_{+}^{2}(\eta) \int d^{3}\vec{q_{1}} \int d^{3}\vec{q_{2}} \;\; \delta_{D}^{(3)}(\vec{k} - \vec{q_{1}} - \vec{q_{2}}) G_{2} (\vec{q_{1}}, \vec{q_{2}}, \eta) \delta_{\vec{q}_{1}}^{in} \delta_{\vec{q}_{2}}^{in}.
\label{eq:Captheta2ndordersoln}
\end{equation}

\noindent
Now, using equations \eqref{eq:deltak2ndordersoln} and \eqref{eq:Captheta2ndordersoln} in equations \eqref{eq:delta2ndorder} and \eqref{eq:Captheta2ndorder}, the evolution equations of $ F_{2} $ and $ G_{2} $ becomes

\begin{equation}
\dfrac{\partial F_{2}(\vec{q}_{1},\vec{q}_{2},\eta)}{\partial \eta} + 2 F_{2}(\vec{q}_{1},\vec{q}_{2},\eta) - G_{2}(\vec{q}_{1},\vec{q}_{2},\eta) = \alpha_{s}(\vec{q}_{1},\vec{q}_{2}),
\label{eq:F2eqn}
\end{equation}

\begin{equation}
\dfrac{\partial G_{2}(\vec{q}_{1},\vec{q}_{2},\eta)}{\partial \eta} + G_{2}(\vec{q}_{1},\vec{q}_{2},\eta) + \frac{3}{2} \dfrac{\Omega_{m}}{f_{+}^{2}} \Big{[} G_{2}(\vec{q}_{1},\vec{q}_{2},\eta) - F_{2}(\vec{q}_{1},\vec{q}_{2},\eta) \Big{]} = \beta(\vec{q}_{1},\vec{q}_{2}),
\label{eq:G2eqn}
\end{equation}

In general, these two coupled differential equations have to be solved numerically. In the matter dominated era approximate analytical solutions can be possible because of the approximation

\begin{equation}
\dfrac{\Omega_{m}}{f_{+}^{2}} \approx 1.
\label{eq:EDSappx}
\end{equation}

\noindent
Except at very low redshifts the assumption \eqref{eq:EDSappx} holds true \cite{LSS2}. Using this assumption, equation \eqref{eq:G2eqn} becomes

\begin{equation}
\dfrac{\partial G_{2}(\vec{q}_{1},\vec{q}_{2},\eta)}{\partial \eta} + \frac{5}{2} G_{2}(\vec{q}_{1},\vec{q}_{2},\eta) - \frac{3}{2} F_{2}(\vec{q}_{1},\vec{q}_{2},\eta) \approx \beta(\vec{q}_{1},\vec{q}_{2}),
\label{eq:G2eqnEDSappx}
\end{equation}

\noindent
The coupled differential equations \eqref{eq:F2eqn} and \eqref{eq:G2eqnEDSappx} can now be solved analytically and the solutions for $F_{2}$ and $G_{2}$ are given by

\begin{eqnarray}
F_{2}(\vec{q}_{1},\vec{q}_{2},\eta) = && \frac{1}{7} \Big{[} 5 \alpha_{s}(\vec{q}_{1},\vec{q}_{2}) + 2 \beta(\vec{q}_{1},\vec{q}_{2}) \Big{]} - \frac{1}{5} \Big{[} 3 \alpha_{s}(\vec{q}_{1},\vec{q}_{2}) + 2 \beta(\vec{q}_{1},\vec{q}_{2}) \Big{]} e^{(\eta_{0} - \eta)} \nonumber\\
&& - \frac{4}{35} \Big{[} \alpha_{s}(\vec{q}_{1},\vec{q}_{2}) - \beta(\vec{q}_{1},\vec{q}_{2}) \Big{]} e^{\frac{7}{2} (\eta_{0} - \eta)},
\label{eq:F2SolnEDS}
\end{eqnarray}

\begin{eqnarray}
G_{2}(\vec{q}_{1},\vec{q}_{2},\eta) = && \frac{1}{7} \Big{[} 3 \alpha_{s}(\vec{q}_{1},\vec{q}_{2}) + 4 \beta(\vec{q}_{1},\vec{q}_{2}) \Big{]} - \frac{1}{5} \Big{[} 3 \alpha_{s}(\vec{q}_{1},\vec{q}_{2}) + 2 \beta(\vec{q}_{1},\vec{q}_{2}) \Big{]} e^{(\eta_{0} - \eta)} \nonumber\\
&& + \frac{6}{35} \Big{[} \alpha_{s}(\vec{q}_{1},\vec{q}_{2}) - \beta(\vec{q}_{1},\vec{q}_{2}) \Big{]} e^{\frac{7}{2} (\eta_{0} - \eta)},
\label{eq:G2SolnEDS}
\end{eqnarray}

\noindent
where we have used the Gaussian initial condition at $ \eta = \eta_{0} $ at a sufficient initial time i.e. at $ \eta = \eta_{0} $ both $ F_{2} $ and $ G_{2} $ vanish. And we take $ \eta_{0} \rightarrow - \infty $ (where $ a \rightarrow 0 $). Putting $ \eta_{0} \rightarrow - \infty $ in equations \eqref{eq:F2SolnEDS} and \eqref{eq:G2SolnEDS}, finally we get approximate analytical solutions both for $F_{2}$ and $G_{2}$ are given by

\begin{equation}
F_{2}(\vec{q}_{1},\vec{q}_{2}) \simeq \frac{1}{7} \Big{[} 5 \alpha_{s}(\vec{q}_{1},\vec{q}_{2}) + 2 \beta(\vec{q}_{1},\vec{q}_{2}) \Big{]} = \frac{5}{7} + \frac{1}{2} \Big{(} \frac{q_{1}}{q_{2}} + \frac{q_{2}}{q_{1}} \Big{)} \hat{q_{1}}.\hat{q_{2}} + \frac{2}{7} (\hat{q_{1}}.\hat{q_{2}})^{2},
\label{eq:F2solnEDSappx}
\end{equation}

\begin{equation}
G_{2}(\vec{q}_{1},\vec{q}_{2}) \simeq \frac{1}{7} \Big{[} 3 \alpha_{s}(\vec{q}_{1},\vec{q}_{2}) + 4 \beta(\vec{q}_{1},\vec{q}_{2}) \Big{]} = \frac{3}{7} + \frac{1}{2} \Big{(} \frac{q_{1}}{q_{2}} + \frac{q_{2}}{q_{1}} \Big{)} \hat{q_{1}}.\hat{q_{2}} + \frac{4}{7} (\hat{q_{1}}.\hat{q_{2}})^{2}.
\label{eq:G2solnEDSappx}
\end{equation}

\noindent
In fact, if we take initial condition at a time enough before matter dominated era instead of the exact era where $ a \rightarrow 0 $, the result will be same (at/after matter dominated era). This can be seen mathematically as $ \eta_{0} \ll \eta $ the exponential terms in equations \eqref{eq:F2SolnEDS} and \eqref{eq:G2SolnEDS} become negligible and only the constant terms survive. This is because that the $\eta$ dependent terms are due to the decaying modes and we exclude these decaying modes.

To be precise, the numerical solution, $ F_{2} $ obtained from the coupled differential eqs. \eqref{eq:F2eqn} and \eqref{eq:G2eqn} has been considered in the subsequent calculations. 

\begin{figure}[tbp]
\centering
\includegraphics[width=.45\textwidth]{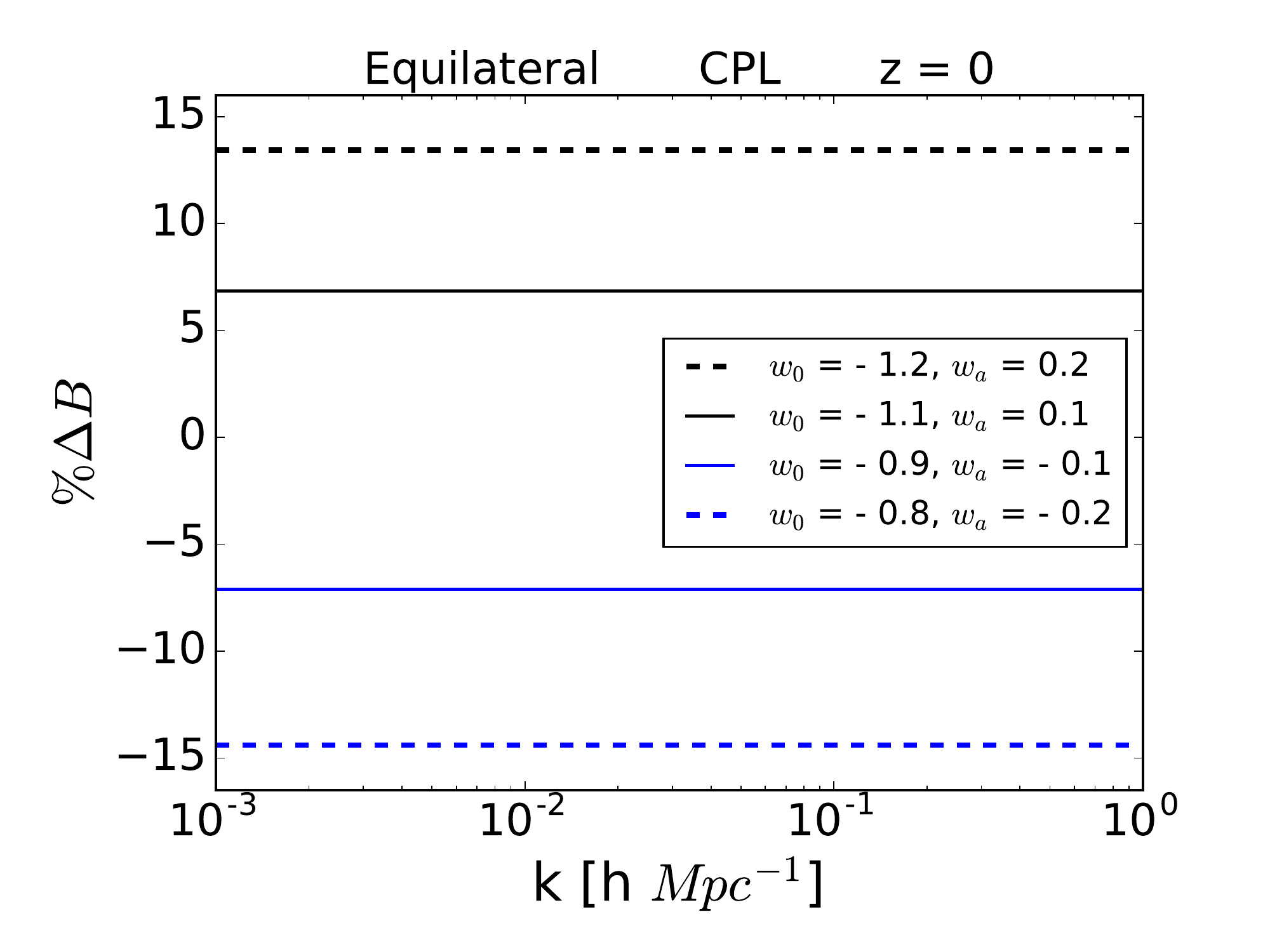}
\includegraphics[width=.45\textwidth]{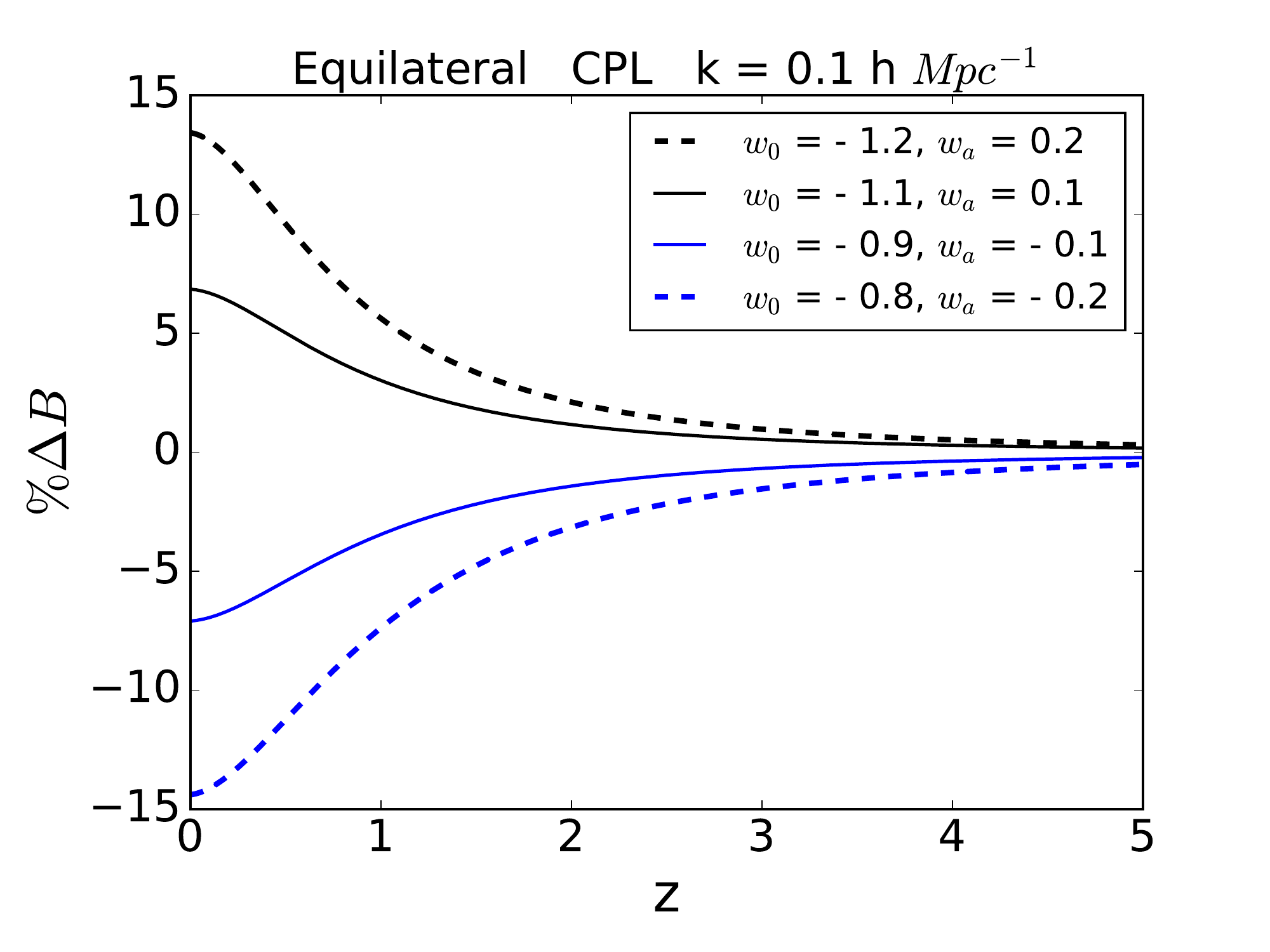}\\
\includegraphics[width=.45\textwidth]{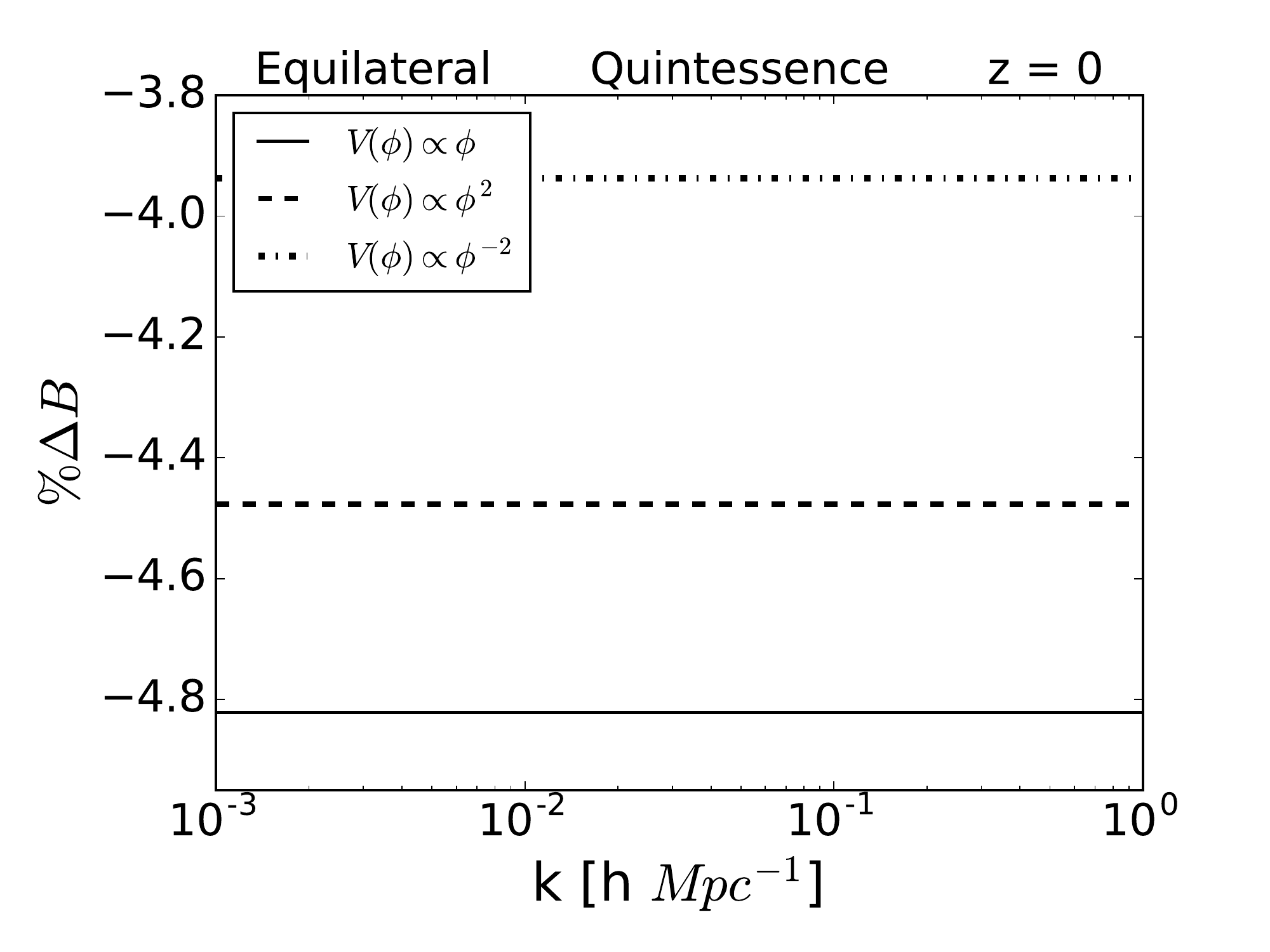}
\includegraphics[width=.45\textwidth]{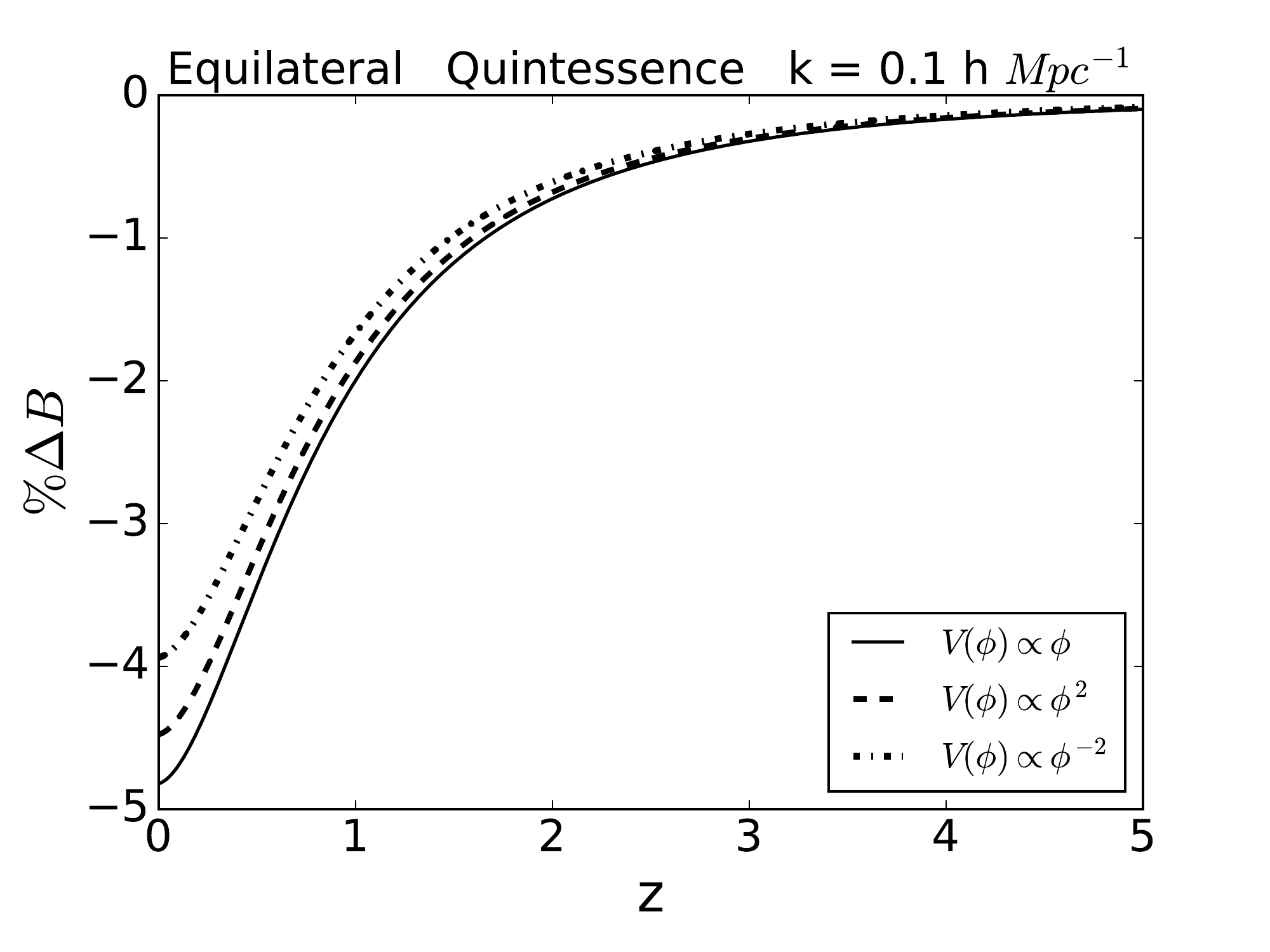}\\
\includegraphics[width=.45\textwidth]{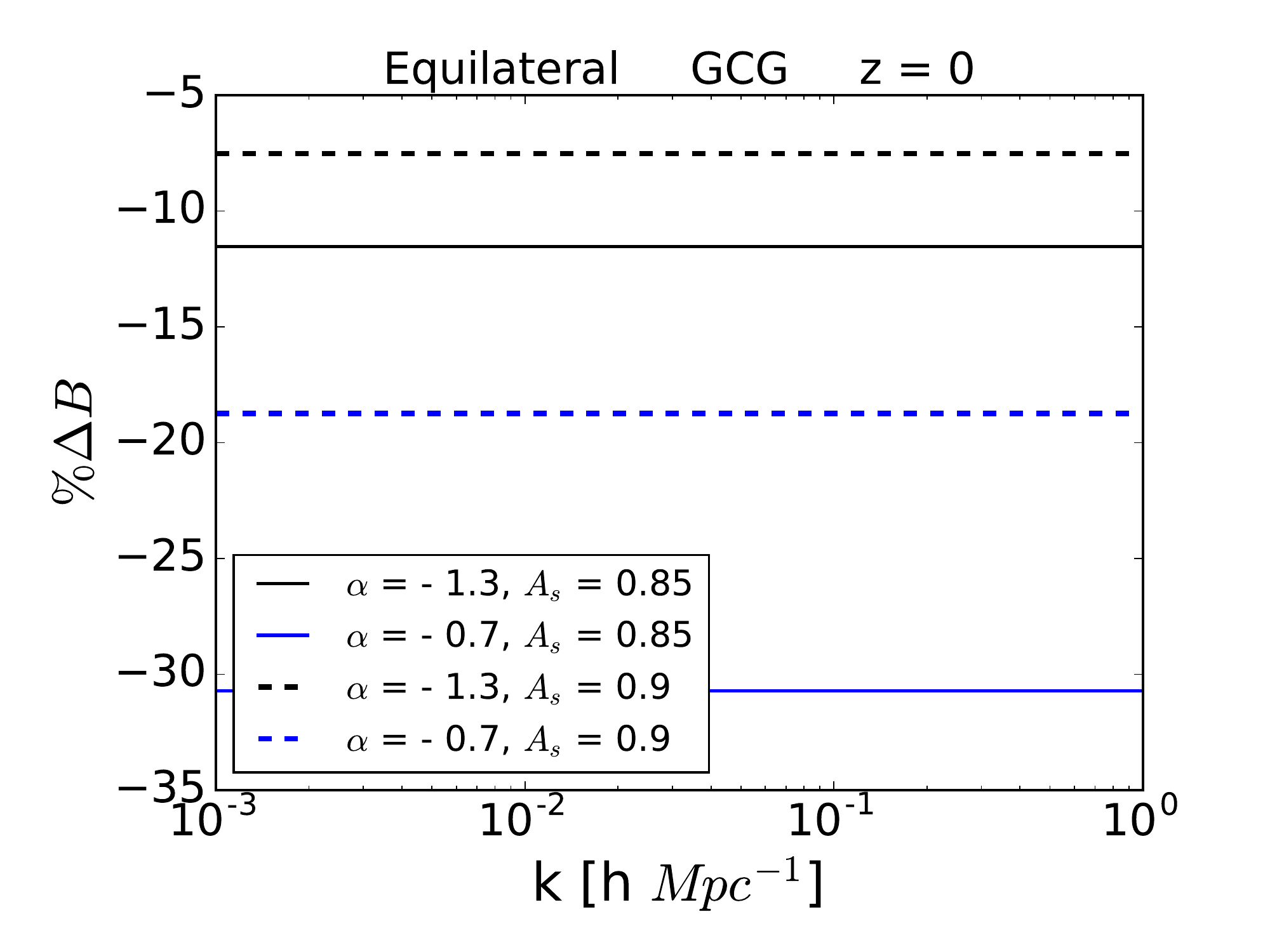}
\includegraphics[width=.45\textwidth]{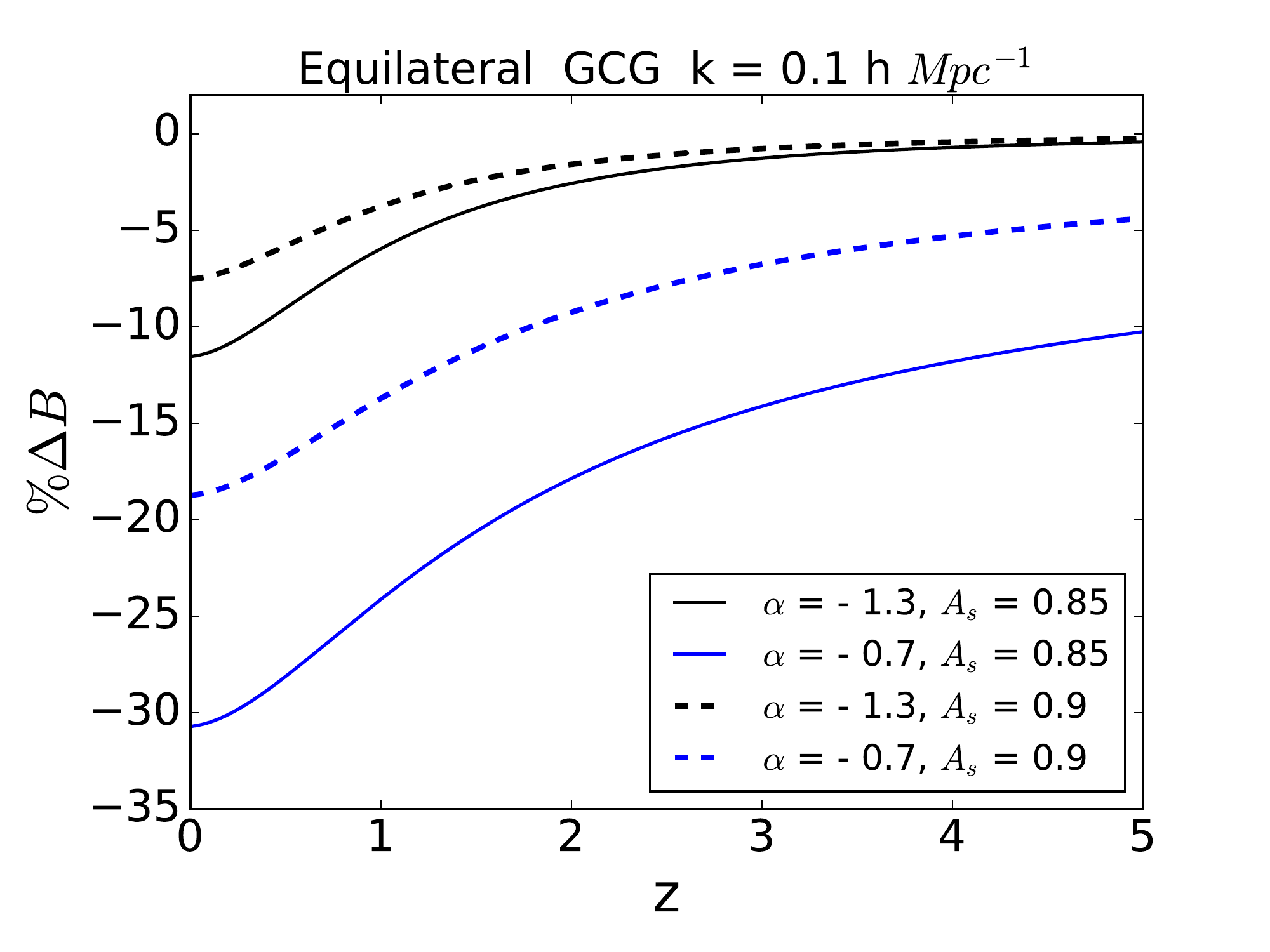}
\caption{\label{fig:bsequilateral} Percentage deviation in the tree-level matter bi-spectrum from $ \Lambda $CDM model for the same models as in Fig.~\ref{fig:eos} for the equilateral configuration.}
\end{figure}

To solve these two coupled differential equations we need initial conditions both for $ F_{2} $ and $ G_{2} $. As because of eqs. \eqref{eq:F2solnEDSappx} and \eqref{eq:G2solnEDSappx} are valid in matter dominated era we take these values at initial time at early matter dominated era (i.e. at $ \eta = 0 $) as initial conditions. Note that although the approximation \eqref{eq:EDSappx} is worse at late times but the approximate solution $ F_{2} $ in equation \eqref{eq:F2solnEDSappx} is a very good approximation at late times too which can be seen from the left panels of Fig.~\ref{fig:bsequilateral}. It is expected that the deviation between the approximate analytical solution and numerical solution is the maximum at $ z = 0 $ compared to the past. If there were any significant deviation then the numerical solution would depend on the cosmological models and this would have a signature on the scale dependency in the left panels of Fig.~\ref{fig:bsequilateral} by using equation \eqref{eq:BStree}. Since there is hardly any scale dependency it is clear that $ F_{2} $ is model independent at least up to present time which further implies that the approximate analytical solution of $ F_{2} $ in equation \eqref{eq:F2solnEDSappx} is indeed a very good approximation. Note that the difference between approximate and numerical solutions is less than $ 0.1\% $ at $ z = 0 $ for all the models considered.

\noindent

\section*{Bi-spectrum:}

The 3-point correlation functions of the matter energy density contrast i.e. the matter bi-spectrum is defined as

\begin{equation}
<\delta_{\vec{k}_{1}}(\eta) \delta_{\vec{k}_{2}}(\eta) \delta_{\vec{k}_{3}}(\eta)> = \delta_{D}^{(3)}(\vec{k}_{1} + \vec{k}_{2} + \vec{k}_{3}) B(\vec{k}_{1}, \vec{k}_{2}, \vec{k}_{3}; \eta).
\label{eq:BSdefn}
\end{equation}

\noindent
Expanding $ \delta_{\vec{k}}(\eta) $ upto the second order the tree-level bi-spectrum becomes \cite{LSS2}

\begin{equation}
B(\vec{k}_{1}, \vec{k}_{2}, \vec{k}_{3}; \eta) = 2 F_{2}(\vec{k}_{1}, \vec{k}_{2}, \eta) P_{lin} (k_{1}, \eta) P_{lin} (k_{2}, \eta) + 2-cycles.
\label{eq:BStree}
\end{equation}

\noindent
Note that here in the argument of the linear matter power spectrum $ \eta $ is used instead of $ \tau $ to be consistent with the argument of the matter bi-spectrum. $ \tau_{in} $ corresponds to the $ \eta_{in} = 0 $.

In Fig.~\ref{fig:bsequilateral} percentage deviation in tree-level matter bi-spectrum for the same models as in Fig.~\ref{fig:eos} from the $ \Lambda $CDM model has been plotted using equation \eqref{eq:BStree} for the equilateral configuration ($ k_{1} = k_{2} = k_{3} = k $). As because the cosmological model dependency on the solution of $ F_{2} $ is very much insignificant the behavior of the deviations in the tree-level bi-spectrum traces the deviation in the linear matter power spectrum (see Fig.~\ref{fig:ps}). And since the combination of the multiplication of the two power spectrum (at same scales for equilateral configuration and at different scales in other configurations) is involved the magnitudes of the deviations in the tree-level bi-spectrum are roughly twice the deviations in the linear matter power spectrum corresponding to the same models for the equilateral configuration. Note that for any other configurations the deviations are almost similar.

\section{Convergence power spectrum and bi-spectrum}

Weak lensing statistics is an important probe to the structure formation. The images of the background galaxies are distorted by the gravitational lensing and these distortion effects are quantified by a quantity called convergence. In Newtonian perturbation theory with the weak lensing limit, the convergence ($ \kappa $) in any particular direction $ \hat{n} $ in the sky can be related to the weighted projection of the three dimensional matter energy density contrast integrated along the line of sight, given by (\cite{wl1},\cite{wl2},\cite{wl3},\cite{wl4})

\begin{equation}
\kappa(\hat{n},\chi) = \int_{0}^{\chi} W(\chi') \delta(\hat{n},\chi') d \chi',
\end{equation}

\noindent
where $ \chi $ is the comoving distance and the weight function $ W(\chi) $ is given by

\begin{equation}
W(\chi(z)) = \frac{3}{2} \Omega_{m}^{(0)} H_{0}^{2} g(z) (1+z),
\end{equation}

\noindent
where $ g/\chi $ is the geometric lensing efficiency factor. $ g $ is given by

\begin{equation}
g(z) = \chi(z) \int_{\chi}^{\chi_{\infty}} d \chi' n'(\chi') \Big{(} 1-\frac{\chi'}{\chi} \Big{)} = \chi(z) \int_{z}^{\infty} dz' n(z') \Big{(} 1-\frac{\chi(z')}{\chi(z)} \Big{)},
\end{equation}

\noindent
where $ g(z) $ is weighted according to the source distribution $ n(z) $ (whose corresponding distribution is $ n'(\chi) $ in $ \chi $ space) with the normalization such that $ \int_{0}^{\infty} n(z) dz = 1 $. Here the source distribution is considered as $ n(z) \propto z^{b_{1}} \exp{\Big{[} - \Big{(} \frac{z}{z_{0}} \Big{)}^{b_{2}}} \Big{]} $ which after normalisation becomes

\begin{equation}
n(z) = \frac{(\frac{1 + b_{1}}{z_{0}})}{\Gamma \Big{(} \frac{1+b_{1} + b_{2}}{b_{2}} \Big{)} } \Big{(} \frac{z}{z_{0}} \Big{)}^{b_{1}} \exp{\Big{[} - \Big{(} \frac{z}{z_{0}} \Big{)}^{b_{2}}} \Big{]},
\label{eq:nOfz}
\end{equation}

\noindent
with the three parameters $ b_{1} $, $ b_{2} $ and $ z_{0} $. Here $ b_{1} = 2 $, $ b_{2} = 1.5 $ and $ z_{0} = 0.9/1.412 $ are considered which are similar to the Euclid Survey (\cite{euclid2},\cite{euclid3},\cite{nz},\cite{pwl1}).

\noindent

\begin{figure}[tbp]
\centering
\includegraphics[width=.55\textwidth]{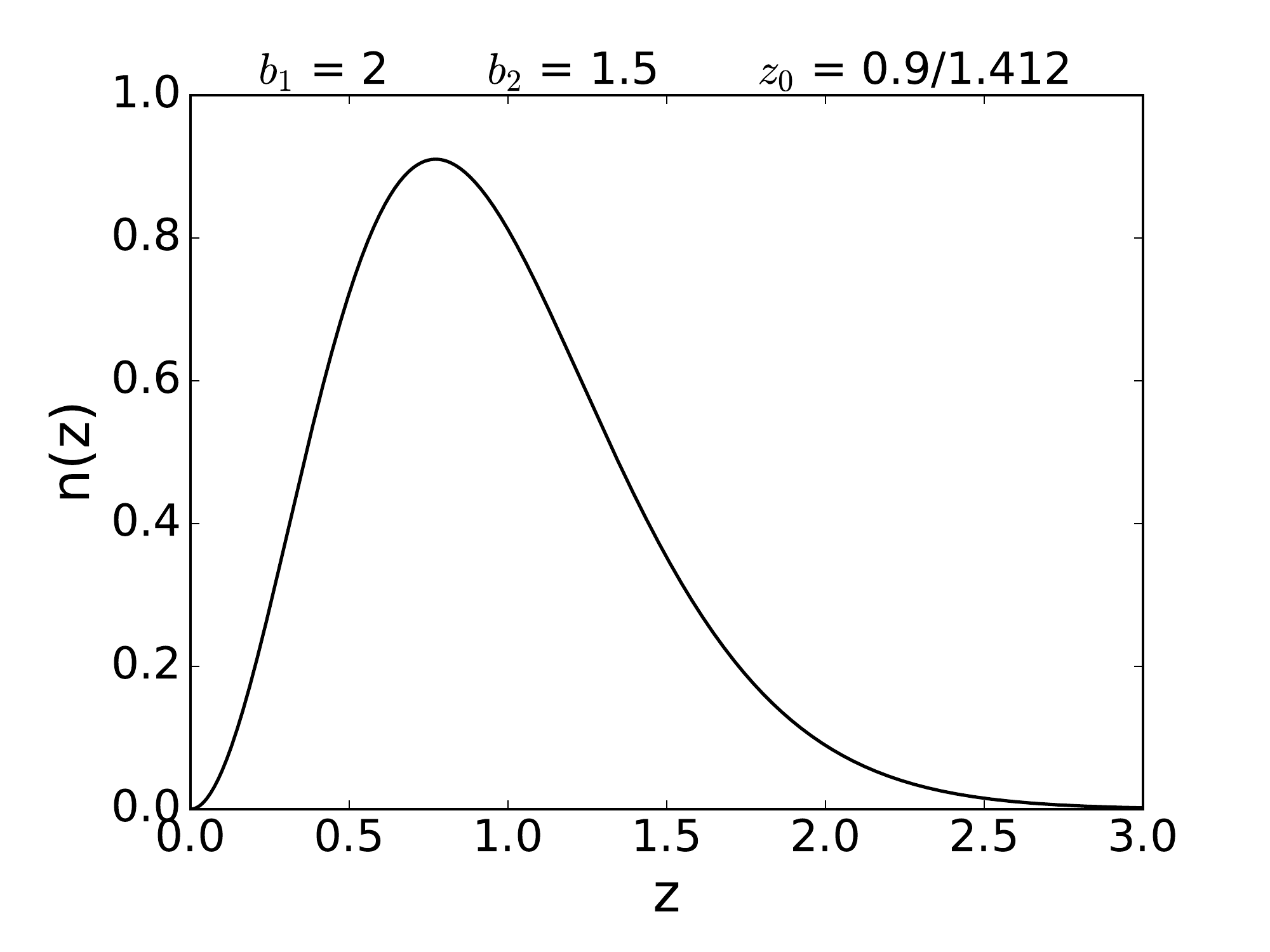}
\caption{\label{fig:nofz} $ n(z) $ vs. $ z $ plot.}
\end{figure}

\noindent
In Fig.~\ref{fig:nofz} the source distribution $ n(z) $ has been plotted with respect to the redshift $ z $ using equation \eqref{eq:nOfz}.
\\
For the purpose to study the statistical correlations of the galaxy shears, the convergence can be transformed in the multipole ($ l,m $) space given by

\begin{equation}
\kappa_{lm} = \int d \hat{n} \kappa(\hat{n},\chi) Y_{lm}^{*},
\end{equation}

\noindent
where $ Y_{lm} $ are the spherical harmonics. Assuming statistical isotropy the convergence power spectrum can be defined as

\begin{equation}
< \kappa_{l m} \kappa^{*}_{l' m'} > = \delta_{l l'} \delta_{m m'} P_{\kappa}(l),
\end{equation}

\noindent
and using Limber approximation the convergence power spectrum becomes (\cite{wl1}-\cite{wl6})

\begin{equation}
P_{\kappa}(l) = \int_{0}^{\chi_{\infty}} d \chi \frac{W^{2}(\chi)}{\chi^{2}} P(\frac{l}{\chi},\chi) = \int_{0}^{\infty} \frac{d z}{H(z)} \frac{W^{2}(z)}{\chi^{2}(z)} P \Big{(} \frac{l}{\chi(z)},z \Big{)}.
\label{eq:PSkappa}
\end{equation}

\noindent

\begin{figure}[tbp]
\centering
\includegraphics[width=.45\textwidth]{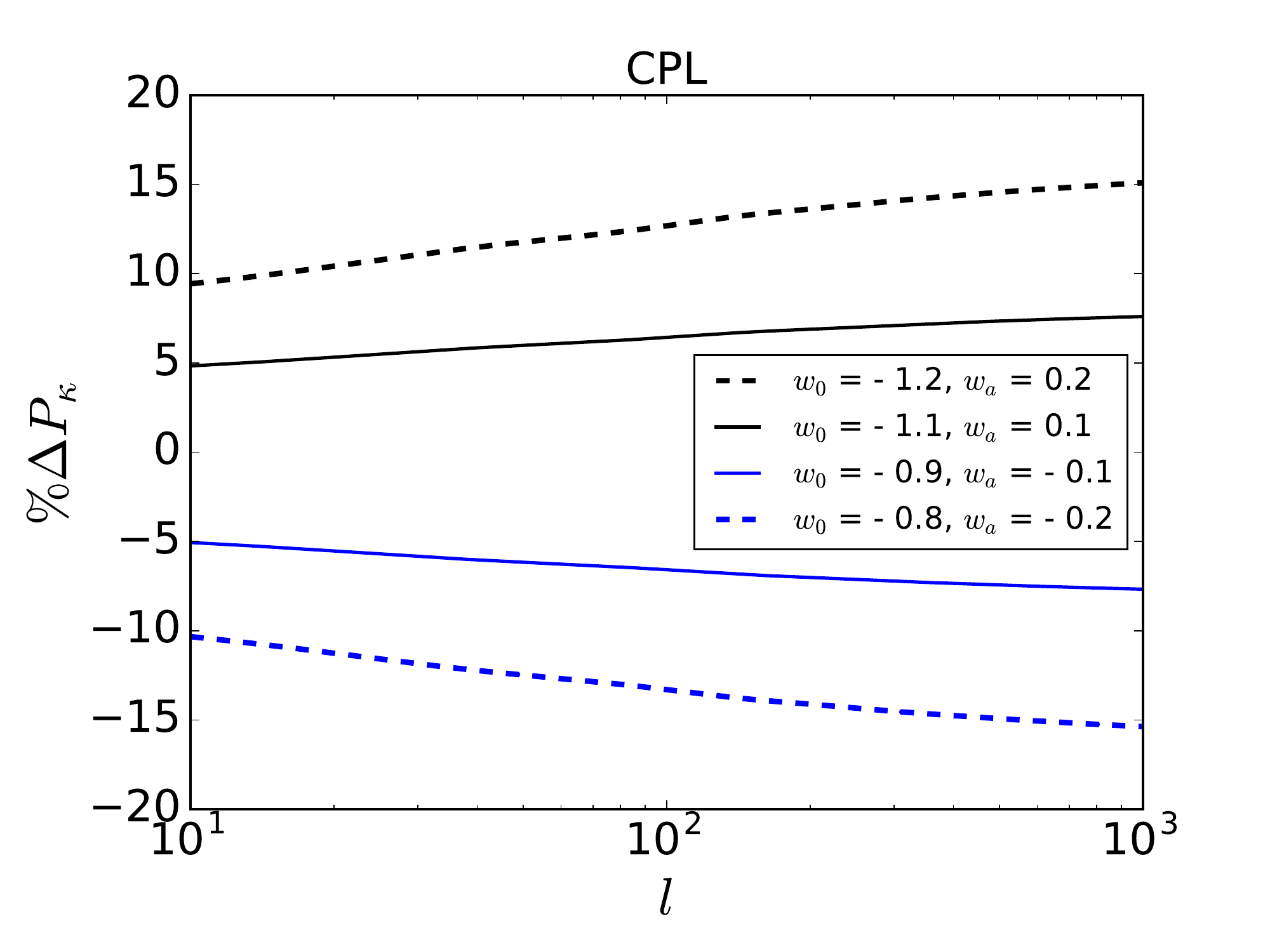}
\includegraphics[width=.45\textwidth]{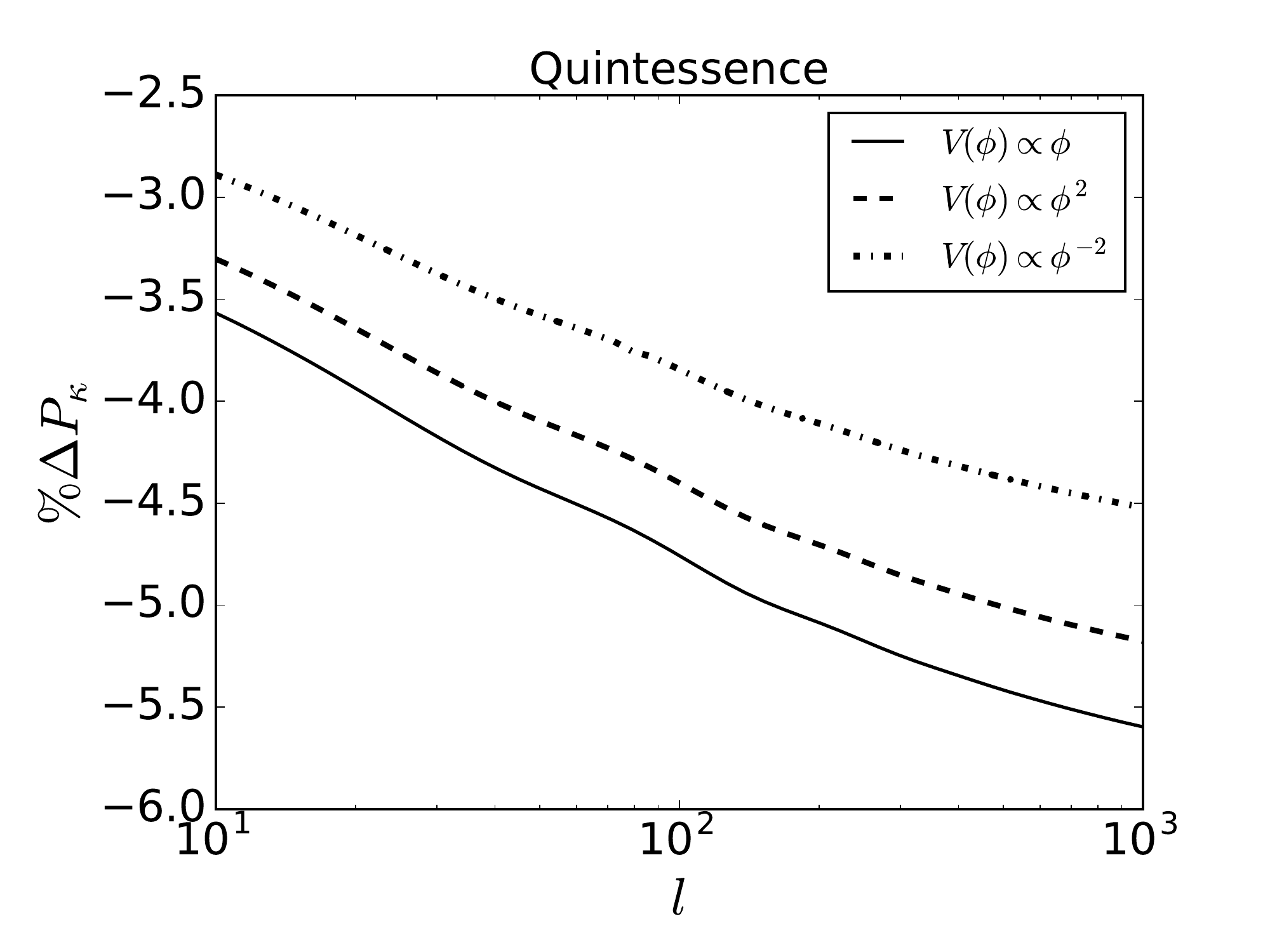}\\
\includegraphics[width=.45\textwidth]{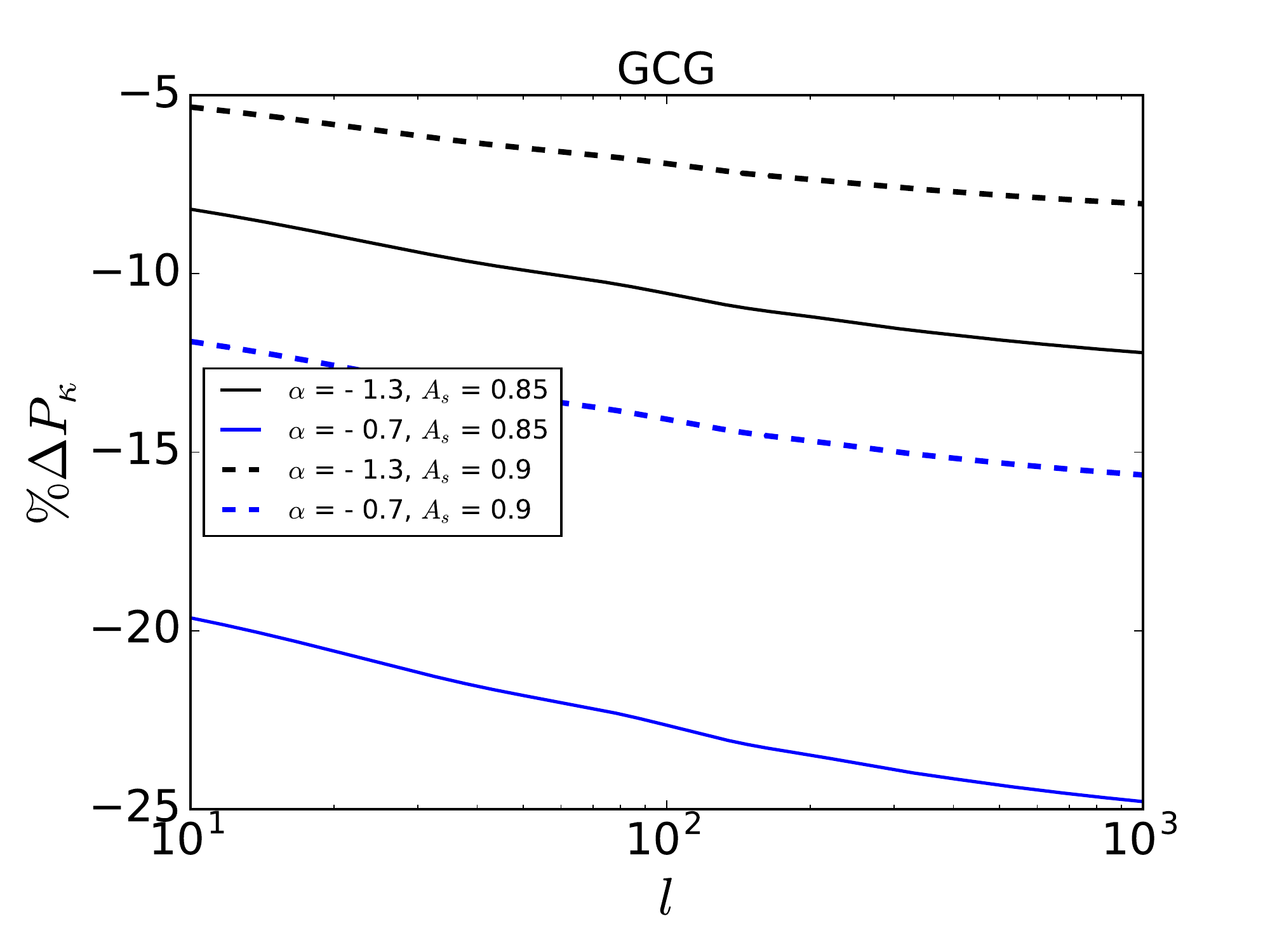}
\caption{\label{fig:wlps} Percentage deviation in convergence power spectrum from $ \Lambda $CDM model for the same models as in Fig.~\ref{fig:eos}.}
\end{figure}

In Fig.~\ref{fig:wlps} percentage deviation in convergence power spectrum for the same models as in Fig.~\ref{fig:eos} from $ \Lambda$CDM model has been plotted using equation \eqref{eq:PSkappa}. Similar to the matter power spectrum the deviations in the convergence power spectrum are positive (negative) for the phantom (non-phantom) models compared to the $ \Lambda $CDM model. The models $ w_{0} = -1.2 $, $ w_{a} = 0.2 $; $ w_{0} = -1.1 $, $ w_{a} = 0.1 $; $ w_{0} = -0.9 $, $ w_{a} = -0.1 $ and $ w_{0} = -0.8 $, $ w_{a} = -0.2 $ in CPL parametrization have the deviations nearly $ 12.5 \% $, $ 6.5 \% $, $ 6.5 \% $ and $ 13 \% $ respectively at $ l = 10^{2} $. The deviations in the thawing quintessence models are nearly $ 3.8-4.8\% $ at $ l = 10^{2} $ and the differences are very small (sub-percentage) between three potentials. The linear potential has the maximum deviation whereas the inverse-squared potential has the minimum deviation. The models $ \alpha = -1.3 $, $ A_{s} = 0.85 $; $ \alpha = -0.7 $, $ A_{s} = 0.85 $; $ \alpha = -1.3 $, $ A_{s} = 0.9 $ and $ \alpha = -0.7 $, $ A_{s} = 0.9 $ in GCG parametrization have the deviations nearly $ 10.5 \% $, $ 22.5 \% $, $ 7 \% $ and $ 14 \% $ respectively at $ l = 10^{2} $. Since all the quintessence and GCG models are always non-phantom the deviations are always negative. Similar to the matter power spectrum the deviations in the convergence power spectrum are significantly larger in the tracker models compared to the thawing models.

For all the dark energy models the deviations increase with the increasing $ l $. This is because of the fact that the comoving distances ($ \chi $) are different for different dark energy models at a fixed redshift. Hence the source matter components lie at different distances for different dark energy models at a fixed redshift z. So, the amplitude of the wave vector $ \vec{k} $ ($ k = \frac{l}{\chi} $) in the argument of the correlation functions differs accordingly for different dark energy models at a fixed $ l $ and at a fixed z. This further implies that although the deviations in the matter power spectrum for different dark energy models compared to $ \Lambda $CDM are scale independent in k-space but at $ l $-space the scale dependency arises. Now, larger the comoving distance in a dark energy model compared to $ \Lambda $CDM, more the intervening matter components and hence more the intervening matter energy density. Hence more the weak lensing signal. Since the phantom models have larger comoving distances compared to $ \Lambda $CDM, the convergence power spectrum becomes larger for phantom models compared to $ \Lambda $CDM. Due to the same reason, the opposite happens for the non-phantom models. And already in k-space, the phantom (non-phantom) models have larger (smaller) convergence power spectrum compared to $ \Lambda $CDM. So, for both the models the deviations add up in l-space. And the deviations enhance for all the models in l-space. Also, since $ \vec{l} $ is conjugate to the angular scale, $ \vec{\theta} $ the distance between two points in the source plane is proportional to $ \frac{\chi}{l} $. So, with the increasing $ l $ the distance decreases and the correlation increases at a fixed redshift. Hence, the deviations in the correlation functions for different dark energy models compared to $ \Lambda $CDM increase with increasing $ l $ at a fixed redshift. Finally, the integral effect of the deviations in the whole late time era gives rise to the enhancement in the deviations in the convergence power spectrum with increasing $ l $ for all the dark energy models compared to $ \Lambda $CDM.

Note that these deviations in the convergence power spectrum are based on the linear matter power spectrum. However, it is important to check that up to which scale the linear result is valid. The nonlinear correction to the matter power spectrum becomes significant on a particular scale depending on different dark energy models. Roughly this scale (denoted by $ k_{nl} $) can be computed by using the fact that on this scale the linear matter power spectrum satisfies the condition given by $ \frac{k_{nl}^{3}}{2 \pi^{2}} P_{lin} (k_{nl},z) = 1 $ (for detailed discussions see equation (2.9) in the reference \citep{LSS10}). For $ k<k_{nl} $ we can safely use the linear matter power spectrum whereas for $ k>k_{nl} $ it is important to take into account the nonlinear correction to the matter power spectrum. So, $ k_{nl} $ denotes the upper limit of the linear scale in k-space up to which linear matter power spectrum is valid. The value of $ k_{nl} $ is typically $ 0.23 $ to $ 0.26 $ $ h Mpc^{-1} $ at redshift $ z = 0 $ for all the models considered. The value of $ k_{nl} $ increases with increasing redshifts i.e. at lower redshifts nonlinearity becomes significant earlier in k-space compared to the higher redshifts.

Now the nonlinear correction to the matter power spectrum induces the nonlinear correction to the convergence power spectrum through \eqref{eq:PSkappa}. The corresponding upper limit in the multipole scale is typically $ l = 30 $ to $ 40 $ where the nonlinear correction in the convergence power spectrum is maximum up to $ 1\% $ for all the models considered.

Now if we want to see how the deviations in the convergence power spectrum changes in the nonlinear regime we have to compute nonlinear matter power spectrum. To compute nonlinear matter power spectrum for the dark energy models with evolving e.o.s strong N-body simulations are required. However, there are few approximate semi-analytical methods to study it. One of such kind is based on the resummation method (for details see \citep{LSS8},\citep{LSS9}). Although in this paper we are not computing nonlinear matter power spectrum to get an idea about the nonlinear matter power spectrum we follow reference \citep{LSS9}. In \citep{LSS9} the authors have shown the deviations in the matter power spectrum both in linear and in nonlinear regimes in figure (5) (in \citep{LSS9}) for different dark energy models compared to the $ \Lambda $CDM (see figure (5) and discussions about it in \citep{LSS9}). They have taken both the smooth (sound speed of perturbation, $ c_{s} = 1 $) and clustering ($ c_{s} = 0 $) dark energy models. Their result shows that the deviations increase with increasing values of k for smooth dark energy models both for phantom and non-phantom models. Since in this paper we have considered smooth dark energy only, the result in \citep{LSS9} should valid for our cases also. So, the deviations in the matter power spectrum are expected to be enhanced in the nonlinear regime compared to the linear regime.

However their method (in \citep{LSS9}) has some limitations such as at redshift $ z = 0 $ the results are within $ 2\% $ and $ 5\% $ errors for $ k\leq 0.25 h Mpc^{-1} $ and $ k\leq 0.6 h Mpc^{-1} $ respectively (see figure (12) and discussions about it in \citep{LSS9}). These errors decrease with increasing redshift. Although these errors are quoted for the $ \Lambda $CDM model we assume that errors should not change drastically for the other dark energy models. After compensating these errors till we can say that the deviations in the nonlinear regime are higher compared to the linear regime and this enhancement decreases with increasing redshifts (see figure (5) in \citep{LSS9}). 

Although these enhancements in the nonlinear regime compared to the linear regime in the deviations of the matter power spectrum decrease drastically with increasing redshifts but due to the source distribution $ n(z) $ (see Fig.~\ref{fig:nofz}) the integration in equation \eqref{eq:PSkappa} is effective at lower redshifts (up to redshift $ 2 $ to $ 3 $). So, we can expect to get larger deviations in the convergence power spectrum in nonlinear regime compared to the linear regime. This result should valid at least up to the multipole scale $ 10^{2} $ order in $ l $ by following the result in figure (5) in \citep{LSS9}. We quote that following the method to compute nonlinear matter power spectrum as in \citep{LSS9} the nonlinear corrections in the convergence power spectrum are up to $ 2\% $, $ 5\% $ and $ 10\% $ at $ l=50 $, $ 75 $ and $ 100 $ respectively.

So, to summarise, we can expect that the deviations in the convergence power spectrum would be larger in the nonlinear regime.

\section*{Convergence bi-spectrum:}

The convergence bi-spectrum can be defined as

\begin{equation}
< \kappa_{l_{1} m_{1}} \kappa_{l_{2} m_{2}} \kappa_{l_{3} m_{3}} > = \left( \begin{smallmatrix} l_{1} \hspace{0.5 cm} & l_{2} & \hspace{.5 cm} l_{3} \vspace{0.5 cm} \\ m_{1} \hspace{.5 cm} & m_{2} & \hspace{.5 cm} m_{3} \end{smallmatrix} \right) B_{l_{1} l_{2} l_{3}}^{\kappa},
\end{equation}

\noindent
where in the r.h.s Wigner 3-j symbol has been used and the above convergence bi-spectrum (which is full-sky) is related to the flat-sky bi-spectrum given by

\begin{equation}
B_{l_{1} l_{2} l_{3}}^{\kappa} = \left( \begin{smallmatrix} l_{1} \hspace{0.5 cm} & l_{2} & \hspace{.5 cm} l_{3} \vspace{0.5 cm} \\ 0 \hspace{.5 cm} & 0 & \hspace{.5 cm} 0 \end{smallmatrix} \right) \sqrt{\frac{(2 l_{1} + 1) (2 l_{2} + 1) (2 l_{3} + 1)}{4 \pi}} B_{\kappa}(\vec{l}_{1},\vec{l}_{2},\vec{l}_{3};z),
\end{equation}

\noindent

\begin{figure}[tbp]
\centering
\includegraphics[width=.45\textwidth]{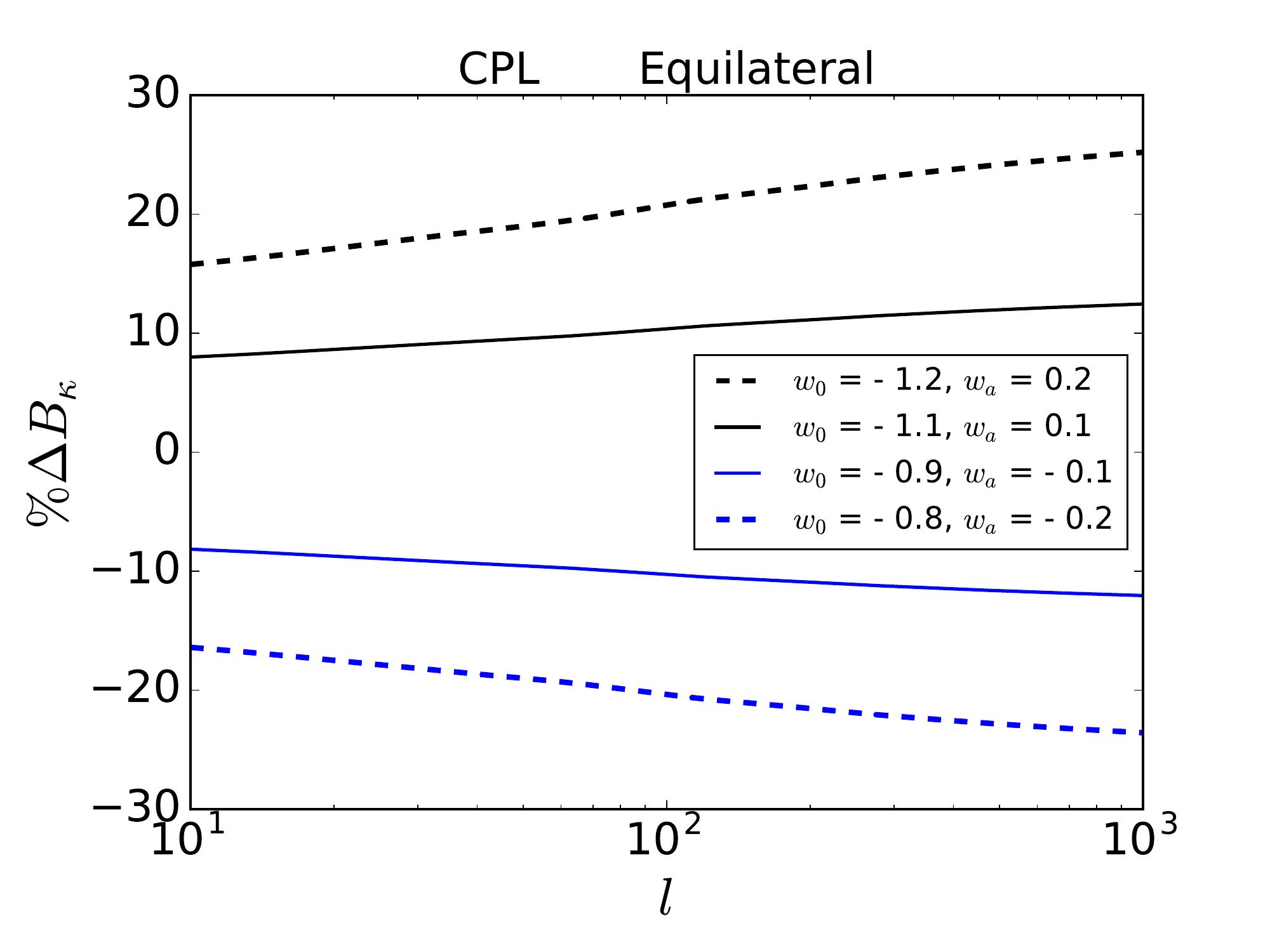}
\includegraphics[width=.45\textwidth]{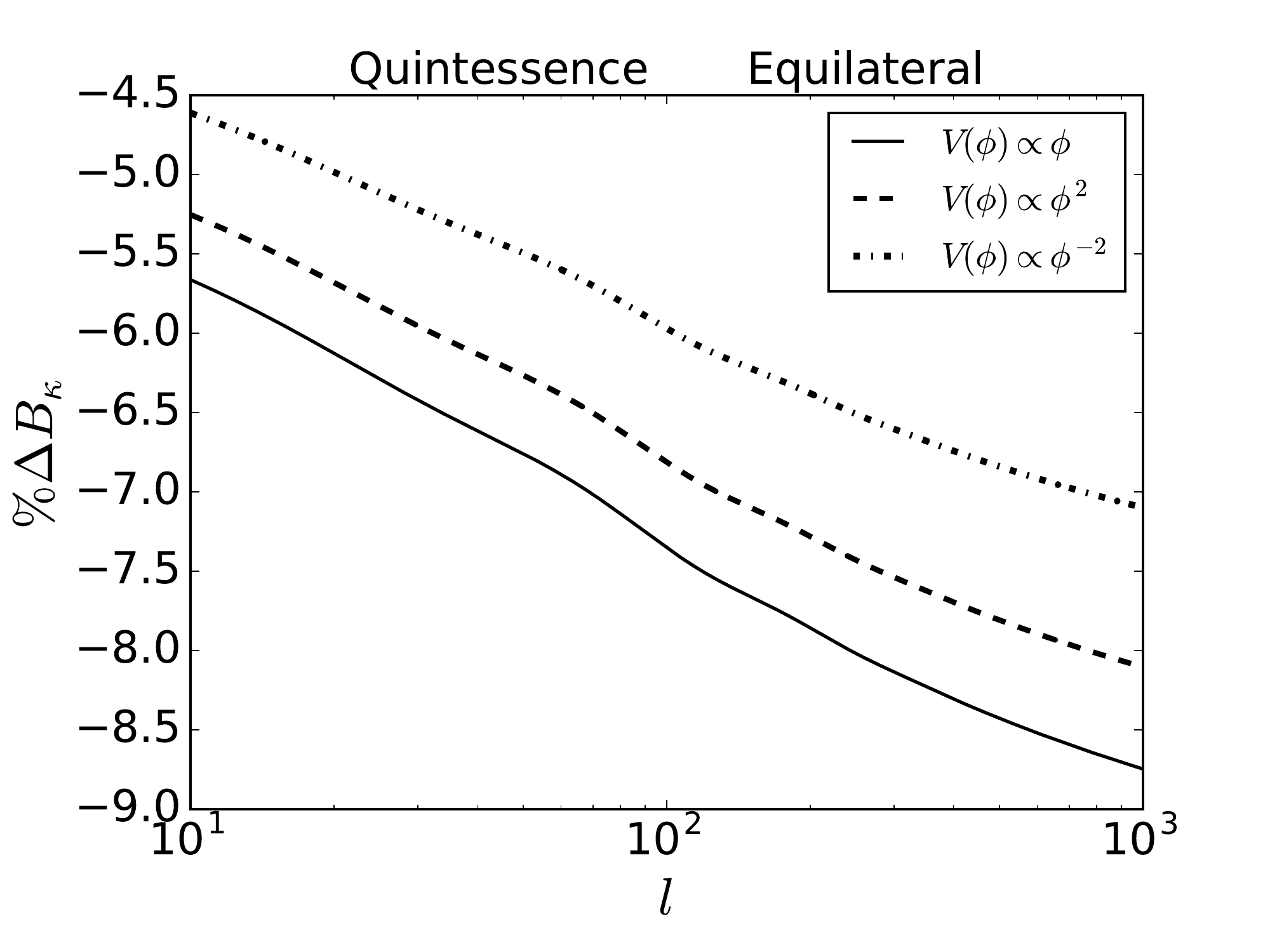}\\
\includegraphics[width=.45\textwidth]{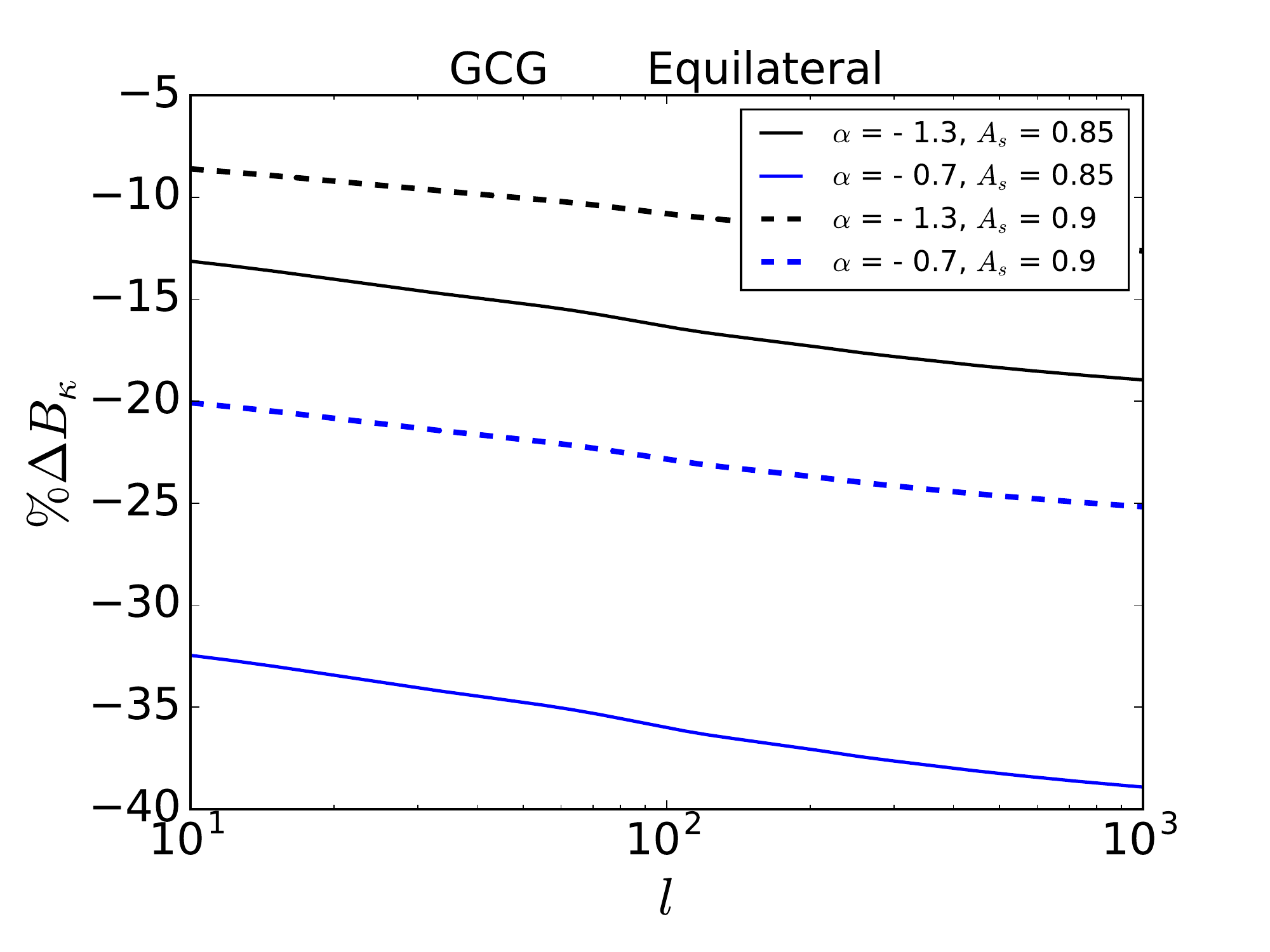}
\caption{\label{fig:wlbsequilateral} Percentage deviation in convergence bi-spectrum from the $ \Lambda $CDM model for the same models as in Fig.~\ref{fig:eos} for the equilateral configuration.}
\end{figure}

\noindent
and similar to the convergence power spectrum, the convergence bi-spectrum (using Limber approximation) can be written as

\begin{equation}
B_{\kappa}(\vec{l}_{1},\vec{l}_{2},\vec{l}_{3};z) = \int_{0}^{\infty} \frac{d z}{H(z)} \frac{W^{3}(z)}{\chi^{4}(z)} B \Big{(} \frac{\vec{l}_{1}}{\chi(z)},\frac{\vec{l}_{2}}{\chi(z)},\frac{\vec{l}_{3}}{\chi(z)};z \Big{)},
\label{eq:BSkappaflatsky}
\end{equation}

\noindent
with $ \vec{l}_{1} + \vec{l}_{2} + \vec{l}_{3} = 0 $. Since, in our subsequent plots, convergence bi-spectrum for different dark energy models have been compared with the $ \Lambda$CDM model, one need not consider the full sky convergence bi-spectrum. Instead of flat sky convergence bi-spectrum in equation \eqref{eq:BSkappaflatsky} is sufficient to compare as the ratio of convergence bi-spectrum between two dark energy models is the same for either full sky or flat sky case.
\\
In Fig.~\ref{fig:wlbsequilateral} percentage deviation in convergence bi-spectrum for the same models as in Fig.~\ref{fig:eos} from the $ \Lambda $CDM model has been plotted using equation \eqref{eq:BSkappaflatsky} for the equilateral configuration ($ l_{1} = l_{2} = l_{3} = l $). The behavior of the deviations are similar as in the convergence power spectrum (see Fig.~\ref{fig:wlps}) but the magnitudes of the deviations become nearly $ 1.5 $ times the deviations in the convergence power spectrum corresponding to the same models. Similar to the convergence power spectrum the deviations in convergence bi-spectrum for all the dark energy models increase with the increasing $ l $. This is because of the same reason described in the context of the convergence power spectrum.

\noindent

\begin{figure}[tbp]
\centering
\includegraphics[width=.45\textwidth]{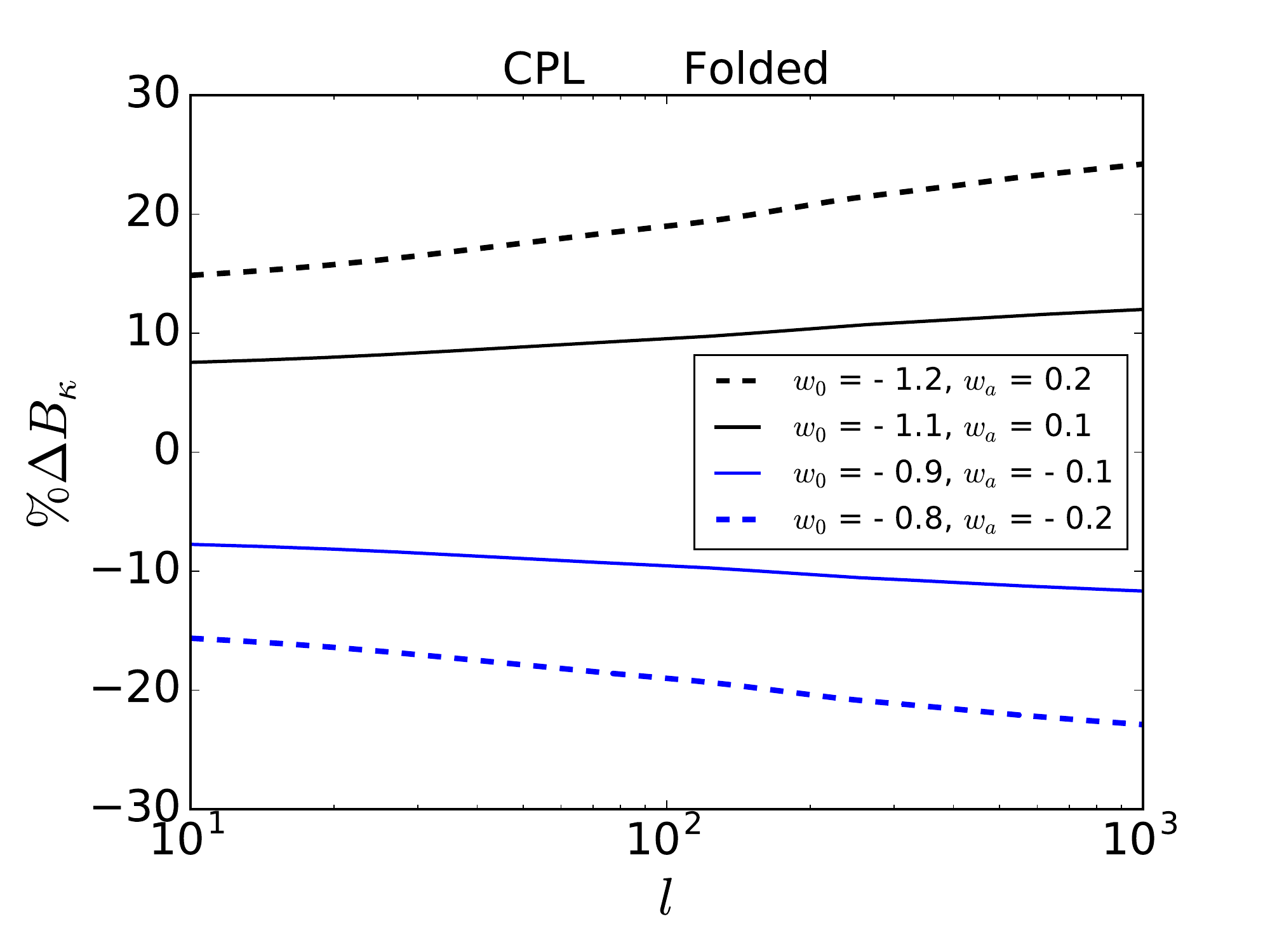}
\includegraphics[width=.45\textwidth]{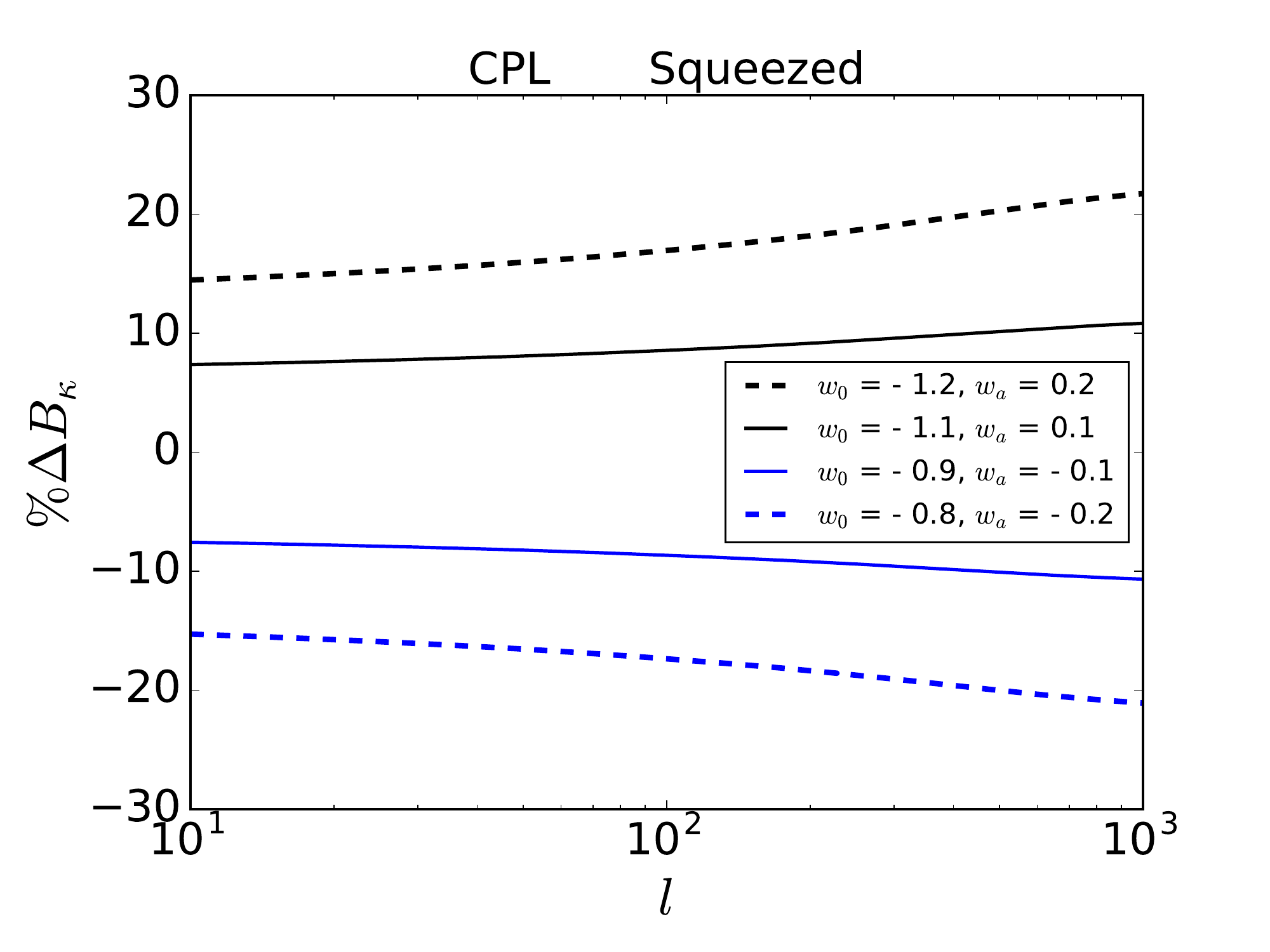}\\
\includegraphics[width=.45\textwidth]{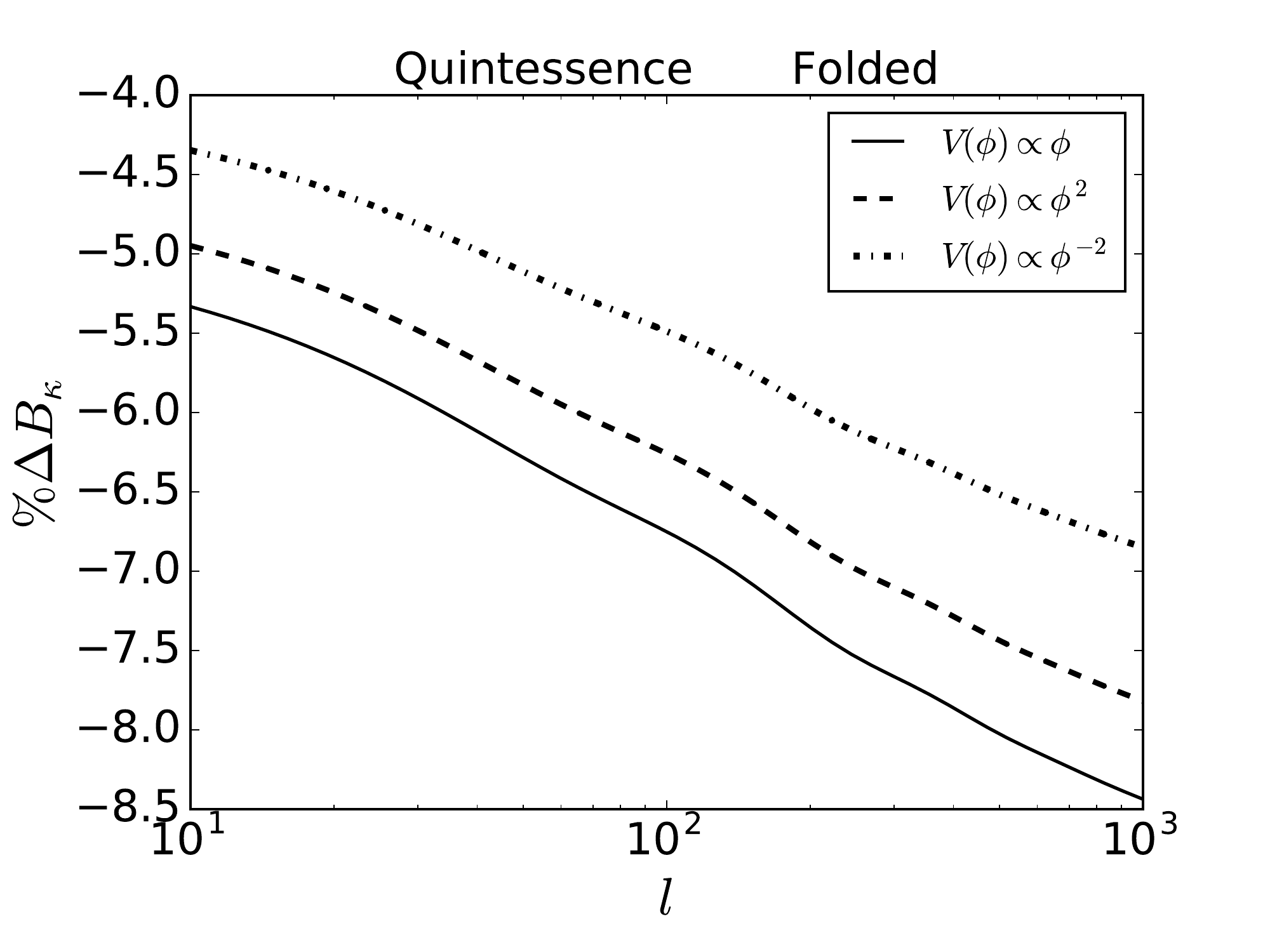}
\includegraphics[width=.45\textwidth]{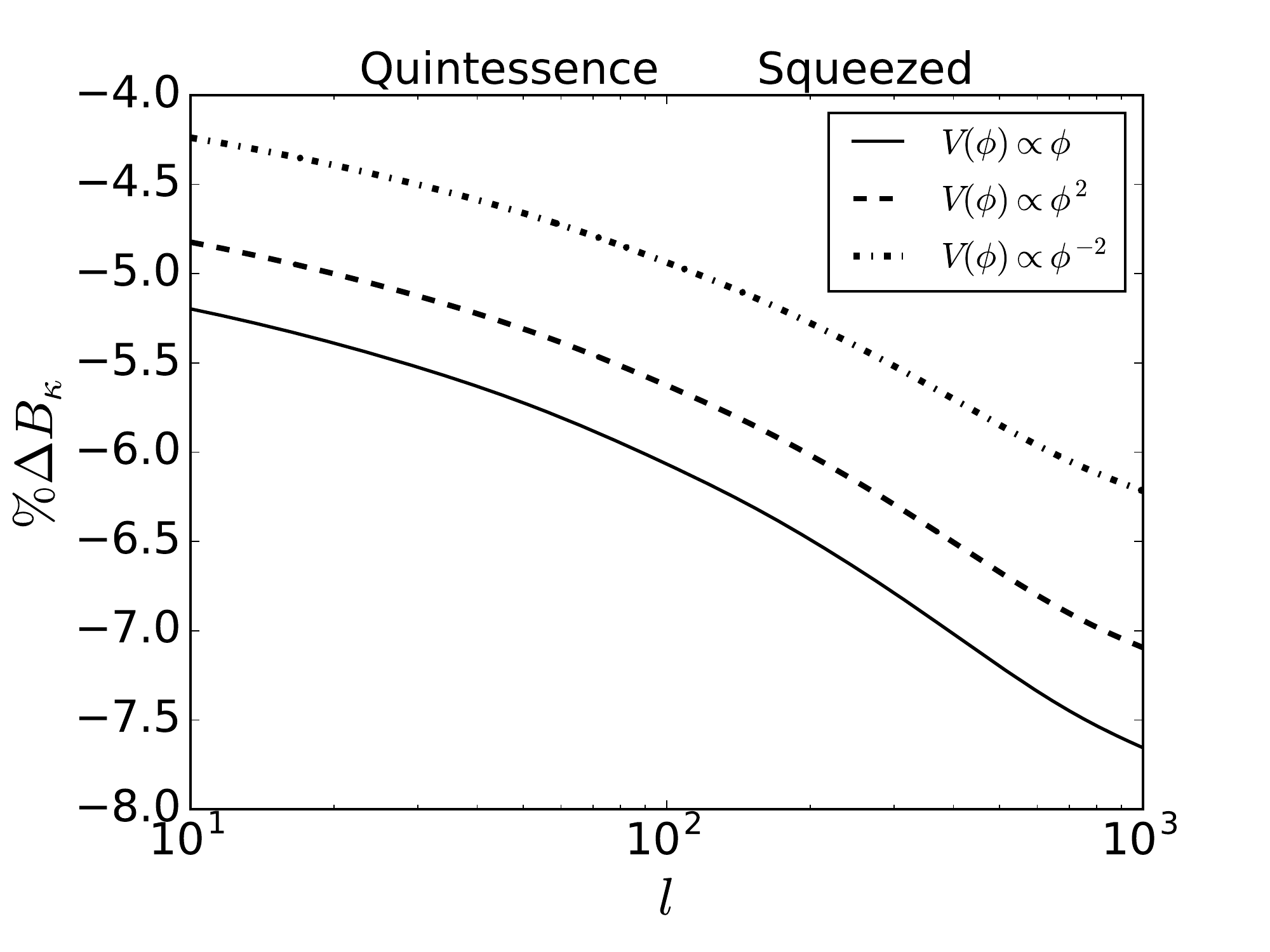}\\
\includegraphics[width=.45\textwidth]{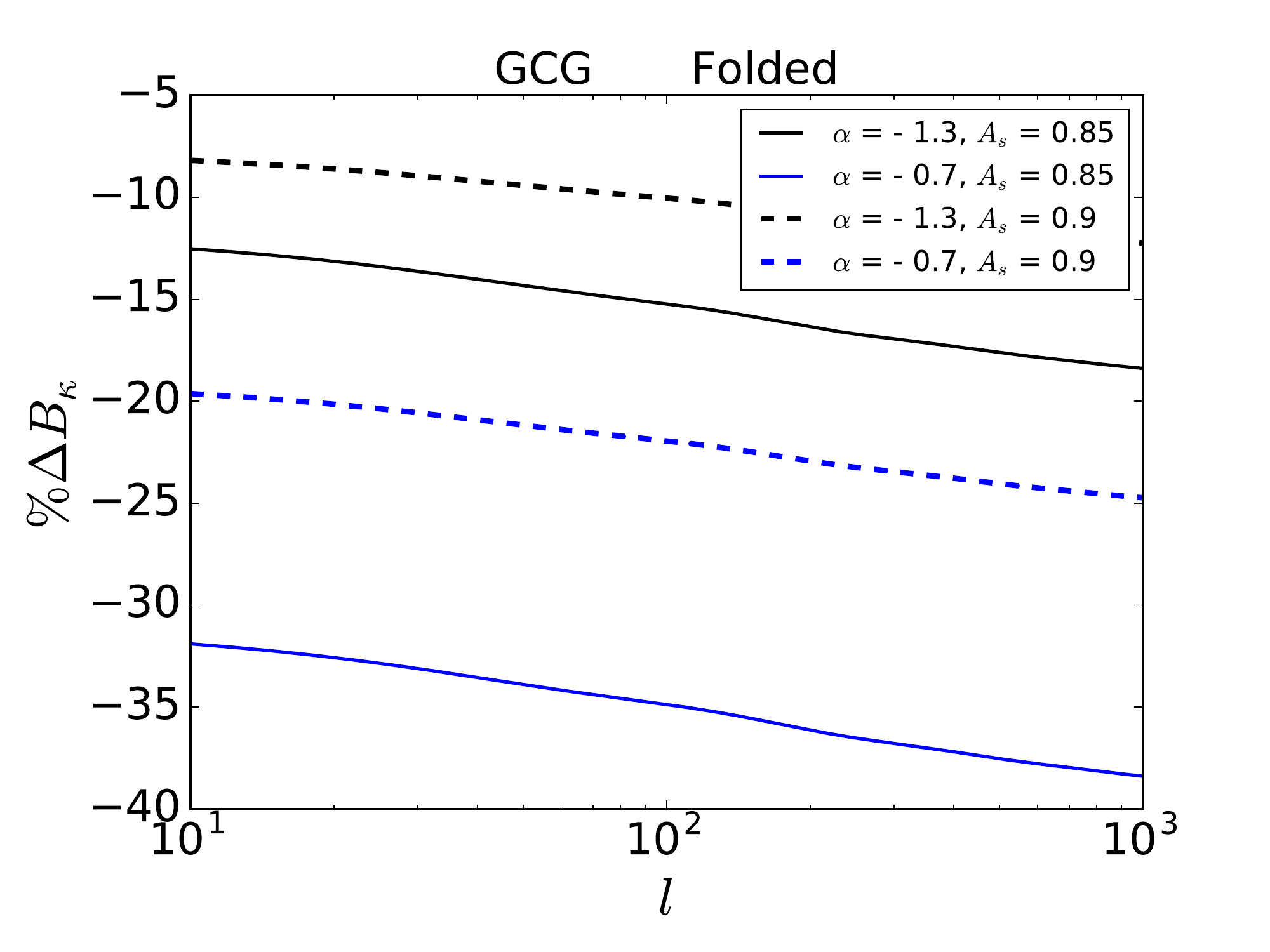}
\includegraphics[width=.45\textwidth]{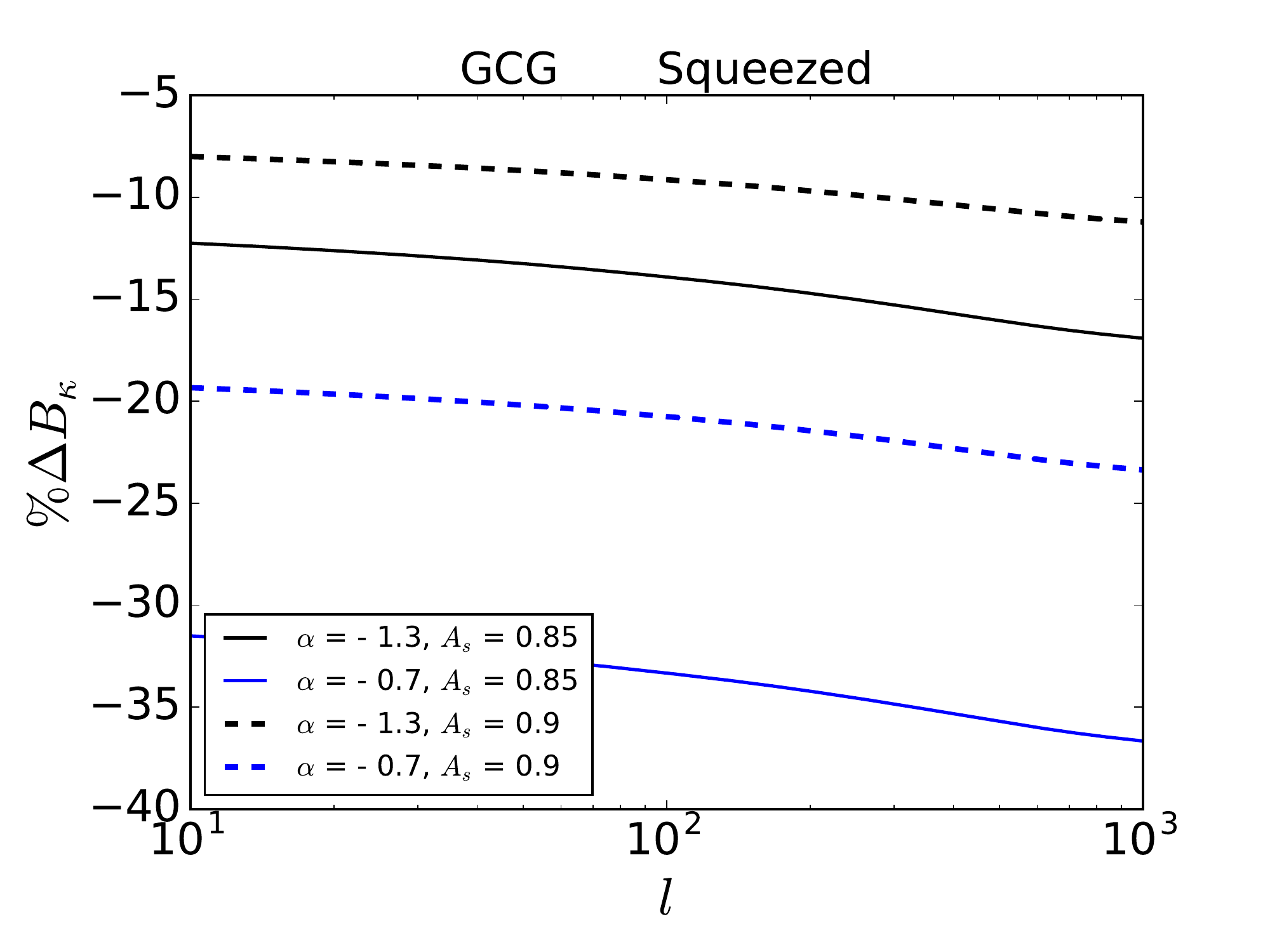}
\caption{\label{fig:wlbsothers} Percentage deviation in the convergence bi-spectrum from $ \Lambda $CDM model for the same models as in Fig.~\ref{fig:eos} for the folded and squeezed configurations.}
\end{figure}

\noindent
So far in all the bi-spectrum plots equilateral configuration has been considered. To see the deviations in other configurations in Fig.~\ref{fig:wlbsothers} two different configurations have been considered for the same models as in Fig.~\ref{fig:eos} and the configurations are folded shape ($ l_{1} = 2 l_{2} = 2 l_{3} = l $) and squeezed shape ($ l_{1} = l_{2} = 20 l_{3} = l $) respectively. Comparing figures ~\ref{fig:wlbsequilateral} and ~\ref{fig:wlbsothers}, it is clear that the deviations are almost same in the three different configurations and this is true for the matter bi-spectrum also. So, it can be concluded that the deviations in the bi-spectrum are highly insensitive to the shape of the bi-spectrum at least at the tree-level.

Similar to the convergence power spectrum, the nonlinear effects are also important in the convergence bi-spectrum. The nonlinear correction in the convergence bi-spectrum comes from the nonlinear correction in the matter bi-spectrum through equation \eqref{eq:BSkappaflatsky}. We have already mentioned that there is a scale, $ k_{nl} $ up to which the linear matter power spectrum is valid. But for the case of matter bi-spectrum the nonlinear correction is significant for the lower k values compared to $ k_{nl} $. This is due to the mode couplings in different triangular configurations in the matter bi-spectrum.

To check the range of the scale where the higher order correction to the tree-level matter bi-spectrum becomes significant we need to compute the higher order matter bi-spectrum. However, in this paper, we are not computing the higher order matter bi-spectrum but to have an idea about the higher order matter bi-spectrum we follow the procedure described in the reference \citep{LSS10} to compute the matter bi-spectrum in the nonlinear regime. Following this reference we can see that the higher order correction to the tree-level matter bi-spectrum in equation \eqref{eq:BStree} is incorporated by two ways: (I) through the correction to $ F_{2} $ in equation \eqref{eq:F2solnEDSappx} and (II) through the correction in the linear matter power spectrum simply by replacing it with the nonlinear matter power spectrum.

There are different parameters which are involved in the correction to the $ F_{2} $ in equation \eqref{eq:F2solnEDSappx} (see \citep{LSS10} for the detailed discussions). Among different parameters in the correction to $ F_{2} $, mainly the $ \sigma_{8} $ parameter is cosmological model dependent (see equation (2.12) with table 2 in \citep{LSS10}). The $ \sigma_{8} $ parameter is proportional to the linear growth function. However because of the small power in $ \sigma_{8} $ ($ \sigma_{8}^{a_{6}} $ with $ a_{6}=-0.575 $) the deviations due to the correction in $ F_{2} $ between different models are not significant. We quote that taking into account only the correction in $ F_{2} $ (i.e. neglecting the second correction) the changes in the deviations in convergence bi-spectrum for all the models considered compared to $ \Lambda $CDM are upto $ 2\% $ and $ 6\% $ at $ l=100 $ and $ 1000 $ respectively for all three configurations. So, the correction in the deviations in the matter bi-spectrum is negligible due to the correction in $ F_{2} $ alone. Only the correction in the linear matter power spectrum (i.e. the second correction) is significant.

Now including the second correction i.e. when we include the correction by replacing the linear matter power spectrum with nonlinear matter power spectrum (computed by the method described in \citep{LSS9}) the total correction in the convergence bi-spectrum is roughly $ 1.5 $ times the correction in the convergence power spectrum. We quote that (following the method to compute nonlinear matter power spectrum as in \citep{LSS9}) the nonlinear corrections in the convergence bi-spectrum are up to $ 4\% $, $ 9\% $ and $ 17\% $ at $ l=50 $, $ 75 $ and $ 100 $ respectively.

We have already mentioned that the deviations in the matter power spectrum for different models compared to $ \Lambda $CDM are expected to be slightly larger in the nonlinear regimes. This incorporates that deviations in the convergence bi-spectrum are also expected to be slightly larger by the correction in the matter bi-spectrum.

In summary, accuracies in the results of convergence power spectrum are $ 2\% $ and $ 5\% $ at $ l=50 $ and $ 75 $ respectively whereas accuracies in the results of convergence bi-spectrum are $ 2\% $ and $ 5\% $ at $ l=40 $ and $ 50 $ respectively. However from the above discussions it is clear that for higher values of $ l $ (at least up to $ 10^{2} $ order) results of this paper are good enough to give a preliminary idea about the detectability of different dark energy models than $ \Lambda $CDM through convergence power spectrum and bi-spectrum since it is expected that at least the deviations should not be reduced in the nonlinear regime compared to the linear regime.

\section{Conclusion}

The weak lensing statistics is a powerful tool to probe the dark energy and structure formation in the Universe. The evolution of the background quantities, as well as the perturbated quantities in any dark energy model, can be measured through weak lensing with high accuracy by the current and future surveys like DES, LSST, Euclid, WFIRST etc.

Firstly, three types of dark energy models with evolving equation of state have been considered to study their detectability in the structure formation through weak lensing statistics. One is the most popular dark energy parametrization named CPL parametrization. The next one is the minimally coupled canonical quintessence scalar field dark energy candidate where thawing class (where initially equation of state of the dark energy is close to $ -1 $ due to the large Hubble damping and at late times Hubble damping decreases due to the expansion of the Universe and the equation of state of the scalar fields increases from $ -1 $ slowly) of scalar field models with linear, squared and inverse-squared potentials have been considered as because a broad class of potentials can give the proper thawing behaviour whereas to get tracker behaviour (where initially scalar fields mimic the background matter density i.e. $ w $ is close to $ 0 $ and at late times the equation of state of the scalar field freezes towards $ -1 $) there are few potentials (e.g. cosine hyperbolic and double exponential potentials etc.) which can give proper tracker behaviour. But to show the thawing versus tracker results next a simple parametrization named GCG parametrization has been considered where $ 1 + \alpha < 0 $ gives thawing behaviour and $ 1 + \alpha > 0 $ gives tracker behaviour (see Fig.~\ref{fig:eos}).

Next, to compare all the dark energy models properly the normalisation has been taken such that at present time the values of background quantities ($ H_{0} $, $ \Omega_{b}^{(0)} $ and $ \Omega_{m}^{(0)} $) are same. In the perturbation level we have taken the same initial matter power spectrum for all the models by fixing $ n_{s} $ and $ A_{s} $.

For all the dark energy models the deviations in the linear matter power spectrum and in the tree-level matter bi-spectrum from $ \Lambda $CDM model are scale independent. The deviations are largest at $ z = 0 $ compared to $ z > 0 $ and the deviations decrease with increasing redshift.

For all the models the deviations in all the perturbed quantities considered (linear growth function, linear matter power spectrum, tree-level matter bi-spectrum, convergence power spectrum and convergence bi-spectrum) compared to $ \Lambda $CDM model the deviations are positive (negative) for phantom (non-phantom) models. Since the quintessence and the GCG models are always non-phantom the deviations are always negative. The quintessence models of thawing class show that the deviations are smaller compared to the other models in CPL and GCG parametrizations and the differences between different potentials are very small (sub-percentage level). It can also be noted that the deviation is the largest for linear potential whereas it is the smallest for the inverse-squared potential. The deviations in the GCG parametrization show that the deviations are significantly larger for tracker models compared to the thawing models.

The deviations in the tree level matter bi-spectrum for all the dark energy models from $ \Lambda $CDM model have the similar behavior as in the linear matter power spectrum but the magnitude of the deviations are approximately twice the deviations in the linear matter power spectrum which is quite obvious from eq. \eqref{eq:BStree}.

The deviations in the convergence power spectrum and in the convergence bi-spectrum for all the dark energy models from $ \Lambda $CDM model increase with increasing $ l $. The deviation in the convergence bi-spectrum from $ \Lambda $CDM model is nearly $ 1.5 $ times the deviation in the convergence power spectrum from $ \Lambda $CDM model corresponding to the same model. The deviations in the convergence bi-spectrum from $ \Lambda $CDM model are almost same for equilateral ($ l_{1} = l_{2} = l_{3} = l $), folded (here $ l_{1} = 2 l_{2} = 2 l_{3} = l $) and squeezed (here $ l_{1} = l_{2} = 20 l_{3} = l $) configurations for all the models accordingly.

To summarise, the tracker models are significantly more probable to be distinguished from $ \Lambda $CDM model compared to the thawing models.

In this paper linear power spectrum and tree level bi-spectrum have been studied. This work can be extended to the nonlinear matter power spectrum, the higher order correction to the tree level matter bi-spectrum and also to the higher order correlation functions (tri-spectrum etc.) by studying the higher order perturbation theory in more details and corresponding to the convergence power spectrum, bi-spectrum etc.

\acknowledgments

The author would like to acknowledge CSIR, Govt. of India for financial support through SRF scheme (No:09/466(0157)/2012-EMR-I). The author would also like to thank Anjan Ananda Sen for exclusive discussions.



\end{document}